\documentclass[useAMS,usenatbib,twocolumn]{mn2e}
\usepackage{epsfig,amsmath,amssymb,bm}
\def\bea{\begin{eqnarray}}
\def\eea{\end{eqnarray}}
\def\be{\begin{equation}}
\def\ee{\end{equation}}

\title{Local Group dSph radio survey with ATCA (II): Non-thermal diffuse emission}

\author[Regis, Richter, Colafrancesco, Profumo, de Blok, Massardi]{{\large Marco Regis$^{1,2}$, Laura Richter$^3$, Sergio Colafrancesco$^4$, Stefano Profumo$^{5,6}$, W.J.G. de Blok$^{7,8,9}$ and Marcella Massardi$^{10}$}\\
$^1$Dipartimento di Fisica, Universit\`{a} di Torino, via P. Giuria 1, I--10125 Torino, Italy\\
$^2$Istituto Nazionale di Fisica Nucleare, Sezione di Torino, via P. Giuria 1, I--10125 Torino, Italy\\
$^3$SKA South Africa, 3rd Floor, The Park, Park Road, Pinelands, 7405, South Africa\\
$^4$School of Physics, University of the Witwatersrand, 2050 Johannesburg, South Africa\\
$^5$Department of Physics, University of California, 1156 High St., Santa Cruz, CA 95064, USA\\
$^6$Santa Cruz Institute for Particle Physics, Santa Cruz, CA 95064, USA\\
$^7$Netherlands Institute for Radio Astronomy (ASTRON), Postbus 2, 7990 AA Dwingeloo, The Netherlands\\
$^8$Astrophysics, Cosmology and Gravity Centre, Department of Astronomy, University of Cape Town, Private Bag X3, \\Rondebosch 7701, South Africa\\
$^9$Kapteyn Astronomical Institute, University of Groningen, PO Box 800, 9700 AV, Groningen, The Netherlands\\
$^{10}$INAF - Istituto di Radioastronomia, Via Gobetti 101, I-40129, Bologna, Italy.\\
E-mail:regis@to.infn.it,laura@ska.ac.za}

\begin{document}

\date{}

\maketitle

\begin{abstract}
Our closest neighbours, the Local Group dwarf spheroidal (dSph) galaxies, are extremely quiescent and dim objects, where thermal and non-thermal diffuse emissions lack, so far, of detection.
In order to possibly study the dSph interstellar medium, deep observations are required.
They could reveal non-thermal emissions associated with the very-low level of star formation, or to particle dark matter annihilating or decaying in the dSph halo.
In this work, we employ radio observations of six dSphs, conducted with the Australia Telescope Compact Array in the frequency band 1.1-3.1 GHz, to test the presence of a diffuse component over typical scales of few arcmin and at an rms sensitivity below 0.05 mJy/beam.
We observed the dSph fields with both a compact array and long baselines. Short spacings led to a synthesized beam of about 1 arcmin and were used for the extended emission search. The high-resolution data mapped background sources, which in turn were subtracted in the short-baseline maps, to reduce their confusion limit. 
We found no significant detection of a diffuse radio continuum component.
After a detailed discussion on the modelling of the cosmic-ray (CR) electron distribution and on the dSph magnetic properties, we present bounds on several physical quantities related to the dSphs, such that the total radio flux, the angular shape of the radio emissivity, the equipartition magnetic field, and the injection and equilibrium distributions of CR electrons. 
Finally, we discuss the connection to far-infrared and X-ray observations.
\end{abstract}

\maketitle


\section{Introduction}
\label{sec:intro}

The cold dark matter (CDM) paradigm consists in postulating a dark matter (DM) component with small velocity dispersion in the early Universe.
A straightforward consequence is the prediction of an abundance of structures on sub-galactic scales.
The CDM model has been collecting enormous successes in explaining large scale observations, over a wide range of redshifts.
On the other hand, a number of tensions have emerged in the description of the smallest scales, such as the innermost regions of galactic DM halos and the Local Group dwarf galaxy satellites.
The CDM controversies include the so-called ``cusp-core'', ``missing satellites'', and ``too big to fail'' problems (see, e.g., \citep{Weinberg:2013aya} for a recent review).

Dwarf spheroidal (dSph) galaxies, and in particular the Milky Way (MW) satellites, are key actors in all of these issues~\citep{Bullock:2009au}.
Indeed, the central density profiles of dSphs have been suggested to be much shallower than predicted in the CDM scenario (see, e.g., \citep{Walker:2012td} and references therein).
On the other hand, while the cusp-core controversy appears to be evident in low surface brightness spiral galaxies (where the profile is derived from rotation curves), some uncertainties in the description of the gravitational potential from the observed dSph velocity dispersion (in particular related to the anisotropy of the stellar velocity) leave the question still open in the case of dSphs.

In cosmological N-body CDM simulations, the formation of MW-like halos preserves a large amount of subhalos (formed in early-time collapses on small scales).
This leads to the prediction of thousands of MW satellites which is at odds with the few tens observed. The disagreement still persists even after the recent discovery of
about 15 new ultra-faint dSphs (UDS). However, taking into account the completeness limits of the SDSS observations, this issue can be alleviated. Indeed, applying luminosity bias corrections, \citep{Tollerud:2008ze} found that few hundreds of UDS should be present within the MW virial radius.

In the same dissipationless simulations, more than about six massive satellites with maximum circular velocity greater than 30 km/sec are predicted~\citep{BoylanKolchin:2011de}, while no similar objects have been observed from the MW or Andromeda's satellites.
This is particularly puzzling since such objects are at the high mass end of the dSph mass spectrum and have the gravitational potential largely dominated by DM (thus simulation results should be robust despite the baryonic contribution is neglected).
Larger galaxies typically show monotonic relation between luminosity and halo circular velocity (or halo mass), while the presence of such massive dark subhalos would strongly violate this relation~\citep{BoylanKolchin:2011de}.

A solution of the aforementioned issues might reside in a departure from the collisionless CDM scheme foreseeing a suppression of small-scale structures either in the primordial power spectrum or due to DM-induced effects during structure formation.
Assuming instead the CDM paradigm to be correct, the solution could lie in baryonic physics, and in particular in connection with supernova feedbacks and low star formation (SF) efficiency.
A variety of studies and simulations have shown that baryonic effects could possibly lead to cores in DM halos and suppress SF
in low mass halos (see \citep{Weinberg:2013aya} for a recent review).

The MW satellites are crucial laboratories for testing the validity of such solutions.
The inefficiency in SF can be explained by a low gas density content in dSphs, below the density threshold for SF.
Different mechanisms have been suggested in order to either prevent gas collection in dSphs (as, e.g., heating of intergalactic gas by the ultraviolet photoionizing background~\citep{Bullock:2000wn}) or removing gas out of the shallow gravitational potential of dSphs (with, e.g., early feedback effects or tidal streams of gas in the dSph orbit around the MW~\citep{Mayer:2001yf}).
Measurements of the presence of gas in dSphs would thus be crucial to discriminate among some of the proposed solutions. 

The injection of energy associated with feedbacks should have left some imprints in the magnetic properties and high-energy cosmic-ray (CR) content of the dSphs. 
Indeed the generation of magnetic fields in galaxies is often associated with dynamo processes, which are sustained by the turbulent energy sourced in turn by supernova explosions. 
The same mechanisms can accelerate low-energy electrons up to TeV-PeV energies (for a recent review of the role of supernovae as CR and magnetic source, see, e.g.~\citep{Blasi:2013rva} and references therein).
Radio observations can probe the synchrotron radiation associated with high-energy electrons spiraling in an ambient magnetic field.

In this project, we performed deep mosaic radio observations of a sample of six local dSphs with the Australia Telescope Compact Array (ATCA). 
We simultaneously collected continuum data (in the 1.1-3.1 GHz band) and HI spectral line data (at 1.4 GHz).
The latter is associated with the atomic transition of neutral hydrogen and will be analysed to constrain the gas mass in dSphs in a future work.
In this paper (Paper II), we discuss the search for a diffuse continuum component. The level of the achieved rms sensitivity is around 0.05 mJy/beam.

A recent attempt in the same direction but making use of single dish observations was performed by \citep{Spekkens:2013ik,Natarajan:2013dsa} with the Green Bank Telescope.

Due to existing bounds on dSph gas density~\citep{Grcevich:2009gt}, thermal emissions are likely to be very dim.
Our search focuses on the possible presence of a non-thermal synchrotron emission associated with high-energy electrons interacting with the interstellar magnetic field.
The expected emission is weak and on relatively large scale (over the dSph size, which is typically several arcmin).
This requires, on one side, a sensitivity enhancement which is provided by wide-band observations, and, on the other side, a wide-field strategy (obtained by means of compact configurations of the ATCA telescope) to access large scales and also mosaicking to map the predicted full extent of the dSph source.
An intrinsic limitation of interferometric observations is that physical scales much larger than the reciprocal of the shortest baseline in the array are not detectable. This limitation can be overcome by including data from a large single antenna, to fill in the zero-spacing region of the visibility plane. Accompanying single dish data were not available for these observations, however. On the other hand, the scales accessible with the interferometric observations cover the size of the expected emission, as we will discuss.

Because of the wide-bandwidth of the ATCA receivers, the beam size varies considerably over the frequency band. The changing beam size manifests as a
changing gain across the band, for off-axis sources. This variation with frequency can be interpreted as structure during imaging, and can result in imaging artefacts for off-axis emissions.
They can be mitigated through the use of frequency-dependent imaging techniques, which simultaneously solve for the spatial and spectral variation of the source. 
Although there are a number of effective methods of wide-band imaging for a single pointing, joint imaging of mosaic pointings is still an unsolved algorithmic problem for the wide-band case. This is due to the frequency-varying primary beam effect over each mosaic panel, which introduces frequency-dependent gains across the image that will differ from panel to panel in overlapping regions.

A significant part of the project has been thus to investigate how state-of-the-art imaging algorithms can deal with such problematics (see also Paper I~\citep{Paper1}), which will become more and more pressing with the next-generation of radio telescopes, and in particular with the Square Kilometre Array (SKA).
We adopted the MFCLEAN algorithm in {\it Miriad} which was found to provide satisfactory results, as shown in the following and in Paper I. 
It implements the algorithm of \citep{Sault:94} to model the source brightness distribution with a linear variation in frequency.
The data reduction and imaging are summarized in Section~\ref{sec:obs}.

The observing setup is composed by both a compact array of five antennas and long baselines involving a sixth antenna. 
Short-spacings are required to detect extended emissions. For the adopted array, we obtain a synthesized beam of about 1 arcmin and a maximum detectable scale of about 15 arcmin.
With such large beam, however, the confusion limit is quickly reached.
Long baselines provide, on the other hand, high-resolution mapping of the small-scale background sources. They are then subtracted from the short-baseline maps to reduce the confusion noise, as explained in Section~\ref{sec:diff}. This is done both directly from the data and in the image plane.
We remind that, since we use an interferometric technique, the data are collected in the so-called visibility (or UV) plane, which is related to the image plane by means of a Fourier transformation.
The median radio source angular size of extragalactic background objects with flux lower than about 100 mJy (namely, sources present in the dSph fields, see Paper I) is below 10 arcsec, see, e.g.,~\cite{Windhorst:1990}.
Clouds within the dSph or in the Galaxy might contribute at few tens of arcsec scales, but their presence is likely to be negligible.
Therefore, although having a complete coverage of the UV plane would be clearly ideal, an observing setup including long baselines to cover scales up to about 10 arcsec and short spacings to measure few arcmin diffuse emissions (as the one employed in this work) can be adequate to infer the presence of a signal.

Starting from the SF history of the observed dSphs (inferred, in particular, through colour-magnitude diagrams), we derived estimates for the expected magnetic field strength and CR content. They are described in Section~\ref{sec:mod} and Table~\ref{tableB}.
The possibility of having high-energy electrons and positrons injected through DM annihilations or decays is investigated in Paper III~\citep{Paper3}.
The CR spatial diffusion and energy losses are modelled with a special care, developing a new numerical solution of the transport equation which is reported in the Appendix.

After introducing the statistical technique, in Section~\ref{sec:ana} we test the presence of a diffuse component.
We report bounds on a variety of physical quantities associated with the expected synchrotron emission.
They include the total radio flux, the angular shape of the radio emissivity, the equipartition magnetic field, and the injection and equilibrium distributions of CR electrons. 
We present a detailed discussion on how the bounds on the CR population depend on the assumption concerning the magnetic properties of the dSphs. 
We also investigate the connection to far-infrared (FIR) and X-ray observations.

Conclusions are in Section~\ref{sec:concl}.

\section{Observations and Data reduction}
\label{sec:obs}

\begin{table*}
{\footnotesize
\centering
\begin{tabular}{|l|c|c|c|c|}
\hline
dSph & Theoretical sensitivity & Measured Sensitivity & Ratio $\frac{Measured}{Theoretical}$ & CLEAN cutoff \\
name & [mJy] & [mJy] & & [mJy]\\
\hline
Carina	 & 44 & 54 & 1.23 & 130 \\
Fornax	 & 46 & 41 & 0.89 & 200 \\
Sculptor & 43 & 40 & 0.93 & 180 \\
BootesII & 31 & 50 & 1.61 & 90 \\
Hercules & 34 & 33 & 0.97 & 105 \\
Segue2	 & 24 & 27 & 1.12 & 70 \\
\hline
\end{tabular}
\caption{Sensitivities (in single panels) for the high resolution images of each target: 
The theoretical sensitivity (Natural weighting), taken from the online ATCA sensitivity calculator and adjusted to assume $33\%$ of data flagged; 
measured off-source RMS noise in the first mosaic panel;
the ratio of the measured RMS noise to the theoretical sensitivity;
the CLEAN cutoff used in imaging ($\gtrsim 3 \times $ the theoretical sensitivity).
}
\label{tab:sensitivities}
}
\end{table*}

\begin{table*}
{\footnotesize
\centering
\begin{tabular}{|l|c|c|c|c|c|c|}
\hline
dSph & \multicolumn{3}{|c|} {$r_{-1}$ map} &  \multicolumn{3}{|c|} {$gta$ map}   \\
name & synthesized beam &average rms   & total flux        &synthesized beam  &  average rms &  total flux   \\
\hline
Carina & $4.0''\times 2.6''$ & 42  $\mu$Jy& 2.6 Jy & $1.3'\times 0.98'$ & 146  $\mu$Jy& 1.6 Jy\\ 
Fornax & $7.8''\times 2.2''$ & 36  $\mu$Jy& 0.8 Jy & $1.4'\times 1.2'$ & 143  $\mu$Jy& 0.6 Jy\\ 
Sculptor & $8.0''\times 2.1''$ & 37  $\mu$Jy& 1.0 Jy & $0.88'\times 0.76'$ & 126  $\mu$Jy& 1.0 Jy\\ 
BootesII & $28''\times 2.1''$ & 39  $\mu$Jy& 0.2 Jy & $1.3'\times 0.94'$ & 145  $\mu$Jy&  0.2 Jy\\ 
Hercules & $27''\times 2.0''$ & 35  $\mu$Jy& 0.3 Jy & $1.3'\times 0.73'$ & 112  $\mu$Jy& 0.2 Jy\\
Segue2 & $17''\times 1.9''$ & 27  $\mu$Jy& 0.4 Jy & $1.5'\times 0.68'$ & 165  $\mu$Jy& 0.4 Jy\\ 
\hline
\end{tabular}
\caption{Main properties of the maps used in this work. Total flux and average rms are quoted for the inner region, namely, within 30 arcmin (20 arcmin) from the center for CDS (UDS).}
\label{tab:obs}
}
\end{table*}

The dSphs considered in this work (Carina, Fornax, Sculptor, BootesII, Hercules and Segue2) were observed (for a total observing time of 123 hours) during 2011 July/August. 
The six 22-m diameter ATCA antennae operating in the frequency range 1.1-3.1 GHz were employed, with the array configuration formed by a core of five antennae (with maximum baseline of about 200~m), and a sixth antenna located at about 4.5 km from the core.
More specifically, the core of the array for the observations of Carina, BootesII, Segue2, and part of Hercules was in the hybrid configuration H214 with maximum baseline of 214~m, while for the observations of Fornax, Sculptor, and the second part of Hercules, it was in the hybrid configuration H168 with maximum baseline of 168~m.
Further details about the observing setup can be found in Paper I.
We will refer to Carina, Fornax, and Sculptor as classical dSphs (CDS), and to BootesII, Hercules, and Segue2 as UDS.

\begin{figure*}
   \centering
 \begin{minipage}[htb]{5.2cm}
   \centering
   \includegraphics[width=0.85\textwidth,angle=-90]{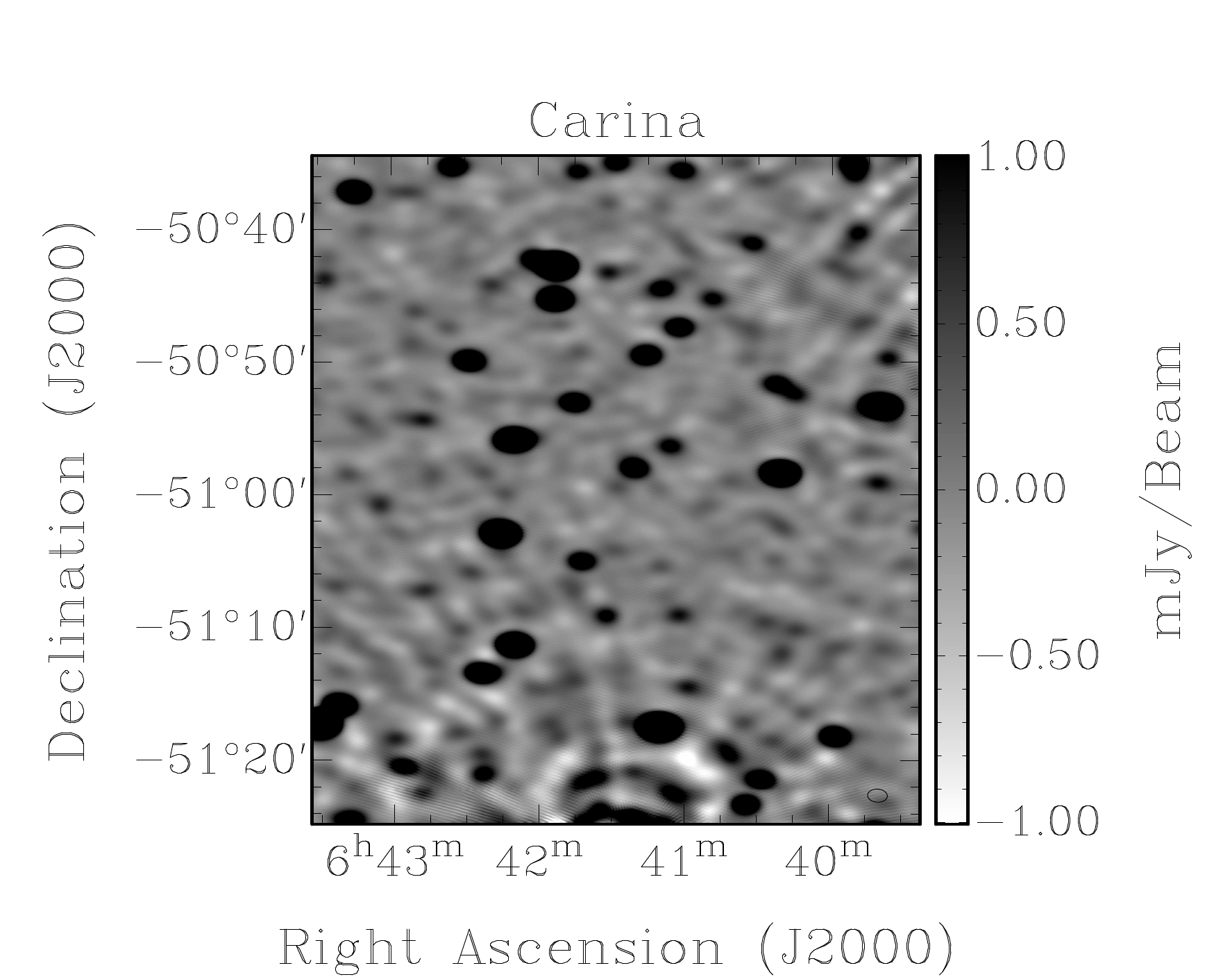}
 \end{minipage}
\hspace{3mm}
 \begin{minipage}[htb]{5.2cm}
   \centering
   \includegraphics[width=0.9\textwidth,angle=-90]{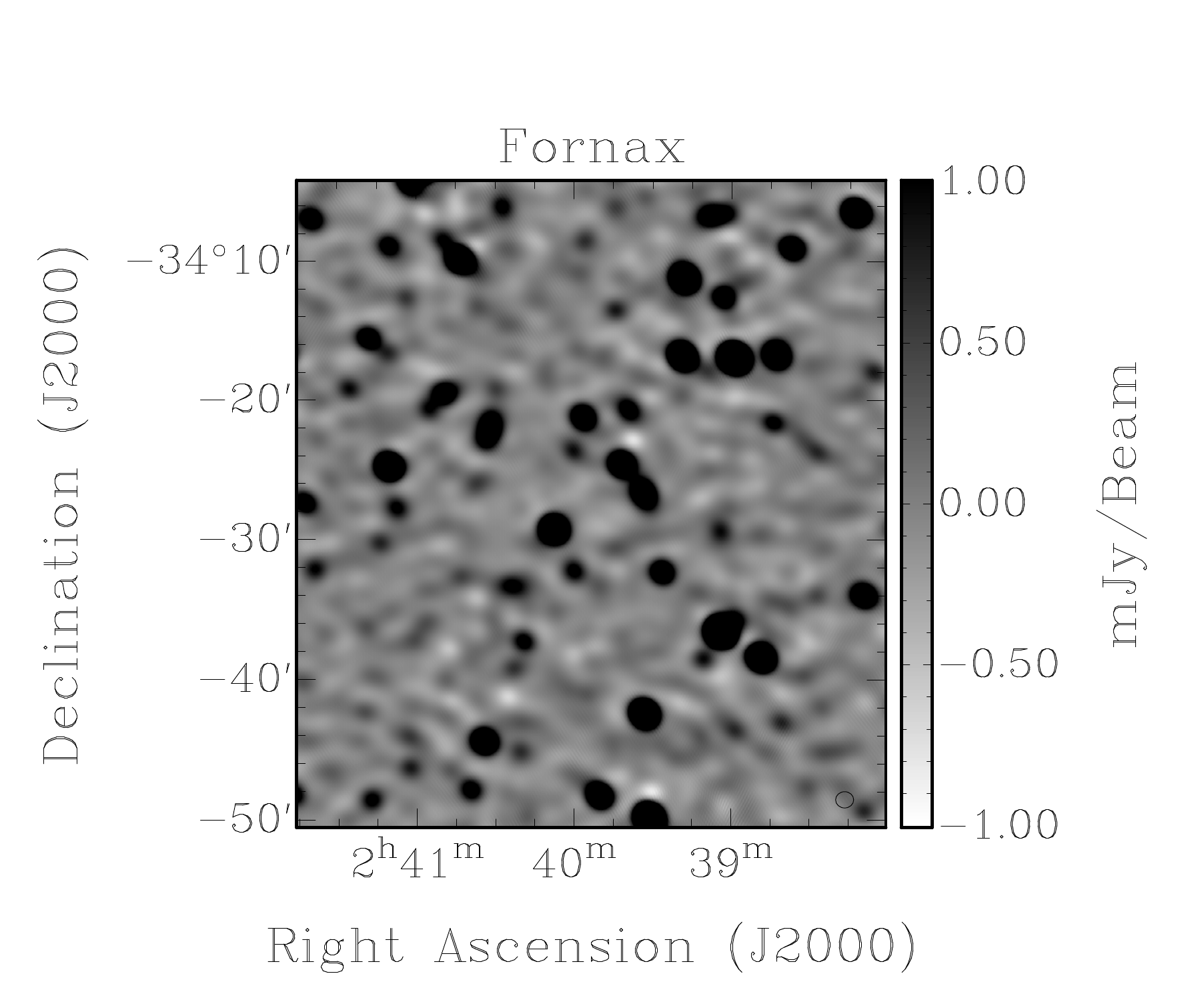}
 \end{minipage}
\hspace{3mm}
 \begin{minipage}[htb]{5.2cm}
\vspace{5mm}
   \centering
   \includegraphics[width=0.75\textwidth,angle=-90]{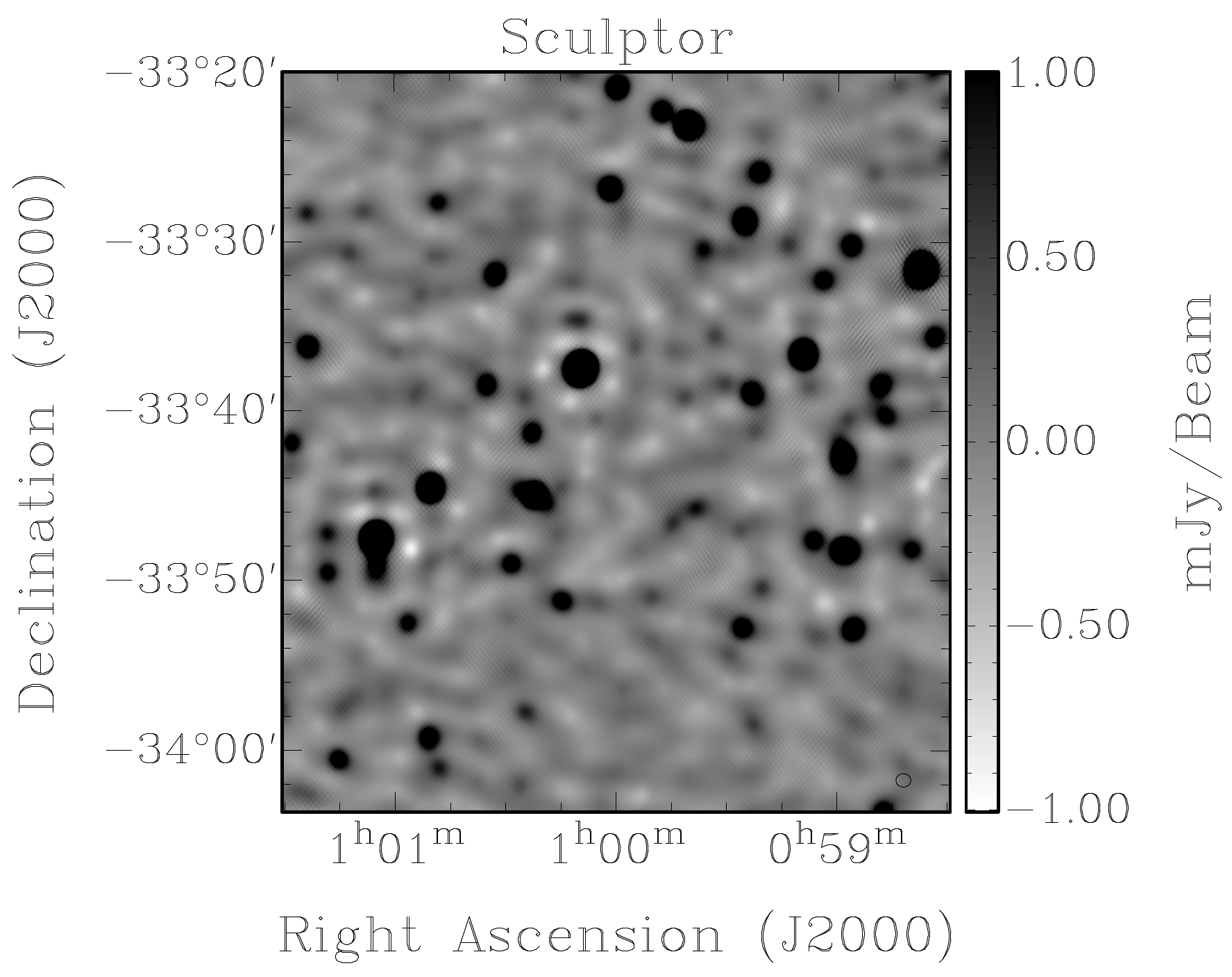}
 \end{minipage} \\
 \begin{minipage}[htb]{5.2cm}
   \centering
   \includegraphics[width=0.85\textwidth,angle=-90]{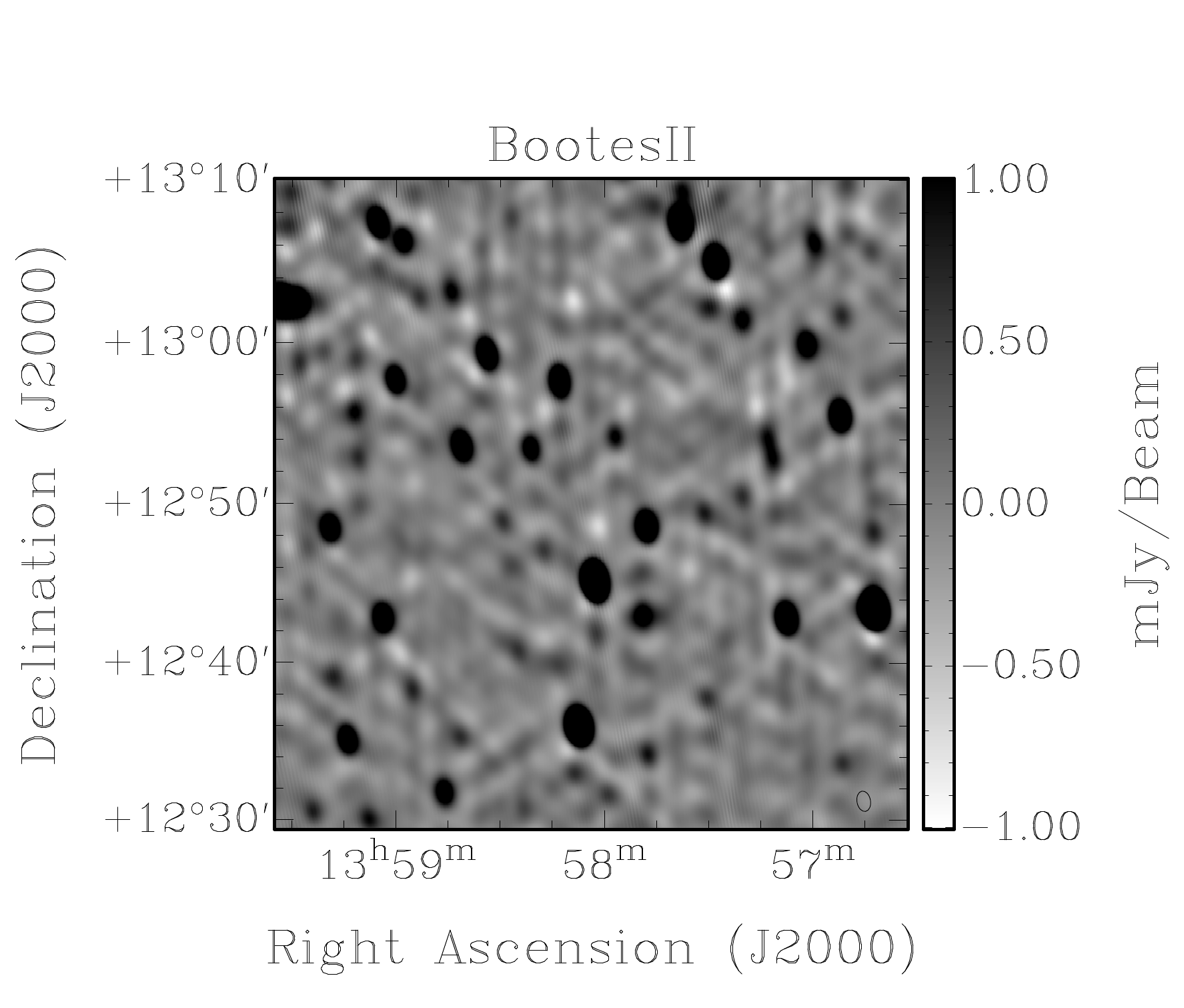}
 \end{minipage}
\hspace{2mm}
 \begin{minipage}[htb]{5.2cm}
   \centering
   \includegraphics[width=0.85\textwidth,angle=-90]{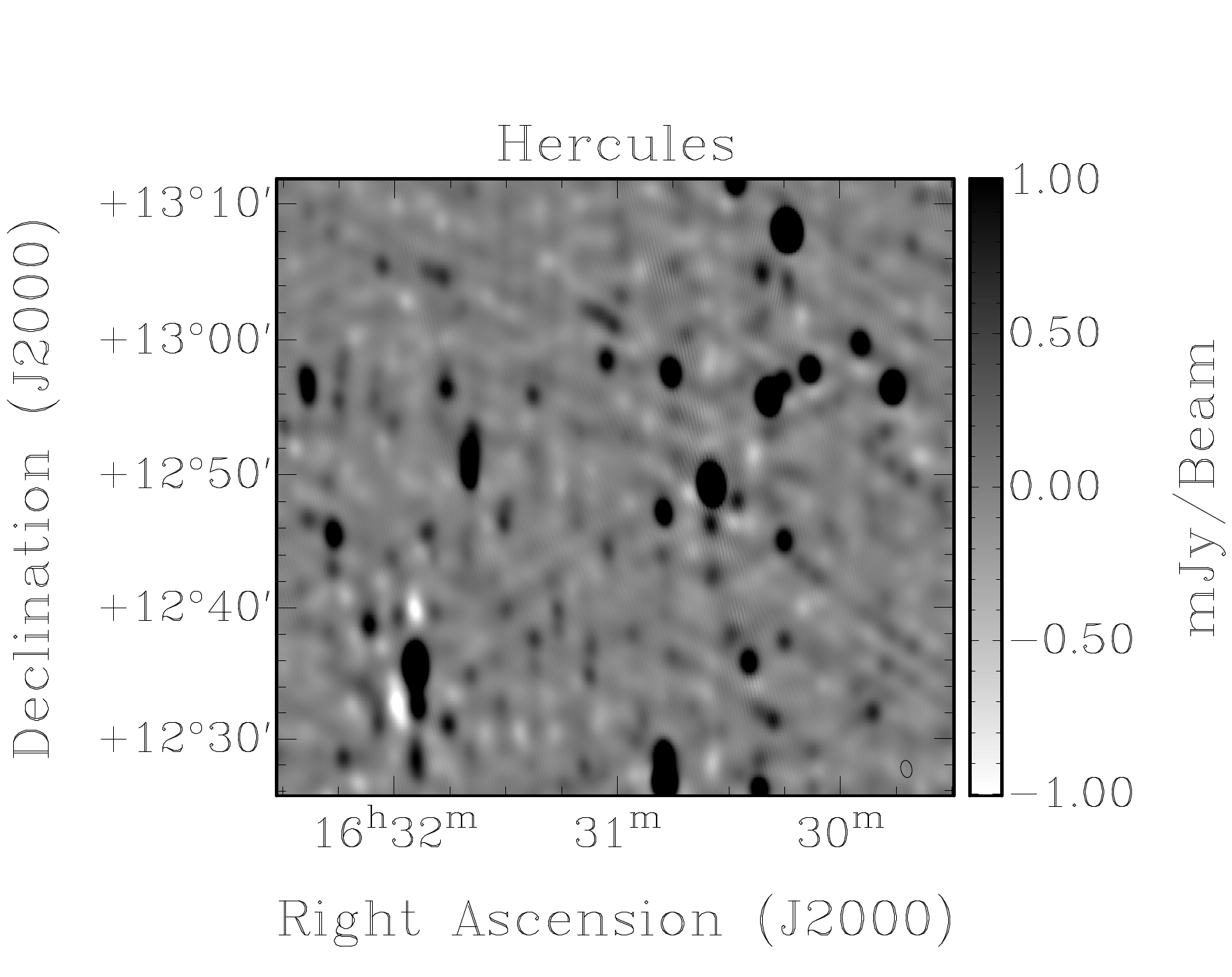}
 \end{minipage}
\hspace{4mm}
 \begin{minipage}[htb]{5.2cm}
   \centering
   \includegraphics[width=0.85\textwidth,angle=-90]{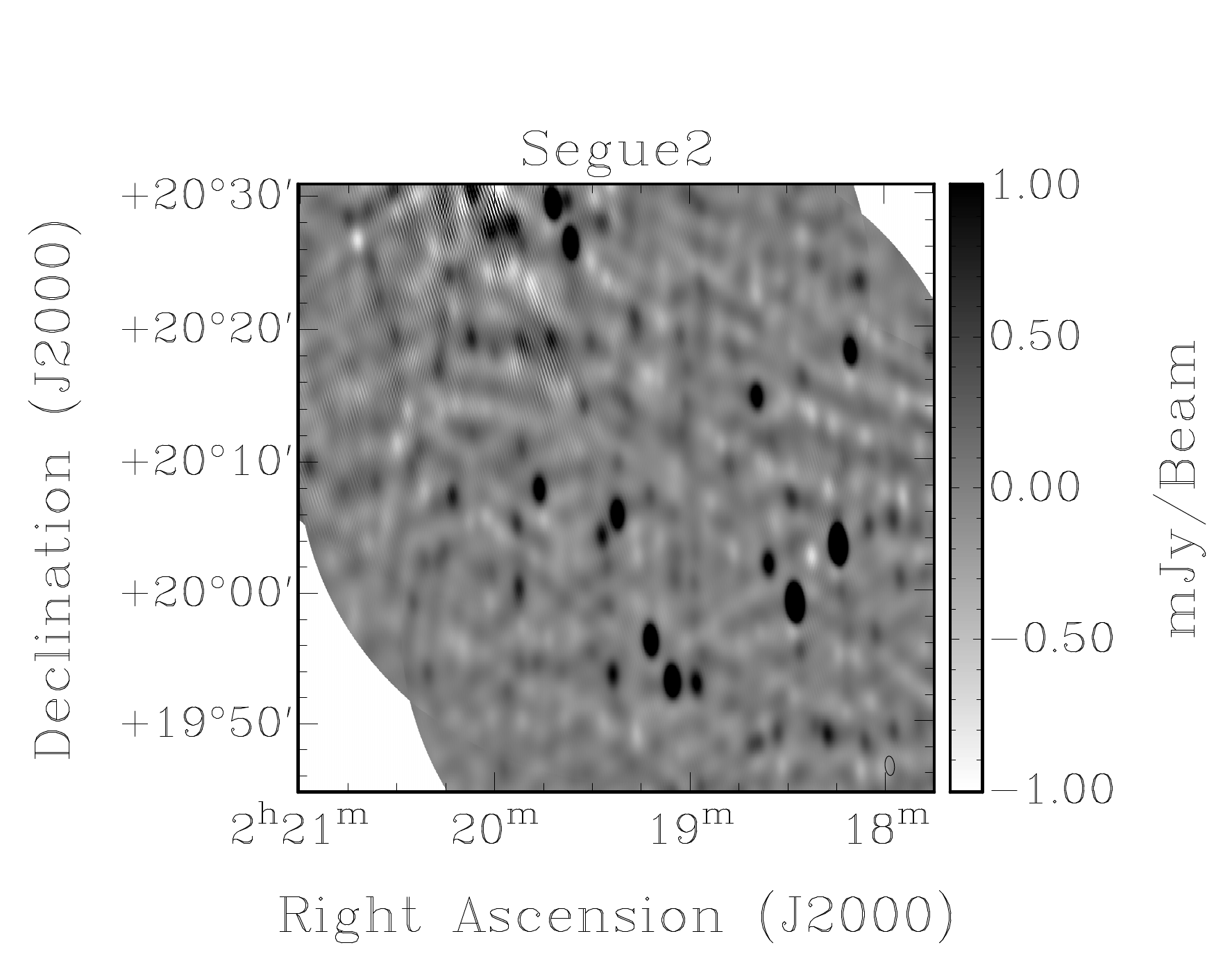}
 \end{minipage}
    \caption{{\bf Observational maps}. Maps obtained by imaging with the robustness parameter set to -1 and applying a Gaussian taper with FWHM=$15''$ before Fourier inversion. Top: Carina, Fornax, and Sculptor FoVs. Bottom: BootesII, Hercules, and Segue2 FoVs. The synthesized beam is shown in the bottom-right corner of each panel.}
\label{fig:map_f15}
 \end{figure*}

\begin{figure*}
   \centering
 \begin{minipage}[htb]{5.2cm}
   \centering
   \includegraphics[width=0.9\textwidth,angle=-90]{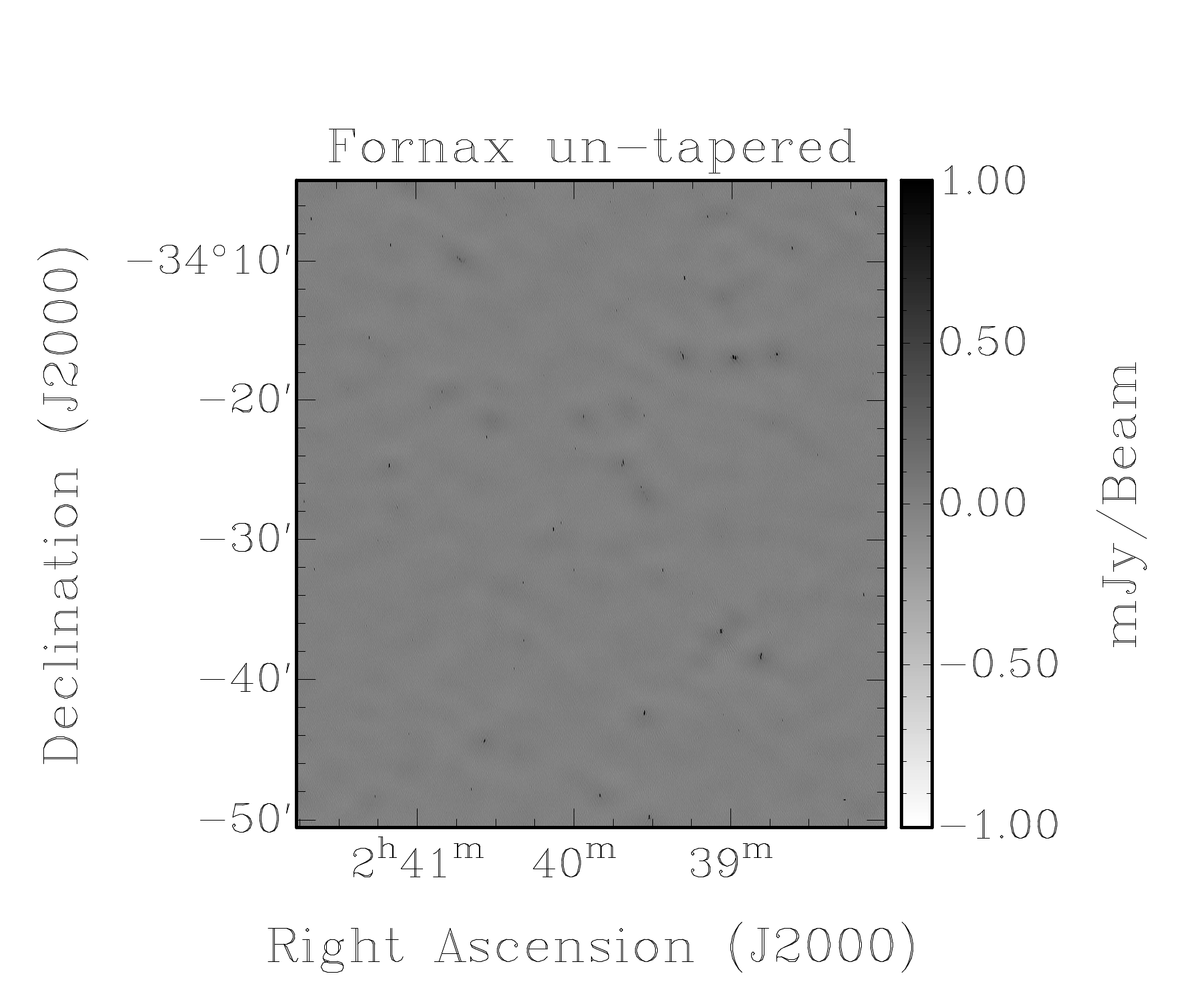}
 \end{minipage}
\hspace{3mm}
 \begin{minipage}[htb]{5.2cm}
   \centering
   \includegraphics[width=0.9\textwidth,angle=-90]{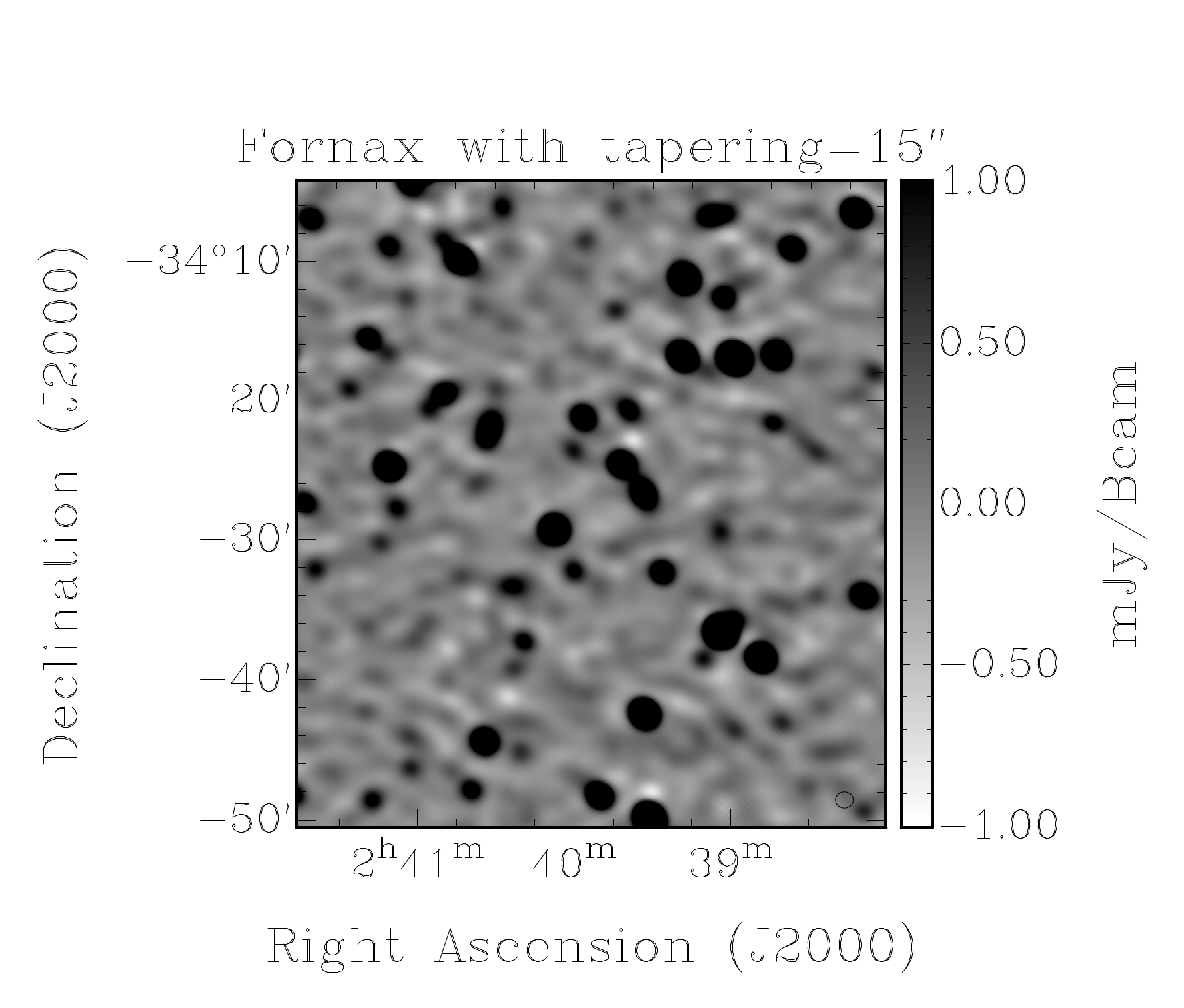}
 \end{minipage}
\hspace{3mm}
 \begin{minipage}[htb]{5.2cm}
   \centering
   \includegraphics[width=0.9\textwidth,angle=-90]{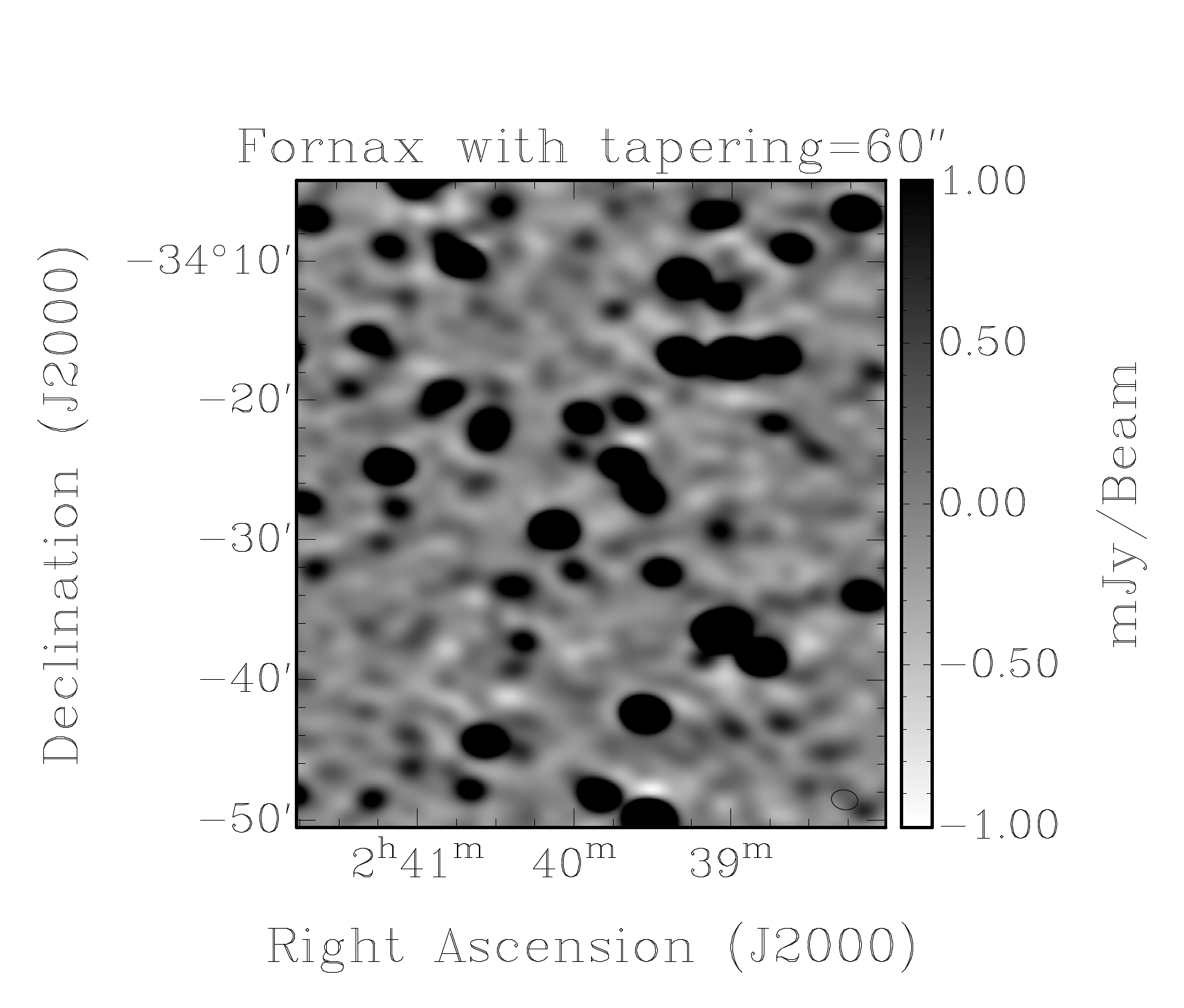}
 \end{minipage}
    \caption{{\bf Imaging}. Example of comparison between the three different types of maps considered, shown for the Fornax FoV. $r_{-1}$ (left) is the map obtained with robust=-1 (which effectively downweights short baselines), $gta$ (central) is the map obtained with robust=-1 and tapered with FWHM=$15''$ (which effectively downweights long baselines), and $gtb$ (right) is the map obtained with robust=-1 and tapered with FWHM=$60''$ (maximizing the sensitivity to large scale emissions). The synthesized beam is shown in the bottom-right corner of each panel.}
\label{fig:map_basel}
 \end{figure*}

The {\it Miriad} data reduction package~\citep{Sault:95} was used for calibration and imaging.
We proceeded producing three maps for each target.
The data were first imaged with the Briggs robustness parameter set to -1 \citep{Briggs:thesis} leading to an high resolution map, where short baselines are effectively down-weighted.
Then we generate a second set of maps, by imaging again with the same robustness parameter, but applying a Gaussian taper to the data before Fourier inversion.
In the following, we will use the label $r_{-1}$ for the map obtained with robust=-1, $gta$ (Gaussian taper $a$) for the map obtained with robust=-1 and tapered with FWHM=$15''$ (which effectively down-weights long baselines), and $gtb$ (Gaussian taper $b$) for the map obtained with robust=-1 and tapered with FWHM=$60''$ (maximizing the sensitivity to large scale emissions).
The $gta$ maps are shown in Fig.~\ref{fig:map_f15} for the various targets, while an example of the three different kinds of maps is reported in Fig.~\ref{fig:map_basel} for the Fornax field of view (FoV).

The theoretical and measured sensitivities in single panels are reported in Table~\ref{tab:sensitivities}. The RMS noise level in the table was measured in an off-source corner section of the first mosaic panel in each field, of size $5\%\times5\%$ of the total image, and considered representative of the sensitivity of all panels.  
The robust -1 images were CLEANed to a cutoff $\gtrsim 3$ times the theoretical sensitivity to avoid CLEAN bias (as reported in Paper I). 
The theoretical sensitivity for the robust -1 images was taken to be the figure given by the ATCA sensitivity calculator \footnote{http://www.narrabri.atnf.csiro.au/myatca/sensitivity\_calculator.html} for Natural weighting. The Natural weighting figure will be an underestimate of the robust -1 sensitivity. Nonetheless, this was chosen as the reference limit, as even Fourier inversion of the images without any CLEANing gave off-source noise floors of less than the robust -1 value given by the calculator, and much closer to the Natural weighting adjusted for 33\% data loss due to flagging (except for the case of BootesII, which is dynamic range limited). The effect of robustness parameter on sensitivity is highly dependent on the UV-distribution of the visibilities \citep{Briggs:thesis}, so the Natural weighting value was taken as lower limit.
The tapered imaged were CLEANED to the same cutoff level as the robust -1 images, as this level was sufficiently larger than the theoretical noise.

The main properties of $r_{-1}$ and $gta$ maps (which will be the two sets used in the analysis) are summarized in Table~\ref{tab:obs}.\footnote{Maps and source catalogue presented in this project can be retrieved at http://personalpages.to.infn.it/$\sim$regis/c2499.html.}
The $r_{-1}$ maps basically probes scales from few arcsec to about 10 arcsec, and have an rms noise of 30-40 $\mu$Jy.
The synthesized beam of the tapered $gta$ maps is instead about 1 arcmin, and the largest scale which can be well imaged is around 15 arcmin (see discussion in Section~\ref{sec:maxsize}). Because of confusion limitation, the rms noise raises up to 0.1-0.15 mJy. Both beam and noise are further increased by about 50\% in the $gtb$ maps.
Again, details about the data reduction can be found in Paper I.

Note that the tapering with FWHM=$15''$ results in a map with a beam of about 1 arcmin and not of about 15 arcsec.
This is because the observations are blind to scales between about 10 arcsec and 1 arcmin, due to the lack of baselines between 250 m and 4 km.
As mentioned in the Introduction, we do not expect neither the signal nor the background sources to significantly contribute at such intermediate scales.
The $gta$ taper thus effectively downweights all the baselines involving the sixth antenna, while keeping the whole signal from the compact array.
After exploring different imaging strategies, the tapering was found to provide cleaner results than, e.g., simply excluding the long baselines.
The $gtb$ taper with FWHM=$60''$ starts instead affecting also the longer baselines within the compact array, and the resulting beam is about 50\% larger.

\section{Estimate of the diffuse component}
\label{sec:diff}

In this Section we describe the estimate of the diffuse component, for which we explored different methods.
Results and comparisons are discussed in Section~\ref{sec:bounds}.

In our analysis, we focus on the inner region, within 30 arcmin (20 arcmin) from the center of the CDS (UDS), motivated by three reasons:  first, this region encompasses the area of the expected emission from sources associated with the dSph stellar component or from the DM halo (being the half-light radius and halo scale radius $\lesssim 20'$ in CDS and $\lesssim 10'$ in UDS). Second, the size of largest structure that can be well imaged through the adopted observational strategy (see Section~\ref{sec:maxsize}) is well below 30 arcmin, so there is no gain in considering a larger area. Finally, in this region we have a uniform coverage and rms, so we can neglect primary beam effect (as verified also with the flux measurements of point sources discussed in Paper I).

Here, we identify the diffuse signal only with extended emissions centered around the optical dSph center, and we do not consider off-center point-like or moderately extended clumps (e.g., associated with clouds or DM subhalos).
This possibility will be investigated in details elsewhere.

\begin{figure*}
   \centering
 \begin{minipage}[htb]{5.2cm}
   \centering
   \includegraphics[width=0.9\textwidth,angle=-90]{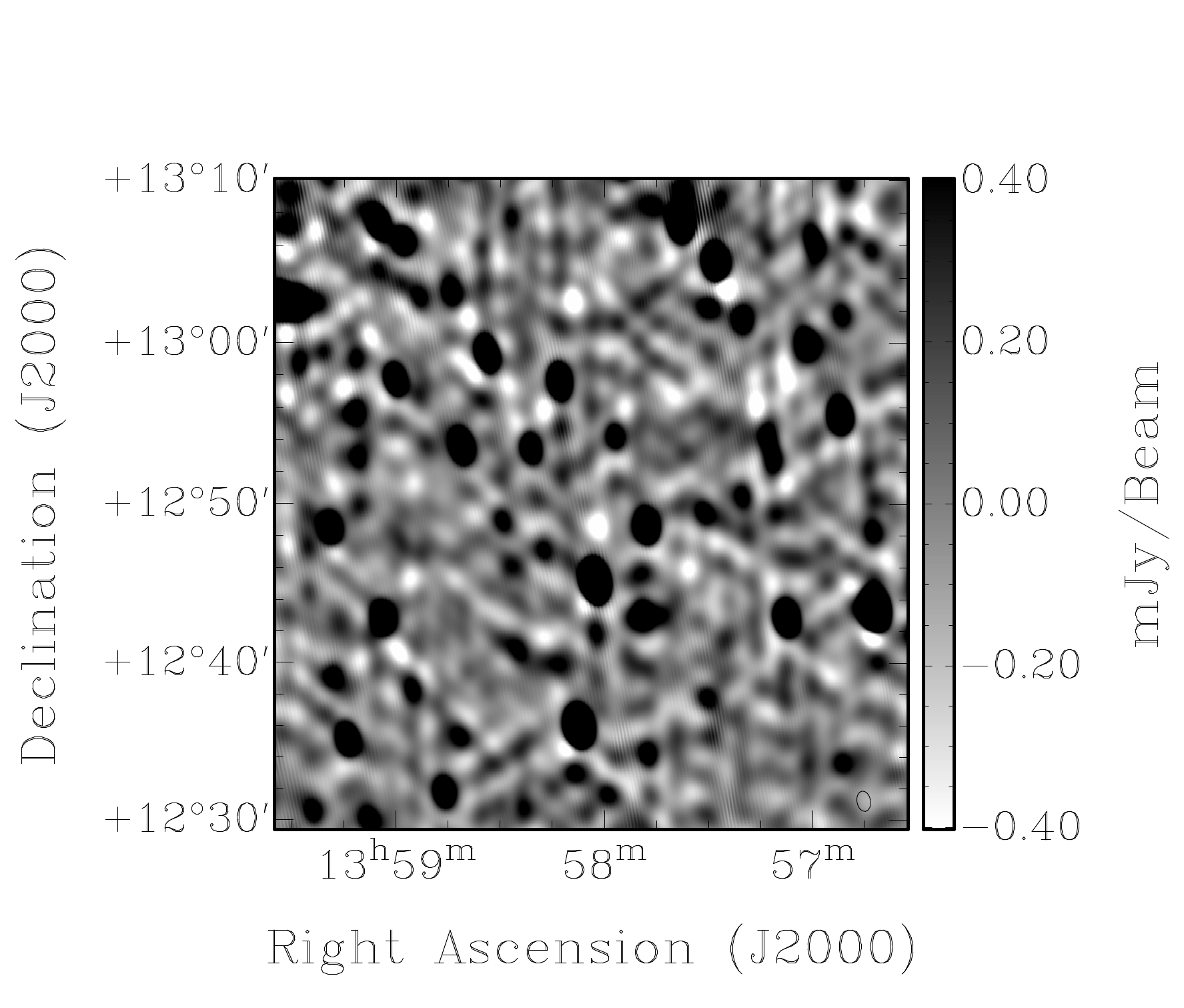}
 \end{minipage}
\hspace{3mm}
 \begin{minipage}[htb]{5.2cm}
   \centering
   \includegraphics[width=0.9\textwidth,angle=-90]{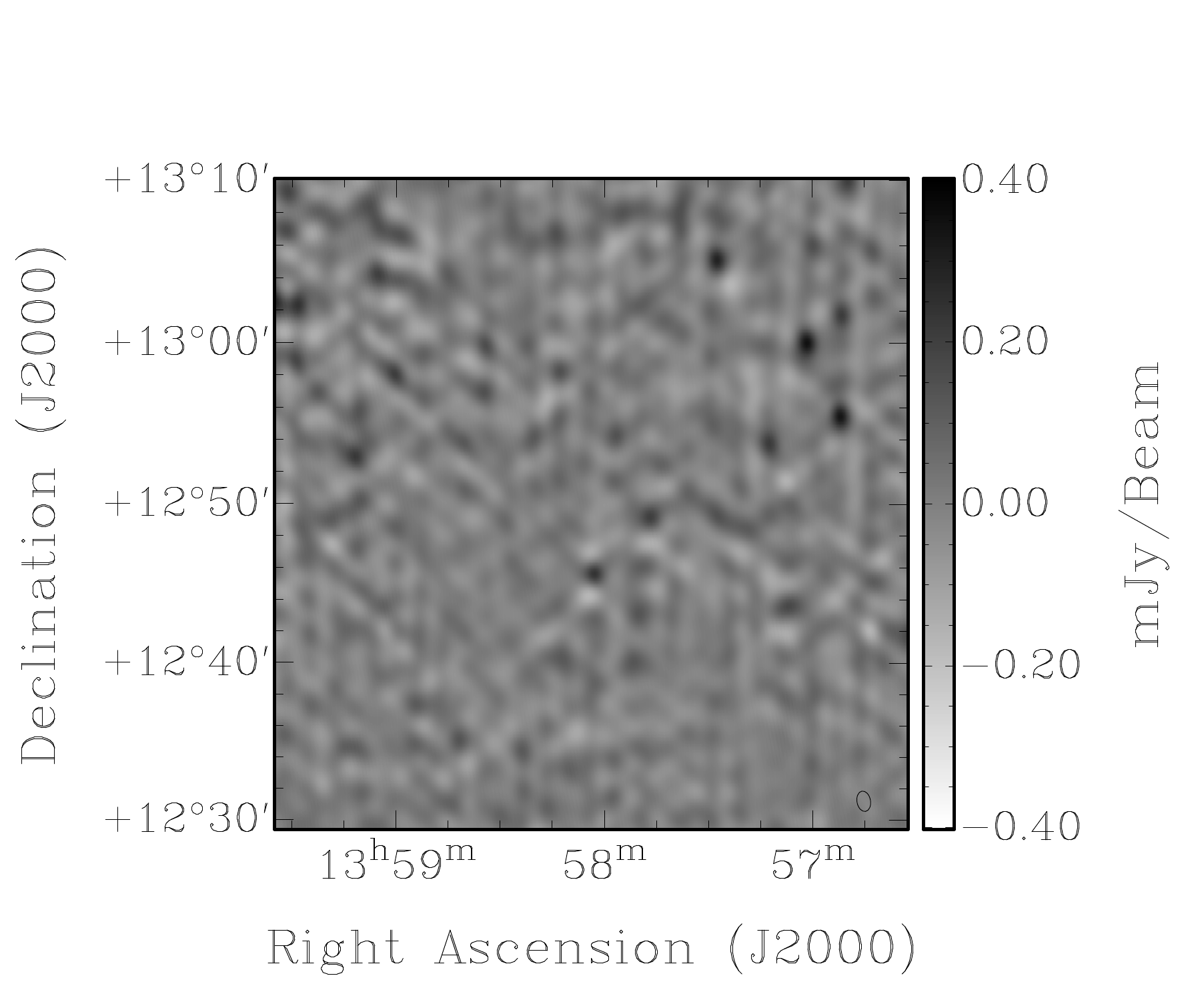}
 \end{minipage}
\hspace{3mm}
 \begin{minipage}[htb]{5.2cm}
   \centering
   \includegraphics[width=0.9\textwidth,angle=-90]{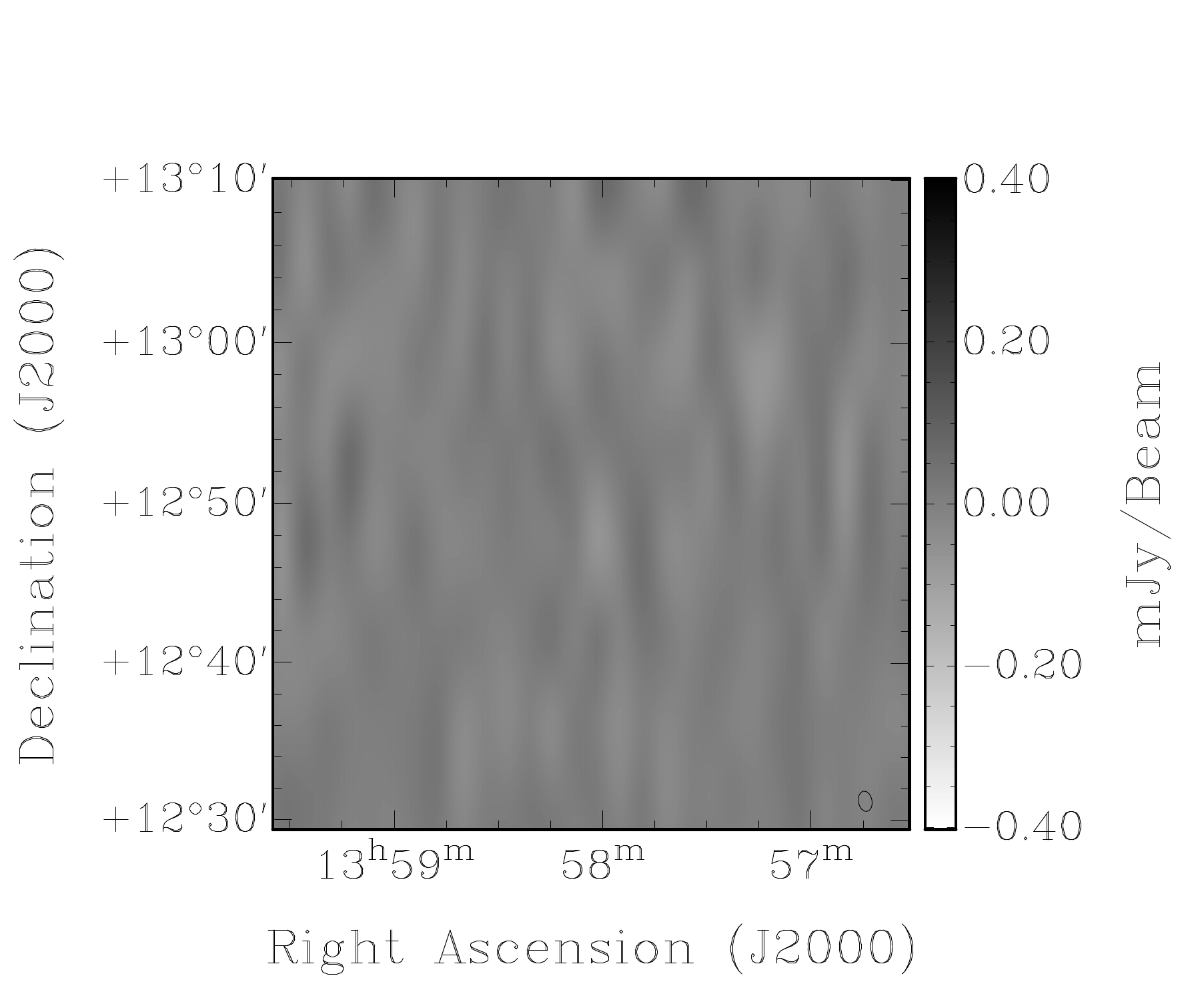}
 \end{minipage}
    \caption{{\bf Source subtraction}. Comparison between the original $gta$ map (left) and the maps obtained after subtracting sources in the UV-plane (central) and in the image-plane (right).
The shown example is for the BootesII FoV. The colour panel is kept constant to easy comparisons (although in this way the original image significantly saturates). The synthesized beam is shown in the bottom-right corner.}
\label{fig:map_sub}
 \end{figure*} 

The search for diffuse components is most successful if performed on short baseline maps.
In this case, the synthesized beam is about 1 arcmin in $gta$ and 1.5 arcmin in $gtb$ (while being of few arcsec in $r_{-1}$), and is more suited to detect a smooth extended emission of few arcmin size (which is the expected size of emission).
The theoretical rms worsens only by a moderate factor with respect to the long-baseline case, i.e. the square root of number of baselines $\sqrt{15/10}=1.2$. In practice, due to limitations from confusion, it actually grows by a factor of few. By means of source subtraction, we can mitigate confusion issues and bring the rms down, closer to the value derived for the $r_{-1}$ maps.

Moreover, if one tries to fit a diffuse component to the original map, the best-fit normalization will be generally different from zero with the no-signal case excluded at a significant statistical level (see discussions and plots in Section~\ref{sec:bounds}). This is obviously fictitious and due to the presence of point-sources.

In order to overcome the above two issues, we estimated the diffuse component by subtracting point-sources either in the UV- and/or image-plane, as we will describe below.
An example of the outcome for the $gta$ map of the BootesII FoV is shown Fig.~\ref{fig:map_sub}.

\subsection{Subtraction of sources on visibilities}
\label{sec:diffsubV}

For the detection of point sources, the inclusion of data from the antenna located at about 4.5 km from the array-core provides superior angular resolution (by a factor of 20) and lower rms than considering only the five antennae of the core, see Table~\ref{tab:obs} (for a comprehensive discussion of source detection in our maps, see Paper I).
Discrete sources are thus characterized by including long-baselines in the $r_{-1}$ maps.
The detected structures vary from few to few tens of arcsec. They can be then subtracted from the short-baseline maps.
The most proper way to do it is to perform the subtraction in the visibility plane.
This has been done with the task UVMODEL in {\it Miriad}.
The resulting visibilities are then reduced and imaged following the same pipeline as for the original maps.

The subtraction of sources in the $r_{-1}$ (used as a cross-check) and $gta$ maps has been performed taking the CLEAN component of the $r_{-1}$ map as the input source model.
In the $gtb$, instead, we first subtracted the $r_{-1}$ CLEAN components and then also the CLEAN components of the subtracted $gta$ map. The latter procedure over-subtracts flux and only emissions on very large scales may survive. However, it is a useful check for our method.

The subtraction procedure came out to be more successful for targets observed with the array configuration H214 (Carina, BootesII, Segue2), rather than with H168 (Fornax, Sculptor, Hercules), because of the better beam reconstruction.
Carina and BootesII are thus the cases showing the lowest rms after source-subtraction (while Segue2 has a larger noise because of imaging issues, partly due to the presence of a very bright source in the field, see Paper I). 

\subsection{Subtraction of sources on images}
\label{sec:diffSE}
To identify pixels in an image which belong to sources and not to the dSph diffuse emission, we use the publicly available tool SExtractor~\citep{Bertin:1996fj}.
In SExtractor, the detection of sources proceeds through segmentation by identifying groups of connected pixels that exceed some threshold above the background. 
The first step for source detection in SExtractor involves the determination of the background and RMS noise maps, since the background is subtracted from the original map, while the thresholds for detection is set in terms of the rms.
In fact the background map can be seen as an estimate of the large-scale diffuse emission. We computed it as follows.

The original map is split in regions of $(3\,{\rm arcmin})^2$, and the mean and the standard deviation of the distribution of pixel values within each region are determined. 
Then the most deviant values are discarded and the computation is re-performed. 
This is repeated until all the remaining pixel values are within 3-$\sigma$ from the mean.
The background in the region is then the mean in the non-crowded case (i.e., if $\sigma$ is changed by less than 20\% per iteration)
and 2.5$\times$median-1.5$\times$mean in the crowded case.
The resulting background map is then a bicubic-spline interpolation between the meshes of the grid, while the standard deviations form the rms map.

We will consider the background map as the source-subtracted map, and the rms will be the one adopted in the statistical analysis.
Another possibility would be to perform the analysis on the original map but masking all the pixels which are occupied by sources. We verified that results do not differ appreciably with respect to considering the source-subtracted map.

\begin{figure*}
\vspace{-2.cm}
   \centering
 \begin{minipage}[htb]{6cm}
   \centering
   \includegraphics[width=\textwidth]{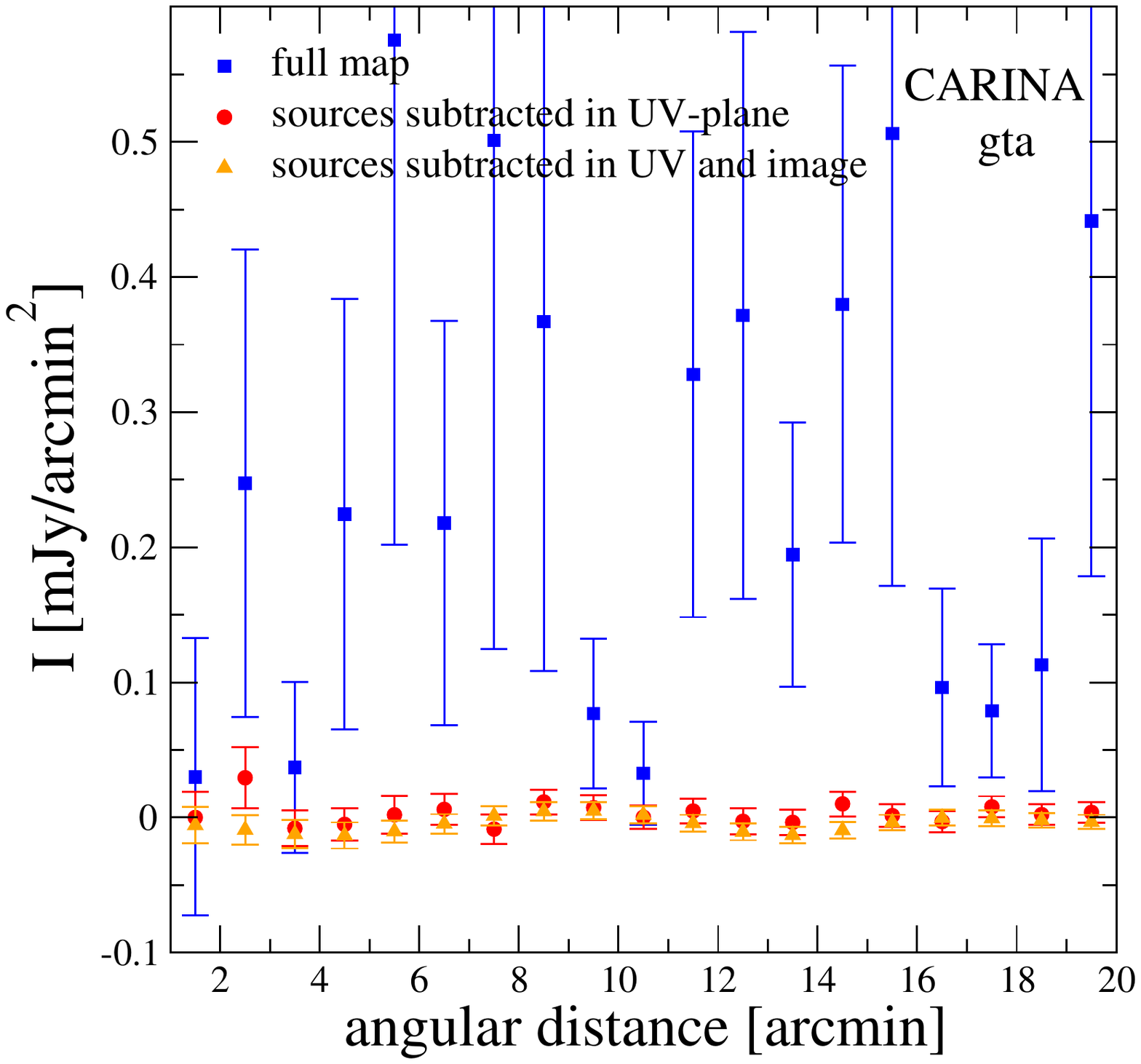}
 \end{minipage}
\hspace{-5mm}
 \begin{minipage}[htb]{6cm}
   \centering
   \includegraphics[width=\textwidth]{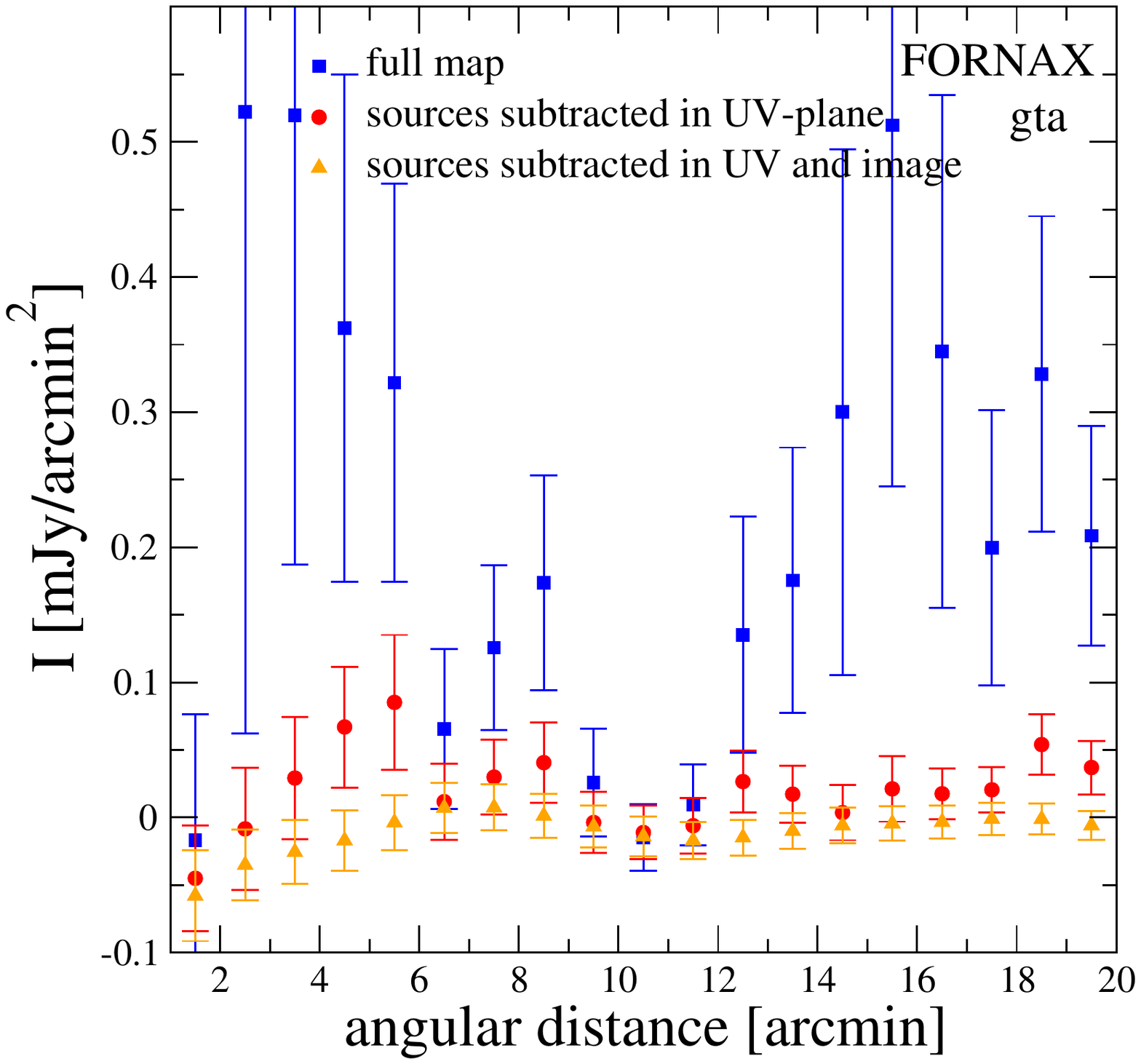}
 \end{minipage}
\hspace{-5mm}
 \begin{minipage}[htb]{6cm}
   \centering
   \includegraphics[width=\textwidth]{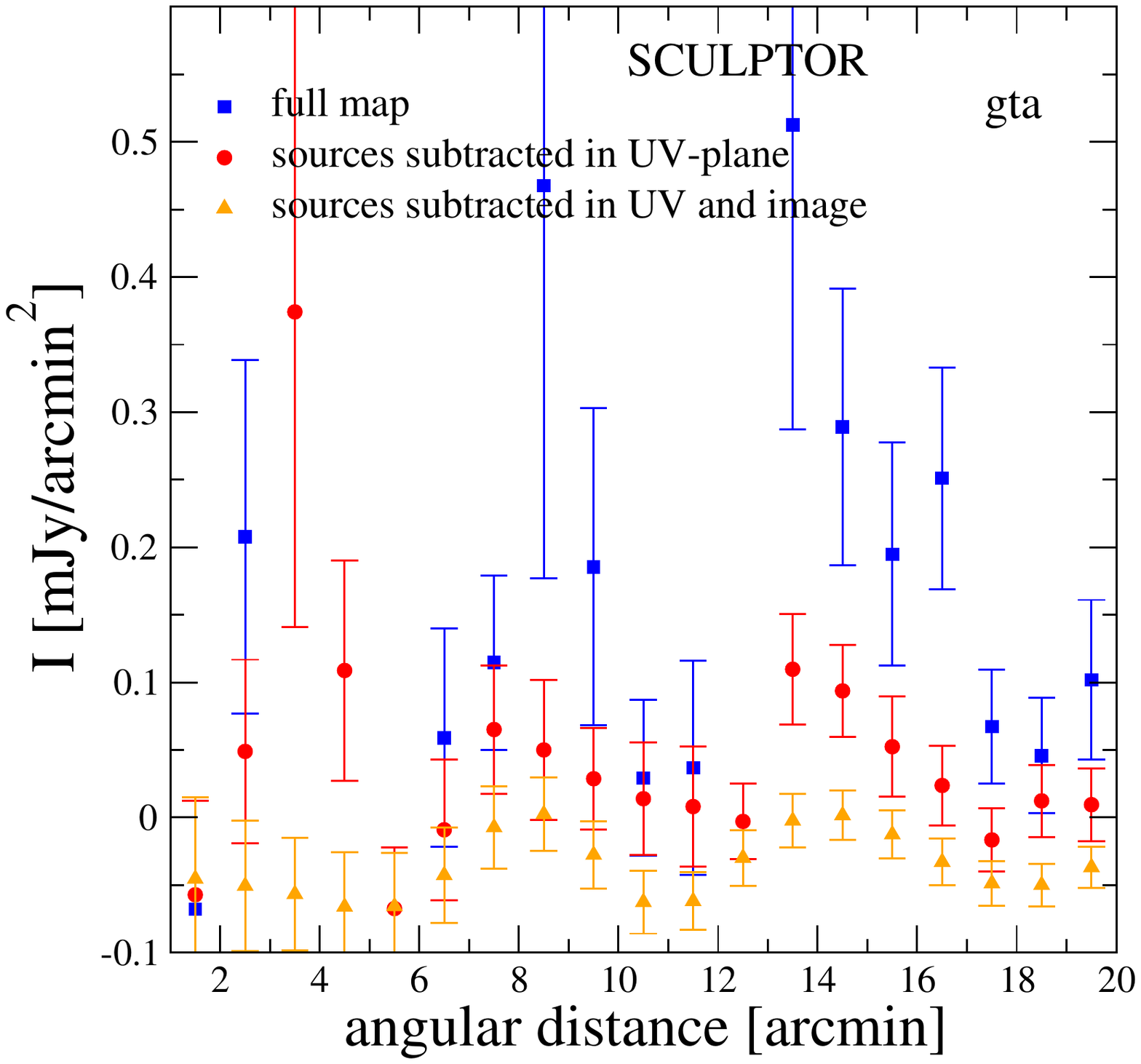}
 \end{minipage} \vspace{-2.5cm}\\ 
 \begin{minipage}[htb]{6cm}
   \centering
   \includegraphics[width=\textwidth]{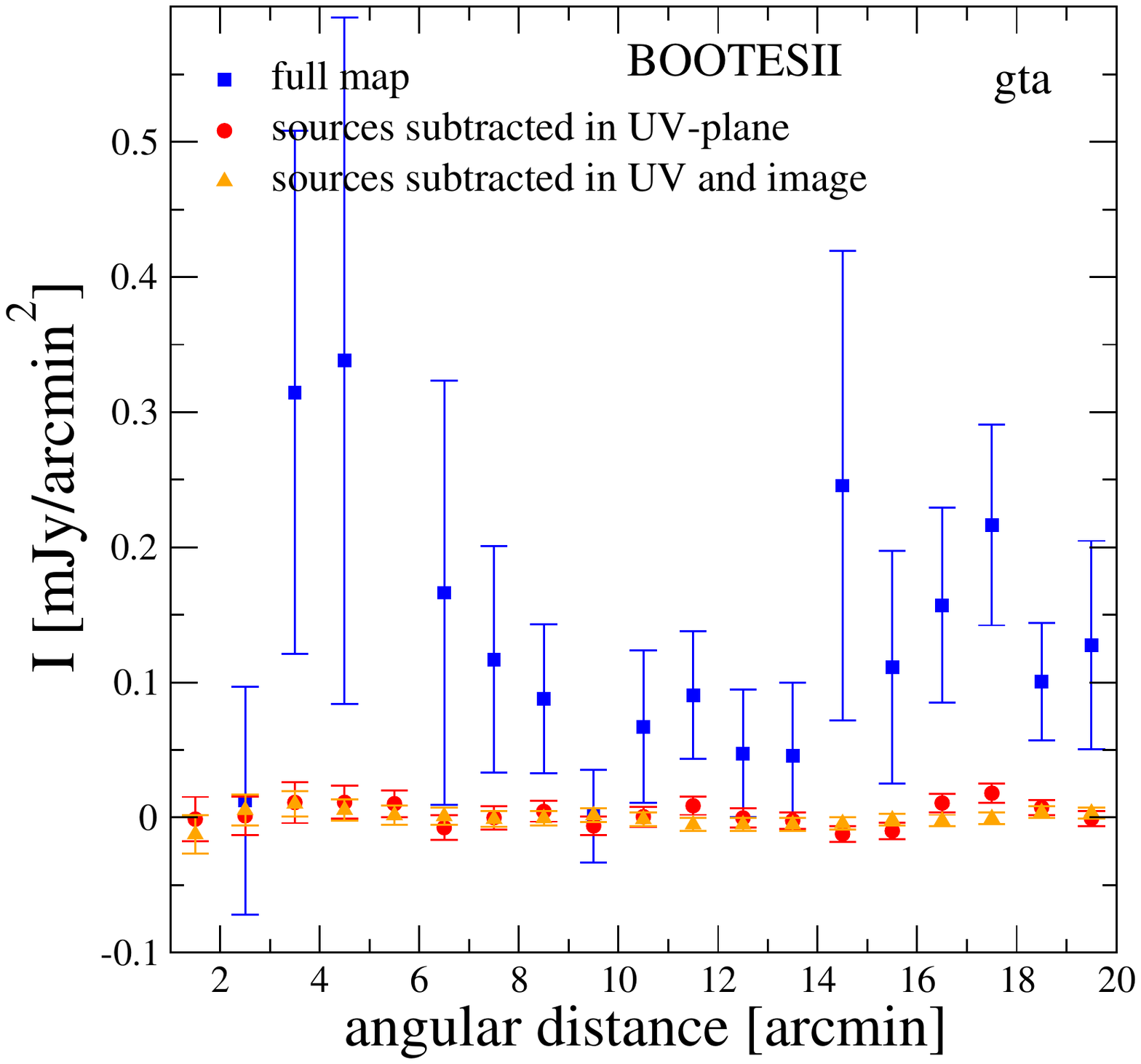}
 \end{minipage}
\hspace{-5mm}
 \begin{minipage}[htb]{6cm}
   \centering
   \includegraphics[width=\textwidth]{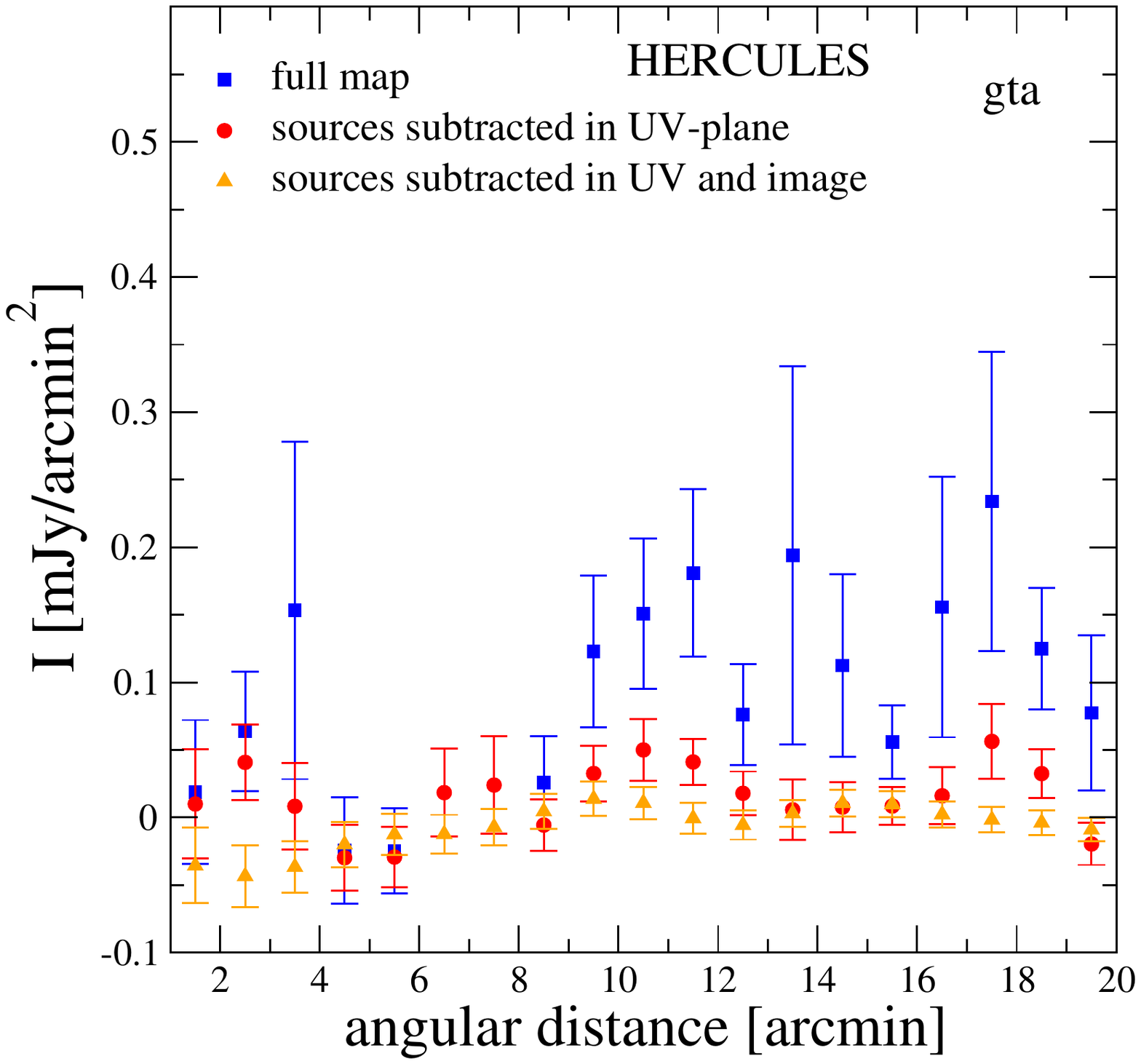}
 \end{minipage}
\hspace{-5mm}
 \begin{minipage}[htb]{6cm}
   \centering
   \includegraphics[width=\textwidth]{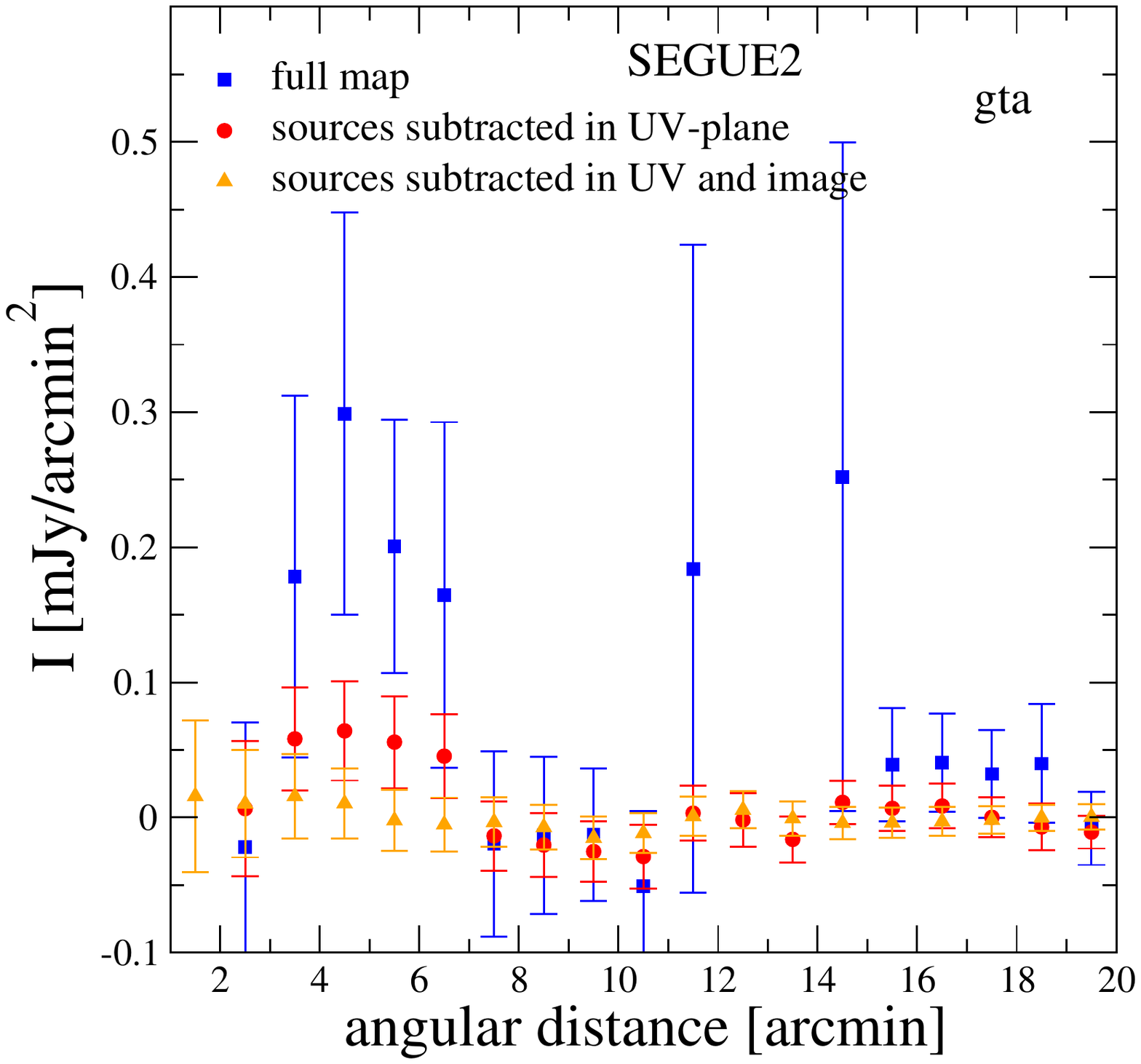}
 \end{minipage}
    \caption{{\bf Spherical profile}. Radial distribution of measured emission in the $gta$ maps, averaged in spherical annuli of 1 arcmin, as a function of the distance from the center.}
\label{fig:annuli_f15}
 \end{figure*}

In Fig.~\ref{fig:annuli_f15}, we show the radial distribution of the observed surface brightness in the $gta$ maps. The points are the average of the emission in spherical annuli of width of 1 arcmin, as a function of the distance from the dSph center.
The error bars are computed by summing in quadrature the average of the rms estimated as described above and the standard deviation of the emission within each annulus.
Blue squares include sources and the emission is not compatible with a null signal.
Red circles show the case with sources subtracted in the visibility plane.
In some cases, they show an evidence of emission. However, the pattern is always similar to the case including sources, although with a much lower amplitude.
This suggests that it is not a truly extended emission but rather a residuals of subtraction. This interpretation is supported also by the fact that after masking the region occupied by sources in the original map, we found that the curves do not show statistically significant deviations from the zero level.
Orange triangles show instead the case with sources subtracted both in the UV and image plane. They are always compatible with a null signal.

\begin{figure*}
\vspace{-2.cm}
   \centering
 \begin{minipage}[htb]{6cm}
   \centering
   \includegraphics[width=\textwidth]{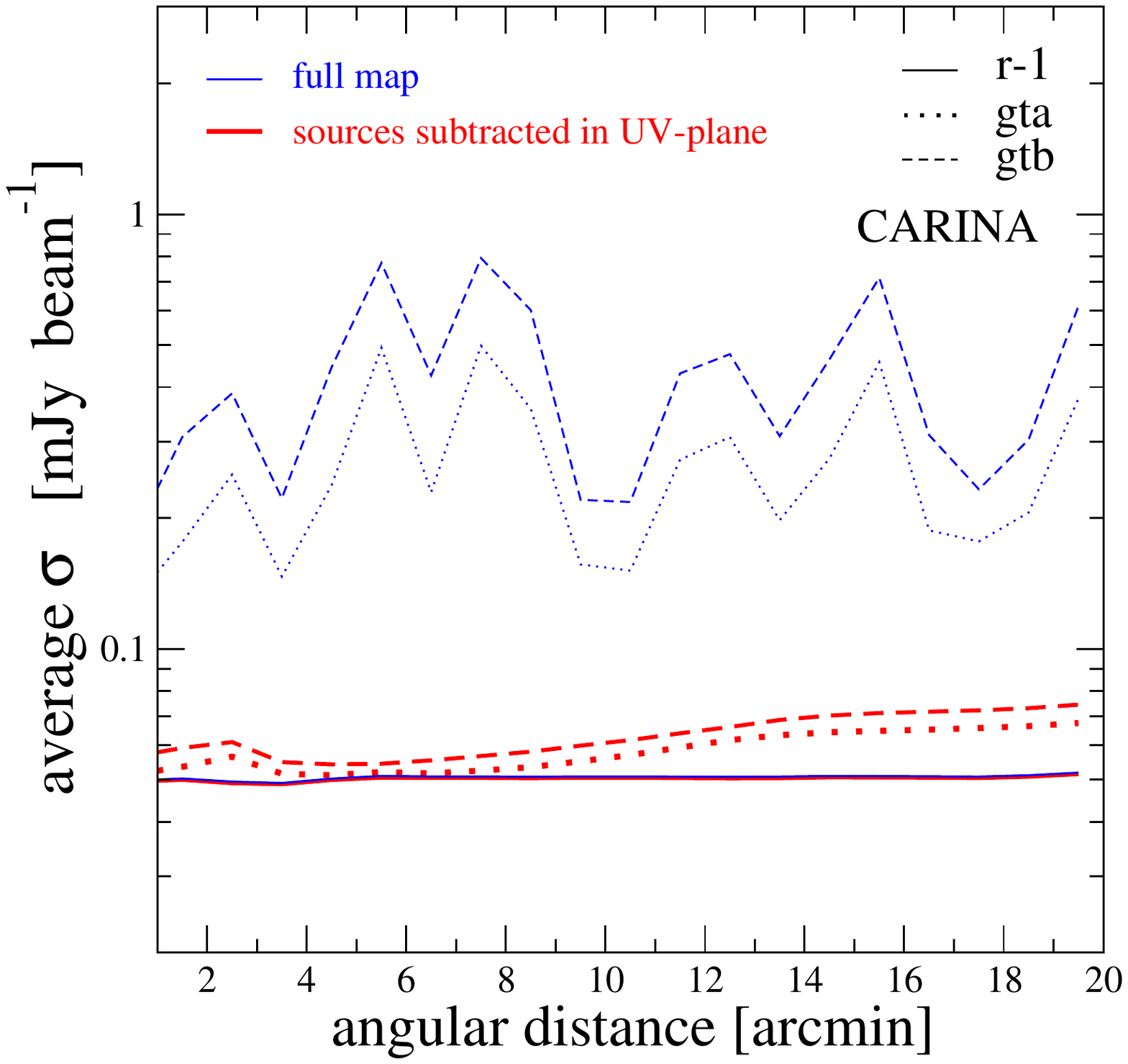}
 \end{minipage}
\hspace{-5mm}
 \begin{minipage}[htb]{6cm}
   \centering
   \includegraphics[width=\textwidth]{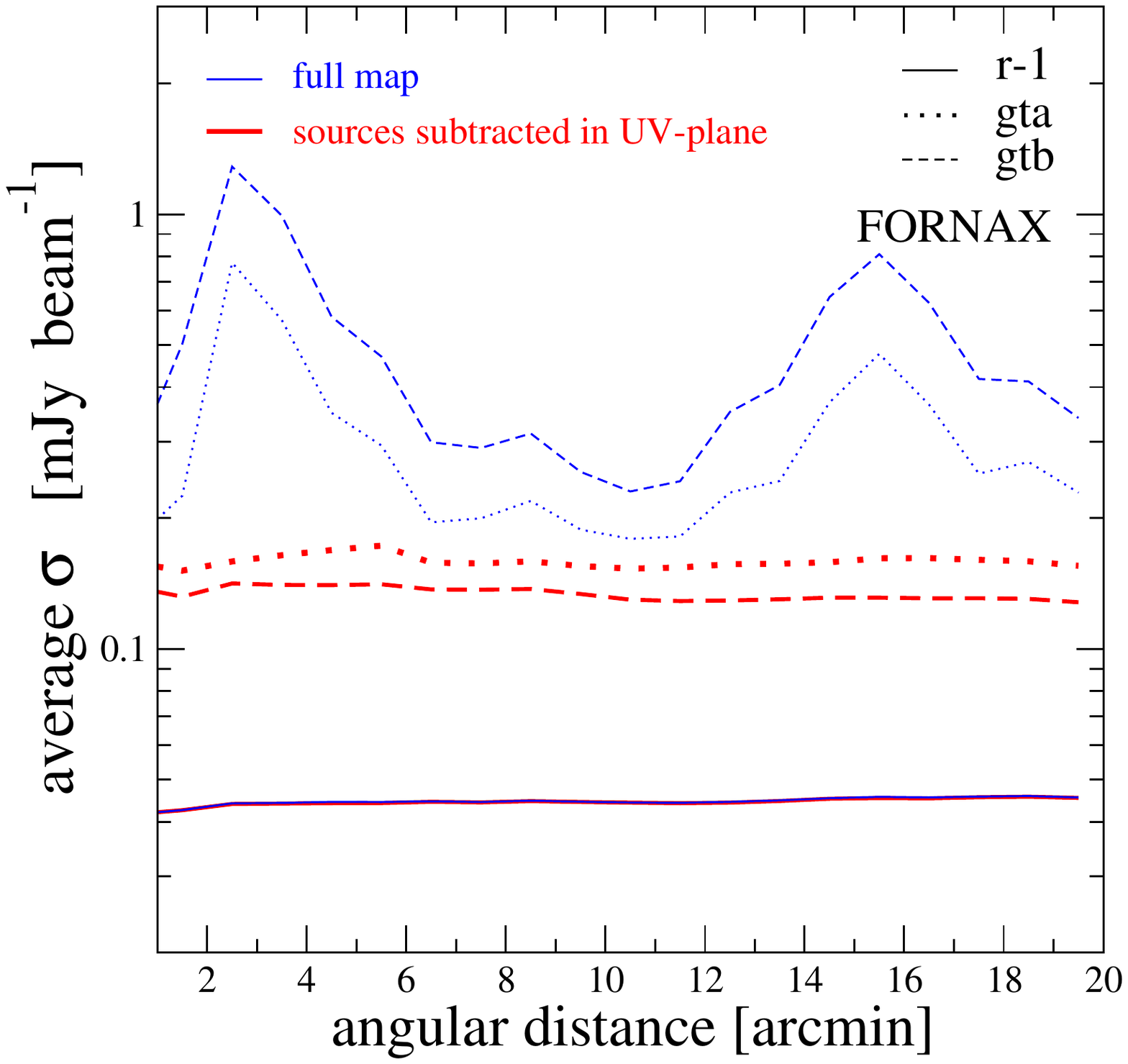}
 \end{minipage}
\hspace{-5mm}
 \begin{minipage}[htb]{6cm}
   \centering
   \includegraphics[width=\textwidth]{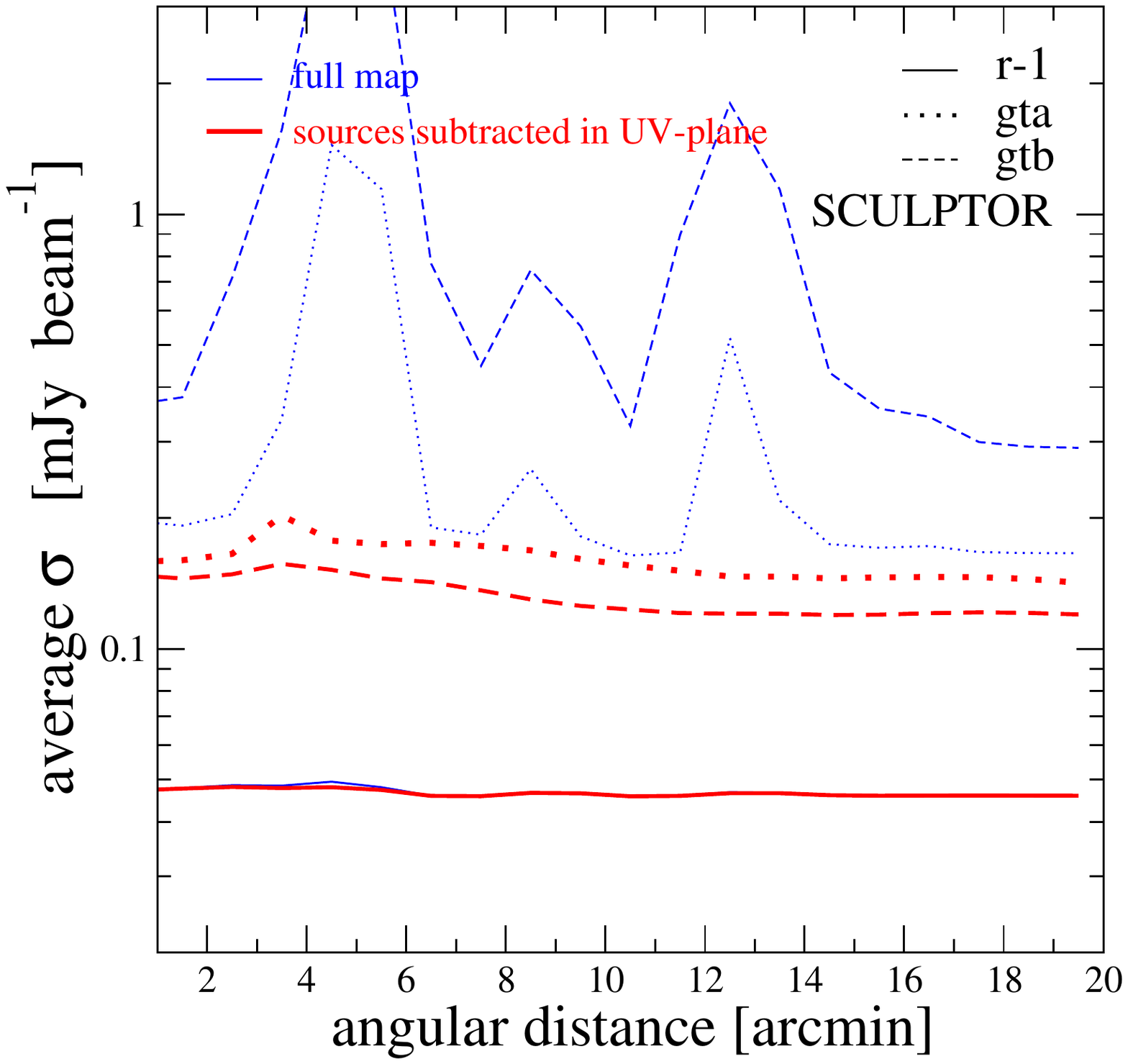}
 \end{minipage} \vspace{-2.5cm}\\ 
 \begin{minipage}[htb]{6cm}
   \centering
   \includegraphics[width=\textwidth]{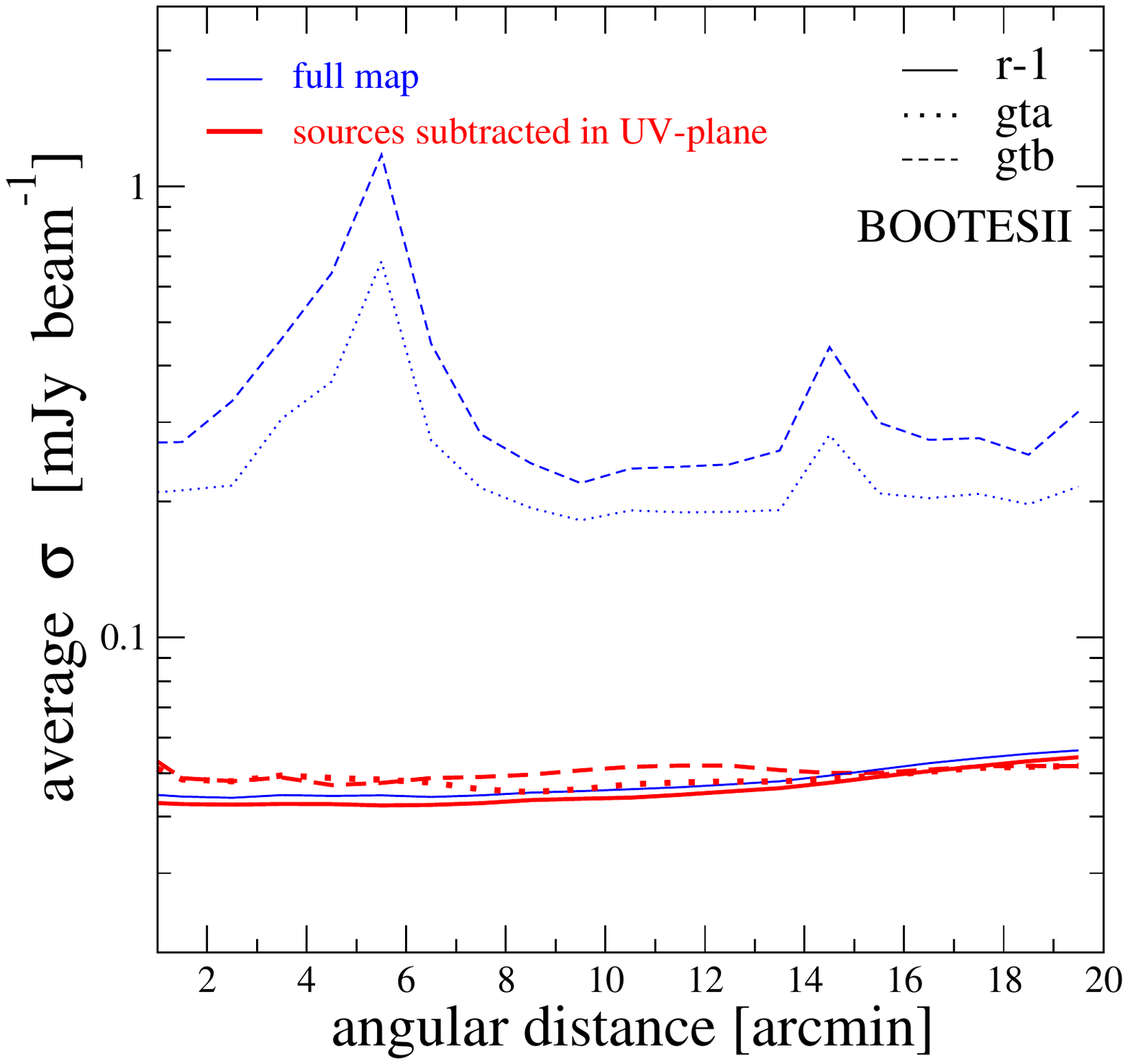}
 \end{minipage}
\hspace{-5mm}
 \begin{minipage}[htb]{6cm}
   \centering
   \includegraphics[width=\textwidth]{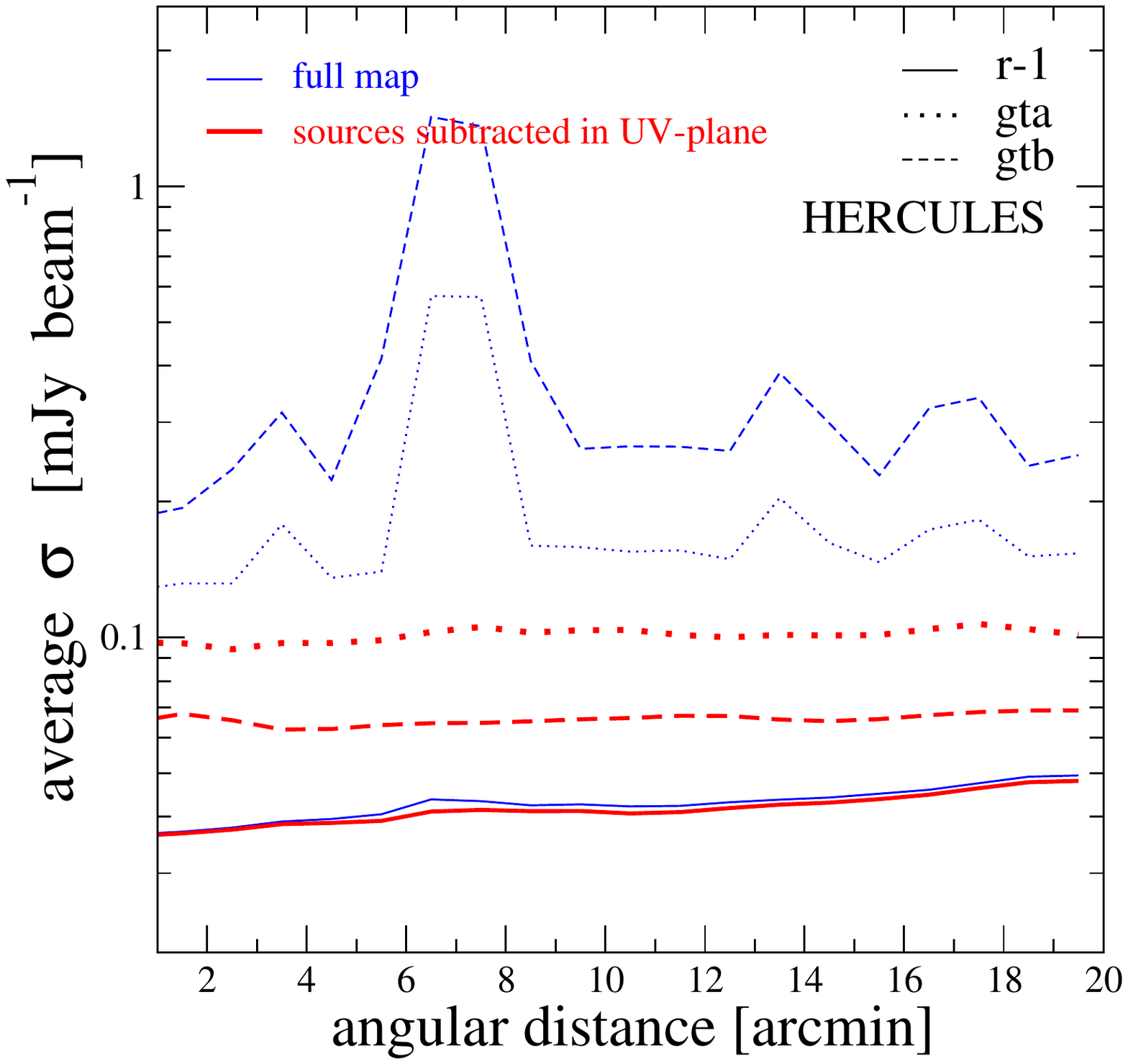}
 \end{minipage}
\hspace{-5mm}
 \begin{minipage}[htb]{6cm}
   \centering
   \includegraphics[width=\textwidth]{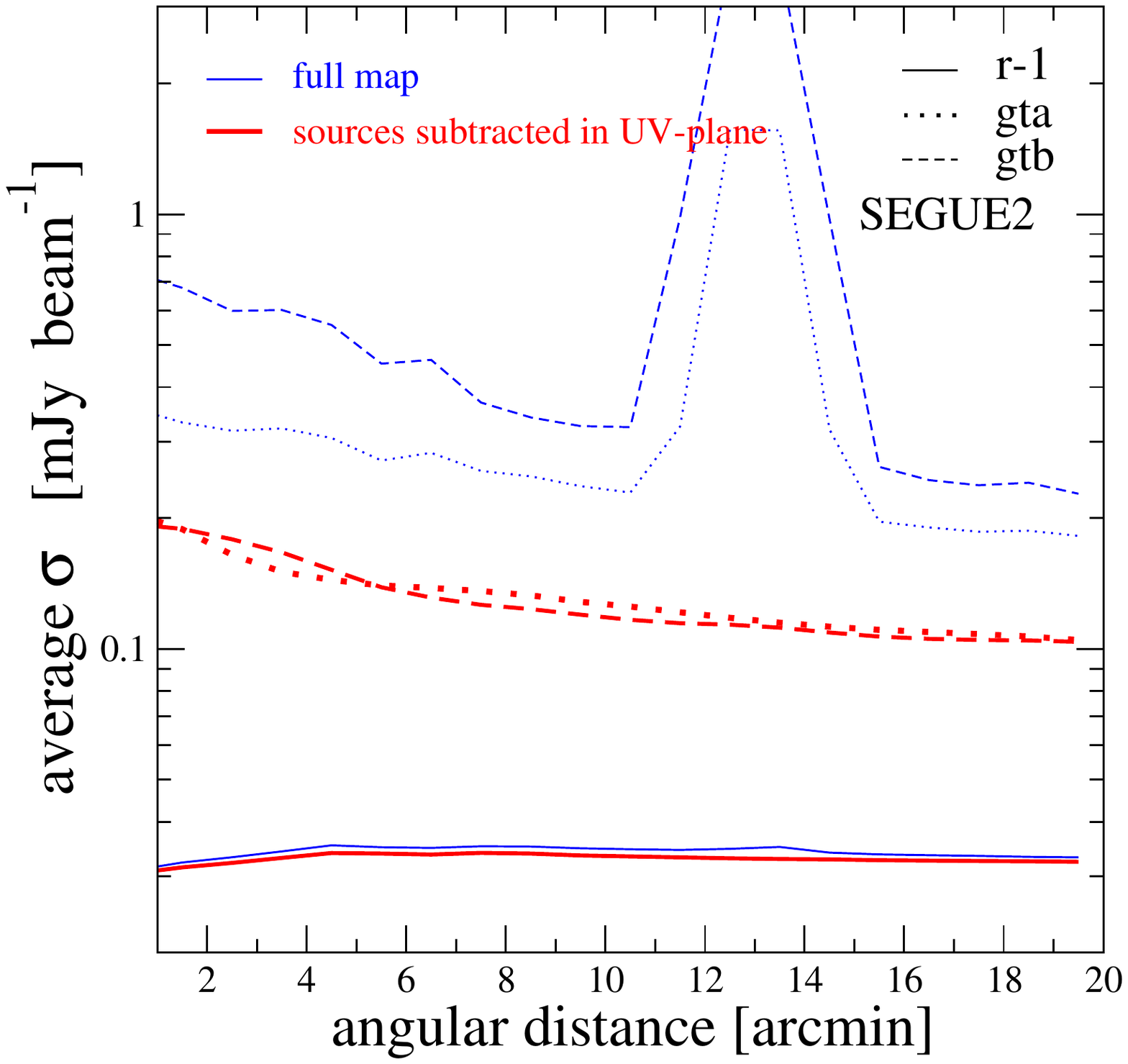}
 \end{minipage}
    \caption{{\bf RMS}. Estimate of the average $\sigma$ in spherical annuli of 1 arcmin, as a function of the distance from the center. It is obtained adding up in quadrature the rms and the standard deviation in each annulus. }
\label{fig:annuli_rms}
 \end{figure*}

In Fig.~\ref{fig:annuli_rms}, we show the radial distribution of the average fluctuations in the maps. It is the sum in quadrature of the rms and the standard deviation of the emission within each annulus. The latter shows pronounced peaks when sources are not subtracted (thin lines) in the $gta$ and $gtb$ maps. This is not the case in the $r_{-1}$ map, because the latter is not confusion limited and so sources always occupy only a small fraction of the annulus.
Once sources are subtracted (thick lines), all curves become smooth. Here it is shown for the UV-subtraction, but this is even more true (in some sense, by definition) when the subtraction is performed also in the image plane.
Note the gain of a factor of few provided by source subtraction in the confusion limited maps $gta$ and $gtb$.

\subsection{Largest well-imaged structures}
\label{sec:maxsize}

As already mentioned, the observations were conducted with the ATCA telescope in the hybrid array configurations H214 (for Carina, BootesII, Segue2, and part of Hercules) and H168 (for Fornax, Sculptor, and the second part of Hercules).
In the array configuration H214 the minimum baseline is $B_{min}=82$ m, while in the H168 case it is $B_{min}=61$ m. 
An estimate of the largest structure which can be well-imaged through a mosaic strategy is $\lambda/(B_{min}-D)$ corresponding to $9.2'$ and $14.1'$ in the H214 and H168 configurations, respectively, at the center of the bandwidth ($\lambda=16$ cm) and taking the antenna dish to be $D=22$ m. Taking the lower end of the bandwidth ($\lambda=27$ cm) we have an upper limit of the size from which we can get a signal, namely  $15.6'$ ($23.9'$) for H214 (H168).
We verified that the shortest UV-distances present in our data approximately match the latter estimates. 

The above numbers obviously apply to the setup including short baselines (i.e., the tapered images). For the long-baselines, $B$ grows to approximately 4.5 km. Therefore, in the case the short-baselines are down-weighted, the largest achievable scale significantly reduces. The $r_{-1}$ map is indeed sensitive only to the smallest scales up to about half arcmin.

The maximum size of well-imaged structures clearly depends, on the other hand, also on a number of observing details and it is not easy to have a precise a priori estimate. 
To overcome such difficulty, we perform few different simulations of detection of large scale emissions.
To this aim, we used the task IMGEN in {\it Miriad} to generate Gaussian emissions of different sizes and fluxes.
They have been converted into mock visibilities and added to the original observational data by means of the task UVMODEL.
The resulting visibilities are then reduced and imaged following the same pipeline as for the original maps.

\begin{figure*}
\hspace{-45mm}
   \centering
 \begin{minipage}[htb]{7cm}
   \centering
   \includegraphics[width=0.85\textwidth,angle=-90]{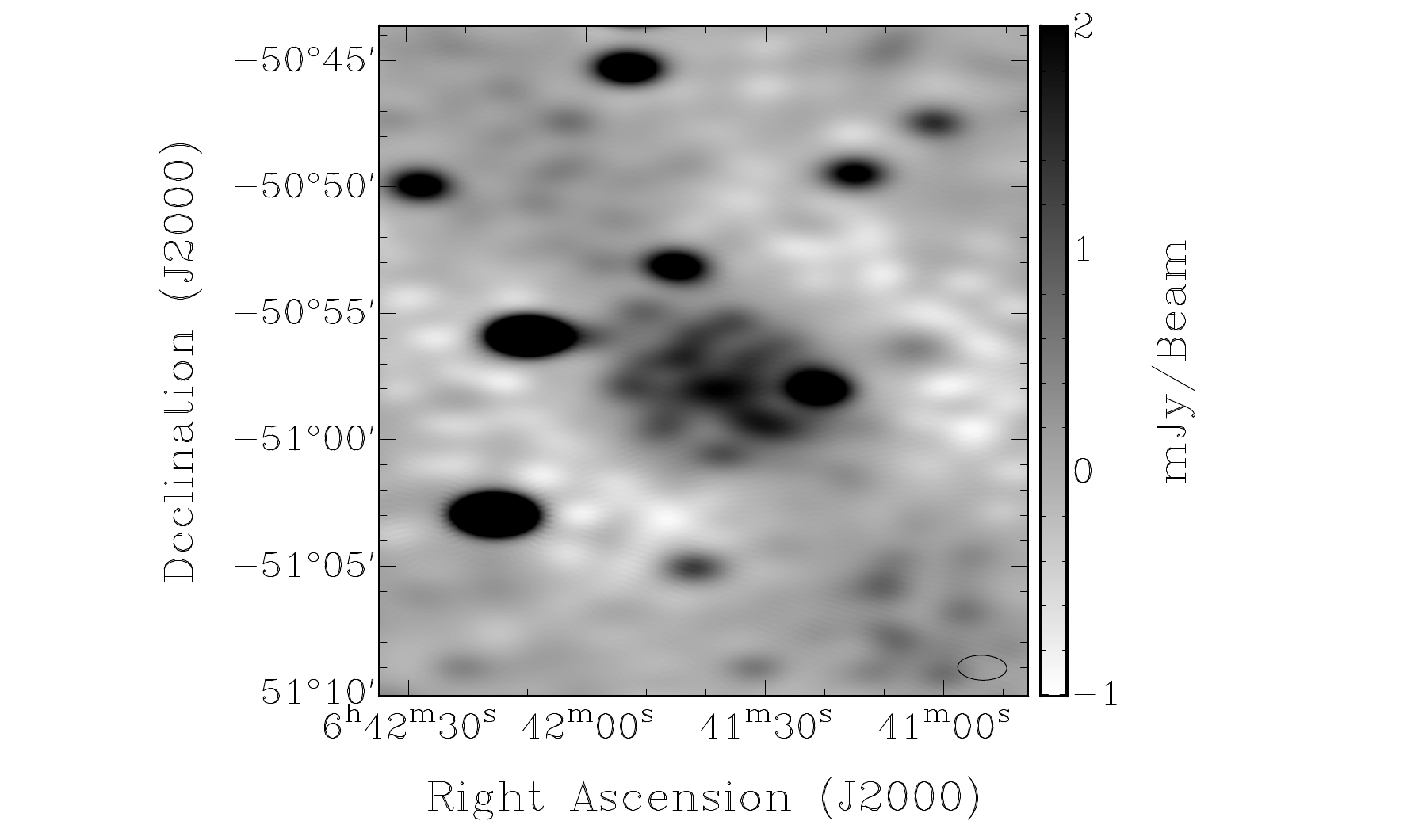}
 \end{minipage}
\hspace{5mm}
 \begin{minipage}[htb]{7cm}
\vspace{-8mm}
   \centering
   \includegraphics[width=1.05\textwidth,angle=-90]{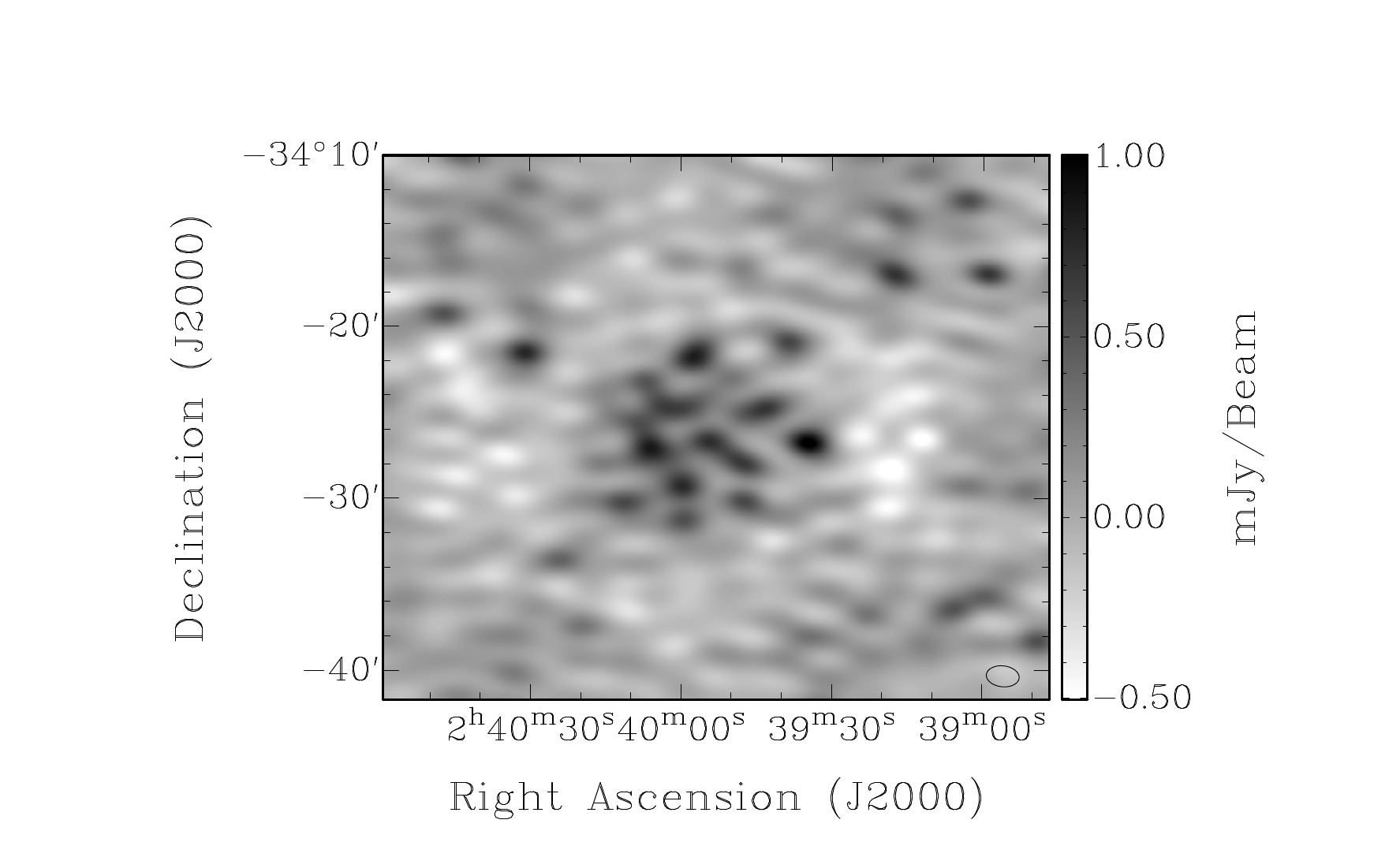}
 \end{minipage}
    \caption{{\bf Largest structures}. 
Examples of extended emission in the Carina (left) and Fornax (right) FoVs. The maps have been obtained from mock visibilities where the Fourier transform of a Gaussian with FWHM=$7.5'$ (left) and FWHM=$11.5'$ (right, with point-sources subtracted) is added to the original visibility data (by means of the task UVMODEL in {\it Miriad}). The synthesized beam is shown in the bottom-right corner of each panel.}
\label{fig:maxsize}
 \end{figure*}

We found that the estimates discussed above are approximately matched, and the reconstructed amplitude starts to decrease for sizes $\gtrsim15'$ ($\gtrsim10'$) in the H168 (H214) configuration.
With the exception of the Fornax dSph which size is comparable to such scale (and so the related bounds might be slightly optimistic), all the other dSphs have expected sizes well-within this scale (see $r_*$ in Table~\ref{tableB}). 
In Fig.~\ref{fig:maxsize}, we show two examples of $gta$ maps obtained with the above procedure, namely after the addition of a mock Gaussian emission. Left panel shows expectations in the Carina case (H214 array) considering FWHM=$7.5'$ and peak amplitude of 1.5 mJy for the mock Gaussian. The right-hand panel is instead a sort of extreme case that can be well imaged, showed for the Fornax FoV (H168 configuration) with FWHM=$11.5'$ and peak amplitude of 0.3 mJy (about $3\times$rms sensitivity).

A full assessment of the sensitivity of the current observations to the models discussed in the following would require the generation of a mock structure for each model.
This is, on the other hand, extremely time consuming. Since in our benchmark examples we find good agreement with expectations, for the sake of simplicity, we will compare models with observations directly in the image plane. We will assume an ideal response up to 15 arcmin, keeping in mind that for the largest scales this might slightly overestimate the sensitivity.

\section{Theoretical models}
\label{sec:mod}

In this section, we describe how we model the GHz-diffuse continuum emission in dSph (the analysis of the possible diffuse HI emissions will be discussed elsewhere contextually to the presentation of the relative data).
Thermal bremsstrahlung from ionized hydrogen clouds (HII regions) and synchrotron radiation from non-thermal electrons are the most notable emissions in galaxies which do not host an AGN.
Thermal re-radiation of starlight by dust becomes important only at frequencies $\gtrsim 100$ GHz, and is not important for our frequency range.

The free-free emission has a pretty flat-spectrum, with index $\alpha\sim 0.1$ (the spectral index is defined by $S\propto\nu^{-\alpha}$, with $S$ being the flux density and $\nu$ the frequency), while synchrotron radiation has a steeper spectrum ($\alpha\sim 0.8$), and typically dominates the radio emission of galaxies up to few tens of GHz.
Moreover, bremsstrahlung emission in dSphs is expected to be very faint given the low gas density.

We therefore focus only on synchrotron emission.
On the other hand, all the bounds on fluxes and emissivities that will be derived in the following can be straightforwardly extended to any thermal emission.

\subsection{Synchrotron emission}
\label{sec:em}

The total synchrotron emissivity at a given frequency $\nu$ is obtained by folding the electron number density $n_e$ with the total radiative emission power $P_{synch}$ \citep{Rybicki}:
\bea
j_{synch}(\nu,r)&=&\int dE\,P_{syn}(r,E,\nu)\, n_e(r,E)\\
&{\rm with}&\;\; P_{synch} (r,E,\nu)= \frac{\sqrt{3}\,e^3}{m_e c^2} \,B(r) F(\nu/\nu_c)\;,\nonumber
\label{eqjsynch}
\eea
where $m_e$ is the electron mass, the critical synchrotron frequency is defined as $\nu_c \equiv  3/(4\,\pi) \cdot {c\,e}/{(m_e c^2)^3} B(r) E^2$,  and $F(t) \equiv t \int_t^\infty dz K_{5/3}(z)$ is the function setting the spectral behaviour of synchrotron radiation.
To obtain the polarized emission, $F$ has to be replaced with $G(t)\equiv t\, K_{2/3}(t)$.
Absorption along the line of sight (l.o.s.) (from the dSph to us) is negligible at these frequencies. Similarly, for the thermal (see also arguments above), self-synchrotron, and self-Compton absorptions within the source which can be disregarded for the (non-compact) cases considered in this work.

The flux density measured by the ATCA telescope can be estimated as
\be
S_{th}(\nu,\theta_0) =\int d \phi\,d\theta\,\sin\,\theta\,\mathcal{G}(\theta,\phi,\theta_0)\int ds\,\frac{j_{synch}(\nu,r(s,\theta,\phi))}{4\pi}\;,
\label{eq:Isynch}
\ee
where $s$ labels the coordinate along the line of sight, $\theta_0$ is the direction of observation, i.e. the angular off--set with respect to the dSph center (the non-circularity of the beam can break the spherical symmetry but this is a very small effect), and we perform the angular integral assuming an elliptical Gaussian response of the detector $\mathcal{G}$ centered at $\theta_0$ and with widths $\sigma_{\theta}$ and $\sigma_{\phi}$ given by the synthesized beam sizes.
To compare theoretical prediction to observations in the case of a mosaic, one should compute $S(\theta_0)=\sum_i P^2(\bar \theta_0^i) S_{th}(\bar \theta_0^i)/\sum_i P(\bar \theta_0^i)$, where $S$ is the actual estimate of the observational flux, $S_{th}$ is the theoretical prediction described in Eq.~\ref{eq:Isynch}, $\bar \theta_0^i$ is the angle with respect to the center of each mosaic panel $i$, and $P(x)=\exp(-4\,\log2\,(x/FWHM)^2)$ is the primary beam pattern (note that $S_{th}^i=S^i P^{-1}$). However for all practical purposes one can identify $S$ with $S_{th}$ of Eq.~\ref{eq:Isynch}. Indeed, we can proceed to two simplifications. First, $S_{th}(\bar \theta_0^i)\simeq S_{th}(\theta_0)$; this is because the maximum difference in terms of radial distance between the case with a l.o.s. $s1$ at a given angle $\theta_0$ from the dSph center and the case with a l.o.s. $s2$ at a given angle $\bar\theta_0^i$ from the center of a panel is $s2/s1=\cos(\theta_{max})$ with $\theta_{max}\lesssim40'$ for our maps. This leads to a mismatch smaller than 0.01\% between the radial distances and so to a negligible difference in the flux computation.
The second simplification consists in neglecting the primary beam weighting. This is because we focus on the central part of the map (as mentioned above). This leads to $S\simeq S_{th}$ with the latter given by Eq.~\ref{eq:Isynch}. 

Since our bandwidth is quite large ($\Delta \nu\simeq 2$ GHz), we need to average the intensity over frequency: $\langle S(\theta_0)\rangle=1/\Delta \nu\,\int^{\nu_2}_{\nu_1}\,d\nu\,S(\nu,\theta_0)$ with $\nu_1=1.1$ GHz and $\nu_2=3.1$ GHz. Note that we can neglect the frequency dependence of the primary beam pattern only because the effect of the latter is negligible in the central part of the mosaic, as mentioned above. 

The synchrotron emission estimate involves the computation of the CR electron and positron equilibrium density $n_e$. We describe it in the limit of spherical symmetry and stationarity, making use of the following transport equation (where convection and diffusive reacceleration are neglected since they are likely to be irrelevant in dSphs):
\be
 -\frac{1}{r^2}\frac{\partial}{\partial r}\left[r^2 D\frac{\partial f}{\partial r} \right] 
  +\frac{1}{p^2}\frac{\partial}{\partial p}(\dot p p^2 f)=
  s( r, p)
\label{eq:transp}
\ee
where $f(r,p)$ is the $e^+-e^-$ distribution function at the equilibrium, at a given radius $r$ and in terms of the momentum $p$, related to the number density in the energy interval $(E,E+dE)$ by: $n_e(r,E)dE=4\pi \,p^2f(r,p)dp$; analogously,  for the source function of electrons or positrons, we have $q_e(r,E)dE=4\pi \,p^2\,s(r,p)dp$. 
The first term on the left-hand side describes the spatial diffusion, with $D(r,p)$ being the diffusion coefficient. The second term accounts for the energy loss of due to radiative processes; $\dot p(r,p)=\sum_i dp_i(r,p)/dt$ is the sum of the rates of momentum loss associated with the radiative process $i$.
Here we consider synchrotron and inverse Compton (IC) on cosmic microwave background (CMB) losses which leads to:\footnote{For simplicity, in this formula, Klein--Nishina corrections for IC are neglected (although they are not in our computations). This is a good approximation for scattering with CMB photons for electron energy up to 10 TeV.}
\be
 \frac{dp}{dt}\simeq 2.7\cdot 10^{-17}\,{\rm \frac{GeV}{s}}\,\big[1+0.095\,(\frac{B}{\mu G})^2\big]\,(\frac{p}{{\rm GeV}})^2\;.
\label{eq:eloss}
\ee
Models for the diffusion coefficient $D$, magnetic field $B$, and source term $q_e$ will be described in the next sections.
Eq.~\ref{eq:transp} is solved numerically making use of the Crank-Nicolson algorithm as described in the Appendix.

\subsection{Cosmic-ray sources}
\label{sec:CR}
To model the synchrotron flux from dSphs we consider two approaches.
First, we directly introduce a functional form for the emissivity $j_{synch}(r)$. 
With this phenomenological approach, we can provide pretty general bounds on the average of the emission $\langle j_{synch}\rangle$ and the spatial extension $r_s$ using some common functions as a Gaussian $j_{synch}=j_0\,e^{-x^2/2}$ and $\beta$-models $j_{synch}=j_0\,(1+x^2)^{-3\,\beta/2}$ where $x=r/r_s$. Note that the latter provides a form which might closely resemble naive expectations (i.e., disregarding possible reshapings due to interactions with the interstellar medium) for the emissions from DM with an isothermal profile ($\beta=4/3$ for annihilating and $\beta=2/3$ for decaying) and from stellar populations with Plummer ($\beta=5/3$) or modified-Plummer ($\beta>5/3$) distributions.

The second (more physical) approach involves instead the modelling of the CR electron density and magnetic properties of the dSph.
High-energy CR electrons are thought to be accelerated in galaxies by SN explosions and so their spatial distribution follows SF regions. Another possible origin (which is extensively discussed in Paper III) is related to DM annihilations/decays. If we assume stars to follow a (modified) Plummer profile and DM to be distributed with an isothermal profile, we can consider again all those cases simultaneously by employing $\beta$-models for the spatial part of the injection electron density $q_e(E,r)=dN_e/dE(E)\,(1+x^2)^{-3\,\beta/2}$ where $x=r/r_s$ and $r_s$ is the core radius of either the stellar or DM profiles. 
The spectrum is taken to be a power-law $dN_e/dE(E)=A_0\,(E/GeV)^{-p_{inj}}$ with the spectral index of injection $p_{inj}\simeq2-3$, which is what is predicted by the theory of first-order Fermi acceleration at astrophysical shocks (in the limit of strong shocks) \citep{Blandford:1987pw}.
We will often refer to the total CR energy density $Q_e(r)=\int_{E_{min}}^{E_{max}}dE\ E\ q_e(E,r)$ (with $E_{min}=100$ MeV and $E_{max}=1$ TeV).

We will also consider the computation of the signal starting from the equilibrium electron density $n_e$, taking the same functional forms for the spatial and spectral distributions outlined above for $q_e$. Diffusion and energy losses typically soften the spectrum and $n_e$ has a spectral index $p_{fin}\simeq3$.

To obtain an order of magnitude estimate of the CR density in dSphs from existing data, one can note that, empirically, a relation between the star formation rate (SFR) and the CR electrons density has been found to hold in galaxies.
If we assume $U_{el}\propto SFR$, take the normalization from a quiescent small galaxy like SMC (for which $SFR\simeq 5\cdot 10^4\,M_\odot\, Myr^{-1}$ and average CR nuclei density $\langle U_{p}\rangle =\langle k\,U_{el}\rangle \simeq1.5\cdot 10^{-10} {\rm GeV/cm^3}$~\citep{Abdo:2010}), and consider $k\simeq100$, we can then compute the average CR electrons density $\langle U_{el}\rangle$ in a dSph (where $U_{el}(r)=\int_{E_{min}}^{E_{max}}dE\,E\,n_e(E,r)$) from the associated SFR estimate. 
Only the SFR at late times is relevant to know the high-energy CR distributions (i.e., the population possibly producing a synchrotron emission at 2 GHz), since they lose energy in a relatively short amount of time, so must have accelerated recently. To compute the late-time SFR we follow~\citep{Orban:2008bb} assuming that about 1\% of the total stellar mass content of dSph is produced in the latest Gyr. 
We use SFR results reported in~\citep{Dolphin:2005mv} (Carina and Fornax), \citep{deBoer:2012dv} (Sculptor), and \citep{Sand:2009ut} (Hercules). 
For a very recent comprehensive study of SFR of dwarf galaxies in the Local Group, see \citep{Weisz:2014}. Their results are consistent with the models adopted here.
Ultra-faint dwarfs represent an observational challenge and currently there are too many uncertainties to infer their SF history. For Segue2 and BootesII, we will simply assume the same $\Sigma_{SFR}$ as in Hercules (which, among our sample, is the dSph which more closely resemble their properties), so the same $\langle U_{el}^{SFR_0}\rangle $. This argument is supported also by \citep{Brown:2013xna} where, analysing a sample of 6 UDS, they found that all cases have very similar and synchronized SF histories.

The derived estimates for $\langle U_{el}^{SFR_0}\rangle$ are reported in Table~\ref{tableB}.

We note that populations of primordial binary stars can actually mimic the signature of recent SF (the so-called, blue straggler problem, see, e.g., \citep{Momany:2007,Mapelli:2007,Mapelli:2009,Monelli:2012}) since binary evolution is typically not accounted for in current models. If such populations are significant in dSphs (something which has been a point of debate), the estimates of SFRs at late times considered in this work should be taken as upper limits.

Magnetic properties, which are the second crucial ingredient of the description, are discussed in the following section.

\subsection{Magnetic Field}
\label{sec:B}
The magnetic properties of dSphs are poorly known and to gain observational insights is very challenging.
The extremely low content of gas and dust makes polarization measurements difficult.
With our data we could attempt to estimate Faraday rotations of background sources (i.e., the rotation of the plane of linear polarization of the background-source waves when going through the dSph ionized medium due to the presence of a magnetic field).

However, the lack of observations of thermal emission in dSphs suggests a very low electron density, most likely well below the thermal density in the MW ($N_e^{MW}\simeq 10^{-2} {\rm cm}^{-3}$~\citep{Cordes:2002wz}) and not far from the cosmological electron density ($N_e^{cosm}\simeq 3\cdot 10^{-7} {\rm cm}^{-3}$).

In principle, a bound on the dSph thermal density can be obtained from null observations in the X-ray band.
The free-free emissivity at keV-frequency can be estimated as \citep{Longair}:
\bea
S_{X}&\simeq& 2\cdot 10^{-31} Z^2(\frac{T}{{10^4\,\rm K }})^{-1/2}\, \frac{g(\nu,T)}{1.2}\frac{N_i}{10^{-6}{\rm cm }^{-3}}  \\ 
&\times&\frac{N_e}{10^{-6}{\rm cm }^{-3}}\,\frac{l_{HII}}{100\,{\rm pc}}\exp(-\frac{h\,\nu}{k\,T})\,{\rm erg} \,{\rm s}^{-1}\, {\rm cm}^{-2}\,{\rm Hz}^{-1}\;,\nonumber
\eea
where $N_i$ and $N_e$ are the number density of the thermal ions and electrons, respectively, $T$ is the temperature of the plasma, $Z$ is the charge, $g$ is the Gaunt factor (typically lying in the range 1.1-1.5), and $l_{HII}$ is the size of the dSph HII region.
From the lack of observation of X-ray bremsstrahlung, one can infer a limit of $S_X\lesssim 4\cdot 10^{-32} (d/100 {\rm kpc})^{-2} {\rm erg}\, {\rm s}^{-1}\, {\rm cm}^{-2}\,{\rm Hz}^{-1}$~\citep{Zang:2001} and in turn on the thermal density of about $10^{-6}{\rm cm }^{-3}$.

The big assumption in this estimate concerns the required temperature. Indeed, in order to emit in the keV range, the thermal plasma has to be heated to temperature up to $10^7$ K, which are probably too high in the case of dSphs.

In any case, an estimate of the expected rotation measure is 
\bea
RM&=&0.81\ \int_0^{l_{dSph}}\  \frac{d{\bf s\cdot B}}{{\rm pc\ \mu G}}\ \frac{n_e}{{\rm cm}^{-3}} \ {\rm rad \, m}^{-2} \nonumber \\
&\simeq& 10^{-2} {\rm rad \, m}^{-2}\, \frac{B}{\mu G} \,\frac{N_e}{10^{-4} {\rm cm}^{-3}}\, \frac{l_{dSph}}{100\,{\rm pc}}\;,
\label{eq:RM}
\eea
and for reasonable assumptions about the thermal density, the result of Eq.~\ref{eq:RM} is well below the sensitivity of our observations.

For similar reasons, other polarimetric surveys do not provide strong bounds as well.

The most promising observational signal of the presence of magnetic fields in dSph stems thus from the detection of a polarized non-thermal radio emission which is the main goal of this project (with past surveys providing only weaker constraints) and will be discussed throughout the paper.

Due to the lack of observational evidences, the magnetic field models will be based on phenomenological/theoretical arguments described in the following.

\subsubsection{Magnetic field strength}
\label{sec:Bstrength}
Star-forming dwarf galaxies typically host a magnetic field of few $\mu$G, which provides an upper limit for the $B$-strength in dSphs.
There is no straightforward lower limit since the cosmological magnetic field could be in principle many orders of magnitude weaker.
However, different physical arguments suggest a strength of the magnetic field that is not too far from the one observed in star-forming dwarf galaxies (within an order of magnitude or so), as motivated in the next subsections.

{\bf Local Group scalings}: The generation of magnetic fields in galaxies is often described in terms of dynamo processes, which are sustained by turbulent energy. The main source of turbulence is often believed to be supernova explosions. Therefore one can expect a correlation between magnetic field and density of SFR $\Sigma_{SFR}$ in galaxies.
\citep{Chyzy:2011sw} analysed the magnetic field in Local Group galaxies, ranging from the MW to $10^7\,M_\odot$ dwarf irregulars.
A high level of correlation between $\Sigma_{SFR}$ and $B$ was found (the correlation coefficient is $r=0.94$) with the scaling well described by a power-law $B\propto\Sigma_{SFR}^{0.3\pm0.04}$.
This agrees well with findings for external more massive spiral and irregular galaxies, suggesting a similar mechanism for the generation of $B$ field at smaller scales.
Assuming that there is no threshold effect in such mechanism with respect to the gas-rich (and larger) systems detected in \citep{Chyzy:2011sw}, we can extrapolate this scaling law down to our dSph sample.
This assumption is also motivated by the fact that dSphs (at least classical ones) experienced a significant SF phase in the past (while being dominated by old stellar population at present) when the conditions for the generation of relevant magnetic fields were present (for a recent review on SF in LG dwarf galaxies, see e.g.~\citep{Tolstoy:2009jb}). 

At the initial stage of evolution, during the first few Gyr of active SF, dSphs and dIrrs show similar photometric properties~\citep{Calura:2008bb}. Then if such progenitors lose their gas, they undergo a change from irregulars to spheroidals (with a transition-type in between)~\citep{Dolphin:2005mv}.
A common progenitor for dSphs and dIrrs is also supported by models. For example, in the so called ``tidal stirring" scenario~\citep{Mayer:2001yf} dSphs originates from late-type, rotationally-supported dwarfs (resembling present-day dIrr galaxies) undergoing interactions with MW-sized galaxies.
Therefore dSphs should have hosted a magnetic field similar to that of dIrrs (i.e., few to ten $\mu G$; for a review of dIrrs interstellar medium see, e.g., \citep{Klein:2012}). 
However, after such initial phase, a large fraction of gas is swept away from dSphs, which then evolve passively.
A key question is thus whether such magnetic field can be sustained until present epoch. 
Since the strength of magnetic field is very low, an extremely low density plasma would suffice to prevent the decay, in absence of turbulences.
Indeed the relevant equation for describing the Ohmic decay of a large scale magnetic field is \citep{Parker:1979}: $\partial B/\partial t =\eta\,c/(4\,\pi)\nabla B$. The estimate for the decay time is then $\tau=4\,\pi\,L^2/(\eta\,c)\simeq 10^{20} \frac{L}{100 {\rm pc}}\frac{n_e}{10^{-6} {\rm cm}^{-3}}$ yr. 

On the other hand, turbulences can destroy the magnetic field in a time-scale much shorter than the age of the galaxy.
Episodes of weak SF (forming a few
percent of the total stellar mass~\citep{Orban:2008bb}) are likely to have occurred at recent time (see e.g. colour-magnitude diagram of Carina~\citep{Dolphin:2005mv,Hernandez:2000qz} and Fornax~\citep{Coleman:2008kk}, with the caveat related to the possible presence of blue straggler populations mentioned at the end of Section 4.2). This implies the presence of some small fraction of ionized medium.
The source of turbulences, however, would also give rise to magnetic field generation via dynamo action, provided the interstellar plasma is sufficiently dense.
In other words, unless a peculiar situation with significant non-thermal processes in a very-low density plasma, we expect the magnetic field in dSph to be around the $\mu$G level. 

To be quantitative, we pursued two approaches. In the first method, we use the relation of \citep{Chyzy:2011sw} mentioned above to link $B$ with $\Sigma_{SFR}$ at each different phase and then take the averaged value over the history of the dSph. In the second, we instead consider the magnetic field to be induced only by the SF in the latest Gyr and assume (following \citep{Orban:2008bb}) that 1\% of the total stellar mass content of the dSph is produced in such range of time.
We consider the same SFR estimates reported in Section~\ref{sec:CR} (i.e., \citep{Dolphin:2005mv} for Carina and Fornax, \citep{deBoer:2012dv} for Sculptor, and \citep{Sand:2009ut} for BootesII, Hercules, and Segue2).

The normalization of $B$ is obtained from Large Magellanic Cloud data~\citep{Gaensler:2005qj}, namely, taking a total magnetic field strength of $B=4.3\,\mu$G for $\Sigma_{SFR}\simeq 4\cdot10^3\,M_\odot\,{\rm kpc^{-2} Myr^{-1}}$, which implies $B=0.35\mu G\,(\Sigma_{SFR}/(M_\odot\,{\rm kpc^{-2} Myr^{-1}}))^{0.3}$.
The corresponding estimates for our dSph sample are reported in Table~\ref{tableB}. It is interesting to note that the two different methods provide similar results, with only a moderate depletion of $B$ when focusing on recent time.

For what concerns the spatial profile, we will assume spherical symmetry and a simple exponential law, $B=B_0\,e^{-r/r_h}$, with $r_h=r_*$ set by the stellar halo scale.

\noindent{\bf Magnetization of MW surrounding medium}:
Galactic outflows typically magnetize the medium surrounding spiral galaxies up to several kpc away from the source-region.
Indeed giant magnetized outflows from the centre of the Milky Way have been recently observed.
They point towards a magnetic field larger than $10\, \mu$G at 7 kpc from the Galactic plane~\citep{Carretti:2013sc}.

These arguments support the idea that a non-negligible magnetic field can be hosted by UDS, which are at about 40 kpc from the Galactic center, even if they have never undergone a significant SF.
Assuming we can adopt a magnetic field with a simple linear scaling $B_{MW}=50\,\mu G\,/(d/{\rm kpc})$ in the limit of large distances $d$ from the center of the MW, we derive $B_{MW}$ at dSph positions  in Table~\ref{tableB}.
For the CDS, $B_{MW}$ would be negligible with respect to the magnetic field generated by the dSph itself and estimated in the previous subsection (and also the extrapolation we adopted is too simplistic at such large distances), while it could indeed be the dominant magnetic term in the UDS. 
Since the dSph size is much smaller than the distance from the MW, we can assume $B_{MW}$ to be constant over the size of the dSph.

\noindent{\bf Equipartition}:
A simple way to avoid the introduction of a magnetic field model is to impose a minimum energy condition for the synchrotron source at each position in the dSph. Taking the energy density of the relativistic plasma to be dominantly provided by CRs and magnetic fields, this condition sets $B$ in terms of the CR density.
The minimum energy required to produce a synchrotron source of a given luminosity is provided by $U_B(r)=B(r)^2/(8\,\pi)=3/4\,U_{CR}(r)$, where $U_B$ and $U_{CR}$ are the magnetic and CR energy density, respectively (see, e.g., \citep{Longair}). As known, this corresponds to (near) equipartition (where here we assume local equipartition). $U_{CR}$ can be written as $(1+k)\,U_{el}$ where $U_{el}$ is the energy density of the synchrotron emitting electrons $U_{el}(r)=\int_{E_{min}}^{E_{max}}dE\,E\,n_e(E,r)$ (we choose again $E_{min}=100$ MeV and $E_{max}=1$ TeV) and $k$ gives the ratio between hadronic and electron CR energy density. Fermi shock acceleration and hadronic interaction models (as well as local CR data) favour $k\sim100$, which will be considered for the estimates of $B$ (slightly less conservative estimates can be obtained in case of leptonic models with $k=0$).

Note that with the assumption of local equipartition the magnetic field is obtained at all positions in the dSph (so we do not need a model for the spatial dependence), and is related to the CR spatial profile. In Table~\ref{tableB} we quote the volume-averaged $B$ over the stellar region $\langle B\rangle=3\,r_*^{-3}\int^{r_*}_0 dr\,r^2 B$.

In the last column of the same Table, we also quote the magnetic field one would obtain assuming equipartition with the CR density estimated as described at the end of Section~\ref{sec:CR}.
It is interesting to note that the strength is very low, in particular lower than $B_{SFR_0}$ (which also relies on late-time SFR).
This is because, in the relations considered to derive the latter, equipartition does not hold. Indeed, we take $U_{el}\propto SFR$ and $B\propto\Sigma_{SFR}^{0.3}$, thus $U_{el}$ is not proportional to $B^2$. If equipartition is assumed to hold, one of the two adopted scalings needs to be revised. On the other hand, although providing a rough estimate of the ball-park for electron and magnetic densities, equipartition is not expected to precisely hold, especially in a peculiar system like dSph (and indeed observationally is found not to hold for many systems).

The impact of the magnetic field model on the final results will be discussed in Sec.~\ref{sec:bounds}.

\begin{table*}
\centering
\begin{tabular}{|l|c|c|c|c|c|c|c|c|}
\hline
dSph & $D$ & $r_*$ & $B_{\overline{SFR}}$ &$B_{SFR_0}$&  $B_{MW}$  & $\langle B_{eq}^{obs}\rangle $& $\langle U_{el}^{SFR_0}\rangle $ & $\langle B_{eq}^{SFR_0}\rangle $ \\
name &[kpc]& $[']$   & $[\mu G]$        &$[\mu G]$  &  $[\mu G]$ & $[\mu G]$& $[10^{-16}{\rm GeV/cm}^3]$& $[\mu G]$  \\
\hline
Carina & 105 & 8.2 & 0.9 & $0.7$ & 0.5 & $<3.8\,(2.5) $ &2.7 &0.03\\ 
Fornax & 147 & 16.6 & 2.0 & $1.2$ & 0.3 & $< 4.2\, (2.0)$ &96 &0.2\\ 
Sculptor & 86 & 11.3 & 1.6 & $1.2$ & 0.6 & $<6.7\,(2.9)$ &23 &0.1\\ 
BootesII & 42 & 4.2 & 0.4 & $0.4$ & 1.2 & $<6.3\,(6.6)$ &0.45 &0.01\\ 
Hercules & 132 & 8.6 & 0.4 & $0.4$ & 0.4 & $< 4.6\, (2.6)$ &0.45 &0.01\\ 
Segue2 & 35 & 3.4 & 0.4 & $0.4$ & 1.4 & $< 7.3\, (10.6)$ & 0.45 &0.01\\ 
\hline
\end{tabular}
\caption{dSph parameters. Columns 2 and 3 show, respectively, dSph distance $D$ and stellar radius $r_*$ (containing half the light of the galaxy), taken from \citep{McConnachie:2012vd} (see references therein). Column 4 and 5 report the magnetic field strength obtained from Local Group scalings discussed in Sec.~\ref{sec:Bstrength}. Column 6 shows the estimate of $B$ we derived from possible magnetization of MW surrounding medium. Column 7 is the equipartition bound obtained from data considering the $gta$ maps with source subtracted from the visibility (and image) plane. Column 8 reports the estimate of the CR density from the dSph SFR discussed in Sec.~\ref{sec:CR}. The associated equipartition magnetic field is in column 9.}
\label{tableB}
\end{table*}

\subsubsection{Turbulence properties}
\label{sec:turb}
As mentioned in Sec.~\ref{sec:em}, we describe the transport of high-energy charged particles in dSphs as a diffusive process. It is governed by the scattering of CR particles with the hydromagnetic waves of the interstellar medium, and so it is set by the turbulence properties of the magnetic field.
In order to account for our poor knowledge of dSph magnetic properties, we consider three limiting cases.

\noindent{\bf Loss at injection-place}: When turbulence is very strong, particles can be considered as being essentially confined at the same place of injection, where they radiate all of their energy. This can be described by Eq.~\ref{eq:transp} neglecting the diffusion term which leads to:
\be
n_e(E,r)=\frac{1}{b(E,r)} \int^{\infty}_E dE'\,Q_e(E',r)\;.
\label{eq:transp1}
\ee

Therefore, in this case we do not need to model $D$.
Note also, from Eq.~\ref{eq:eloss}, that the typical loss time is below hundreds of Myr, so the description concerns recent time (and it is reasonable to neglect time-evolution).

\noindent{\bf Free-escape}: The opposite limit with respect to the above picture is when turbulences are extremely weak and there is no other efficient mechanism for confinement. In this case, particles can easily escape the object and they are subject only to energy losses along their way. The latter can be however neglected in the computation of the equilibrium distribution since they only mildly affect the electron energy. Indeed, for GeV electrons we have $dE/dt\lesssim10^{-16}$ GeV/s which means on dSph-scale ($\lesssim 1$ kpc) $\Delta E\lesssim 0.01$ GeV. 
Thus the electron density can be simply found through the equation:
\be
n_e(E,r)=\frac{1}{4\,\pi\,c} \int\,d\phi\,d\theta\,\sin\,\theta\,\int\,ds\,Q_e(E,r'(s,\theta,\phi,r))\;,
\label{eq:transp2}
\ee
with $r'=\sqrt{s^2+r^2-2\,r\,s\,\cos\theta\,\cos\phi}$.\footnote{Since, even in this idealized picture, electrons do not escape with straight trajectories but rather spiralling around magnetic field lines, the velocity $c$ in above equation should be replaced by an effective velocity $v_p=c\,\cos\,p$, where $p$ is the pitch angle. However, under reasonable assumptions (e.g., isotropic distribution of pitch angles), the factor $\cos\,p(r')$ averaged over the l.o.s. is $\mathcal{O}(1)$, and in the following, we take $v_p=c$.}
Note that the negligible impact of synchrotron radiation (which is the signal we aim to detect) on the computation of $n_e$ means also that most of the CR electrons power is actually carried out the dSph, and so this scenario will be much less promising than the above one (where instead all the power is radiated within the dSph).

A free-escape picture is somewhat too pessimistic. Indeed CRs cannot stream along a magnetic field much faster than the Alfv\`en speed because they generate magnetic irregularities which in turn scatter them (see, e.g., \citep{Cesarsky:1980pm} for a review). 
On the other hand, if we assume that the only ionized medium in the dSph is in fact the cosmological population (on top of the CR component), then the typical associated Alfv\`en speed will be very large, a fraction of the speed of light.
This means that the confinement time would be just a factor of few larger with respect to the free-escape case and so the synchrotron flux a factor $\mathcal{O}(1)$ larger. The possibility of an accurate modelling is limited by uncertainties in the density of ionized gas and magnetic field. Moreover, it would require a description in terms of convection rather than diffusion.
Therefore, we will still keep free-escape as the most conservative case, but keeping in mind that, even in the absence of turbulence, some confinement is expected, with bounds being at least a factor of few stronger than for the reported free-escape scenario.

\noindent{\bf Standard diffusion}: In between of the above two cases, turbulences can play a major role but allowing particle diffusion lengths on scales comparable to the object-scale. This is the typical scenario for the MW and other LG galaxies.
At the energies of interest, $\mathcal{O}(\rm{GeV})$, and in the quasi linear approximation, the diffusion tensor can be simplified to a scalar with the form:
$$ D=\frac{v\,r_g}{12\,\pi}\frac{B^2}{k_{res}\,P(k_{res})}=\frac{v\,r_g^\alpha}{3\,(1-\alpha)}\frac{B^2}{k_L^{1-\alpha}\,\delta B_L^2}\,,$$ where $r_g=1/k_{res}=R/B$ is the gyroradius (with R=particle rigidity), $P(k)\propto k^{\alpha-2}$ is the turbulence power spectrum, $k_L$ is the wavenumber of choice for the normalization of the random magnetic field (i.e., at $k_L$ it takes the value $\delta B_L$ and we normalize the power spectrum through $\int^{\infty}_{k_L}dk\,P(k)=\delta B_L^2/4\pi$), and for the spectral index $\alpha$ we assume, for simplicity, a Kolmogorov power-spectrum $\alpha=1/3$ (note that for electrons $D\propto E^\alpha$).

With this formalism, and once the total magnetic field strength is set (see previous section), the computation of the diffusion coefficient $D_0$ (with $D=D_0\,(E/{\rm GeV})^\alpha$) requires only to know the ratio between coherent and turbulent components $B/\delta B$, with $D_0\propto B^{2-\alpha}/\delta B_L^2$.
For typical values of such ratio in galaxies of the LG one finds $D_0\sim3\cdot10^{28}{\rm cm^2/s}$.
For the spatial dependence of $D$ we consider a profile related to the $B$ shape, namely, $D(r)\propto e^{r/r_*}$.

The numerical solution of Eq.~\ref{eq:transp} for this scenario is described in the Appendix.
The requirement of numerical convergence puts bounds on the minimal and maximal allowed diffusion timescales. This translates into bounds on the diffusion coefficient. They are similar to the bounds one would obtain from physical arguments, e.g., requiring to have a diffusion time ($\tau_d\sim L^2/D\sim 10^{15}s\,(L/{\rm kpc})^2(10^{28}{\rm cm^2/s}/D)$) shorter than the age of the Universe ($\sim4\,10^{17}$ s), which provides a lower bound (typically, $D_0\gtrsim10^{26}{\rm cm^2/s}$ at 1 GeV), below which we have the loss at injection place scenario, and to have a diffusion velocity ($v_d=L/\tau_d\sim 10^{-4}c\,(D/10^{28}{\rm cm^2/s})/({\rm kpc}/L)$) smaller than the speed of light, which provides an upper bound (typically, $D_0<10^{32}{\rm cm^2/s}$ at 1 GeV), corresponding to the free-escape limit.

\begin{figure*}
\vspace{-2.cm}
   \centering
 \begin{minipage}[htb]{6cm}
   \centering
   \includegraphics[width=\textwidth]{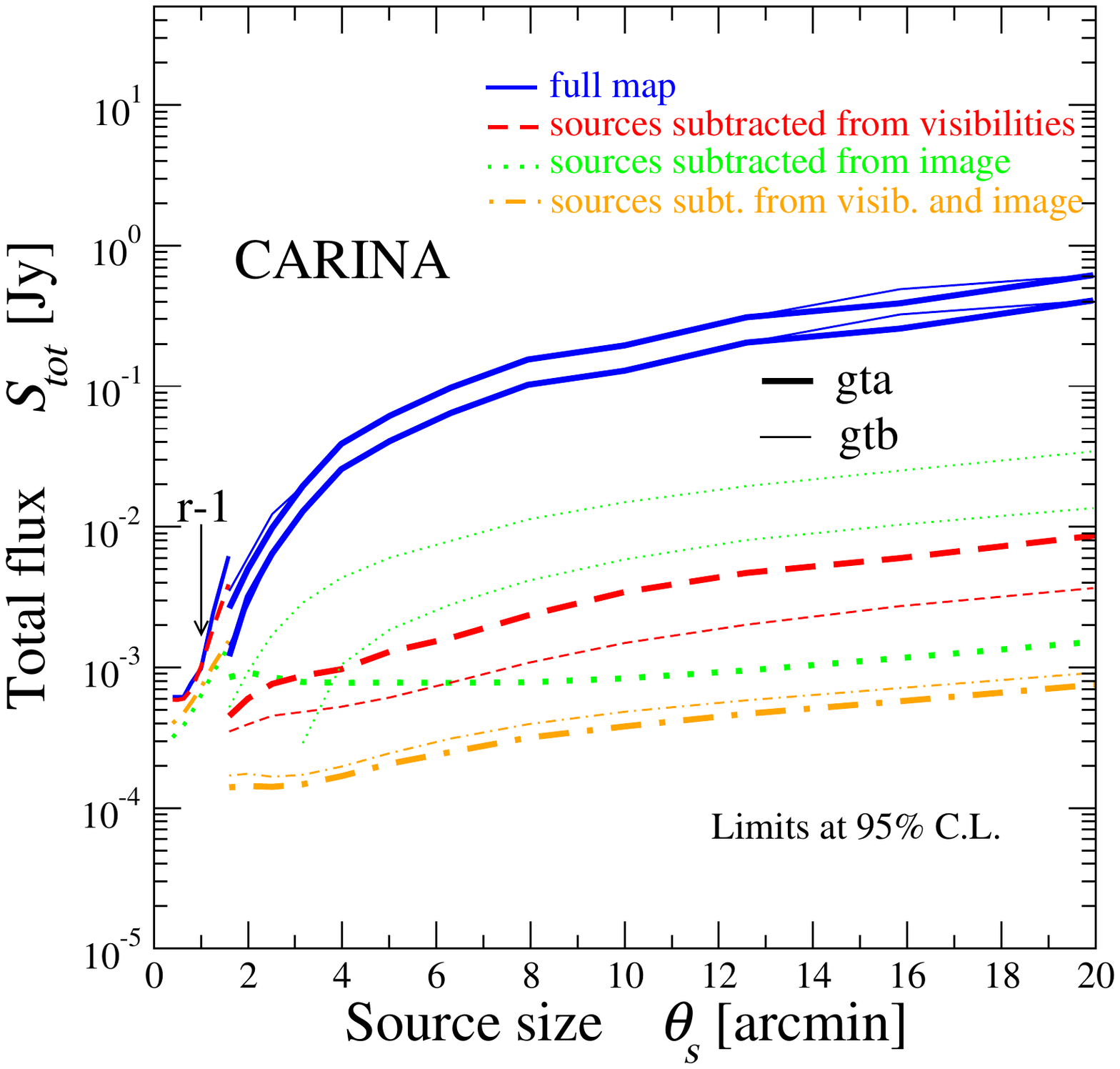}
 \end{minipage}
\hspace{-5mm}
 \begin{minipage}[htb]{6cm}
   \centering
   \includegraphics[width=\textwidth]{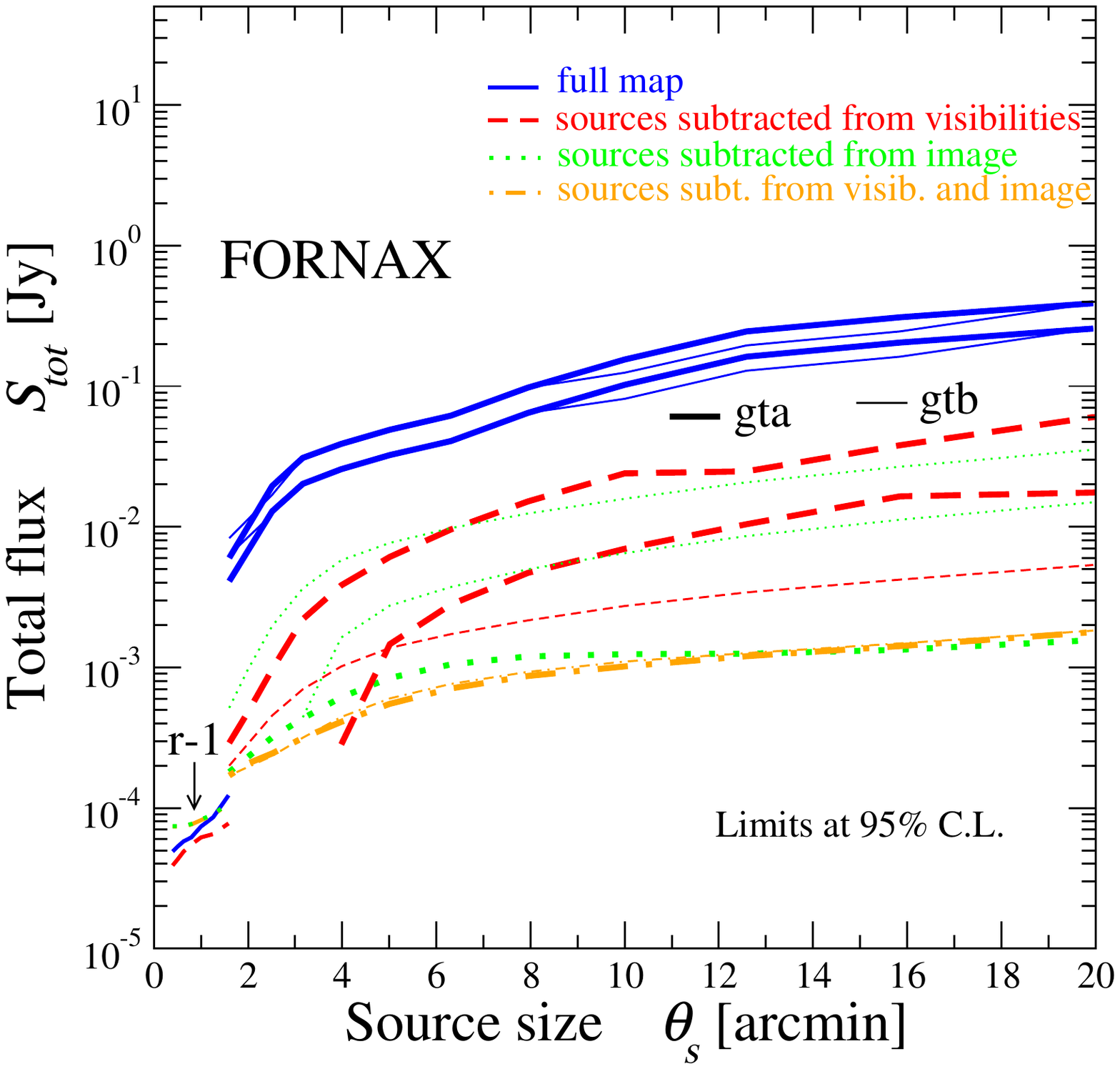}
 \end{minipage}
\hspace{-5mm}
 \begin{minipage}[htb]{6cm}
   \centering
   \includegraphics[width=\textwidth]{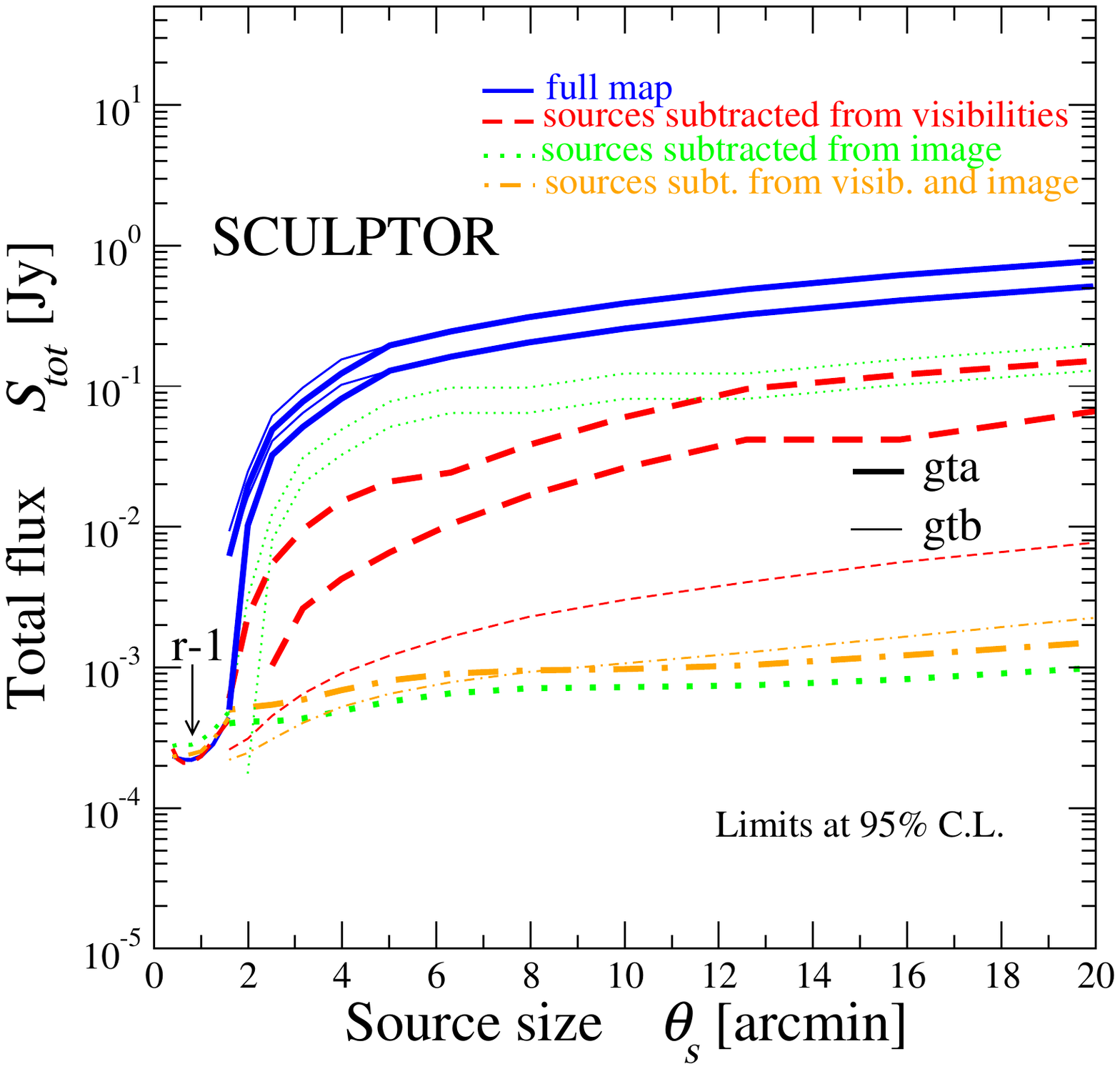}
 \end{minipage} \vspace{-2.5cm}\\ 
 \begin{minipage}[htb]{6cm}
   \centering
   \includegraphics[width=\textwidth]{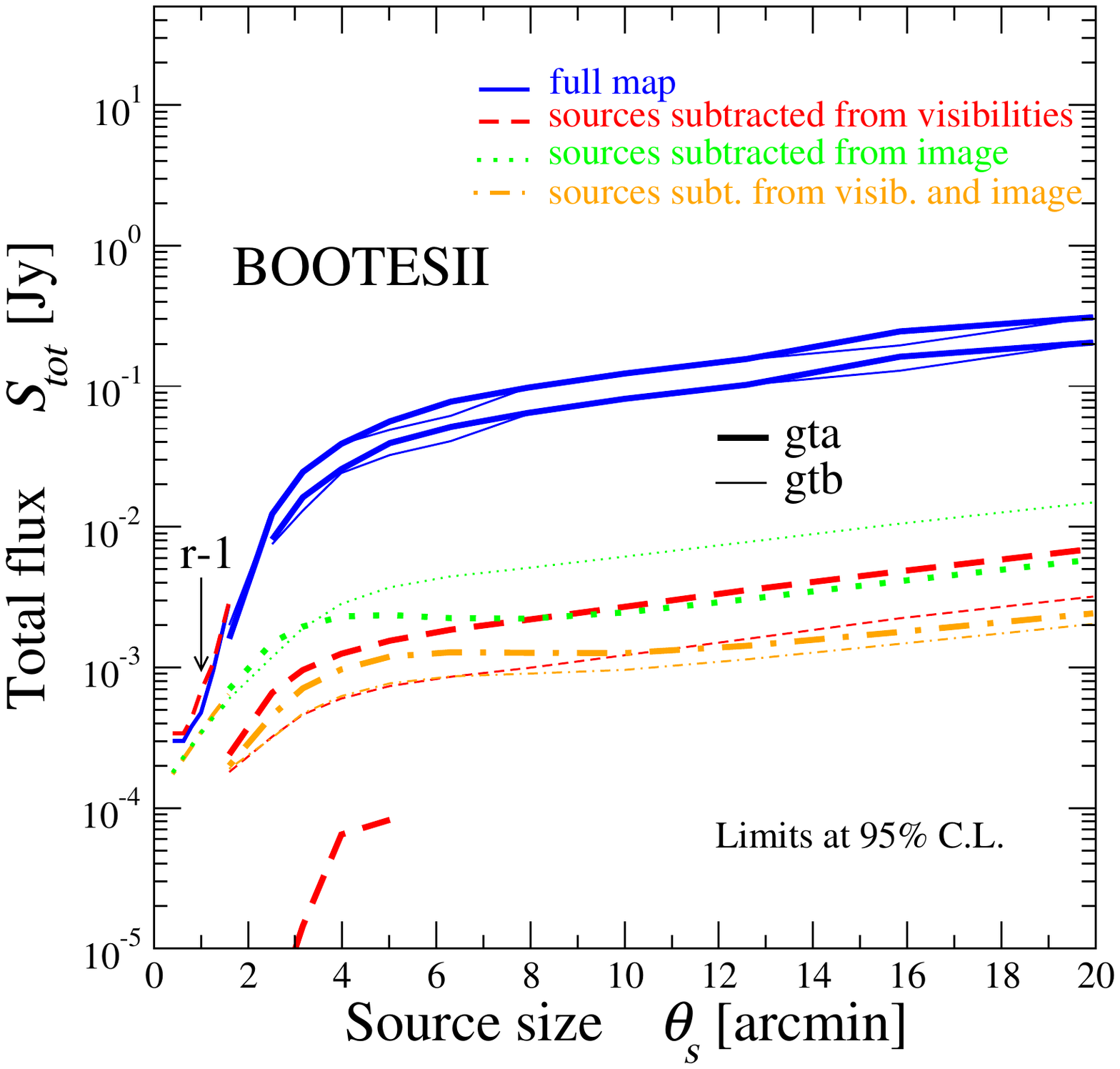}
 \end{minipage}
\hspace{-5mm}
 \begin{minipage}[htb]{6cm}
   \centering
   \includegraphics[width=\textwidth]{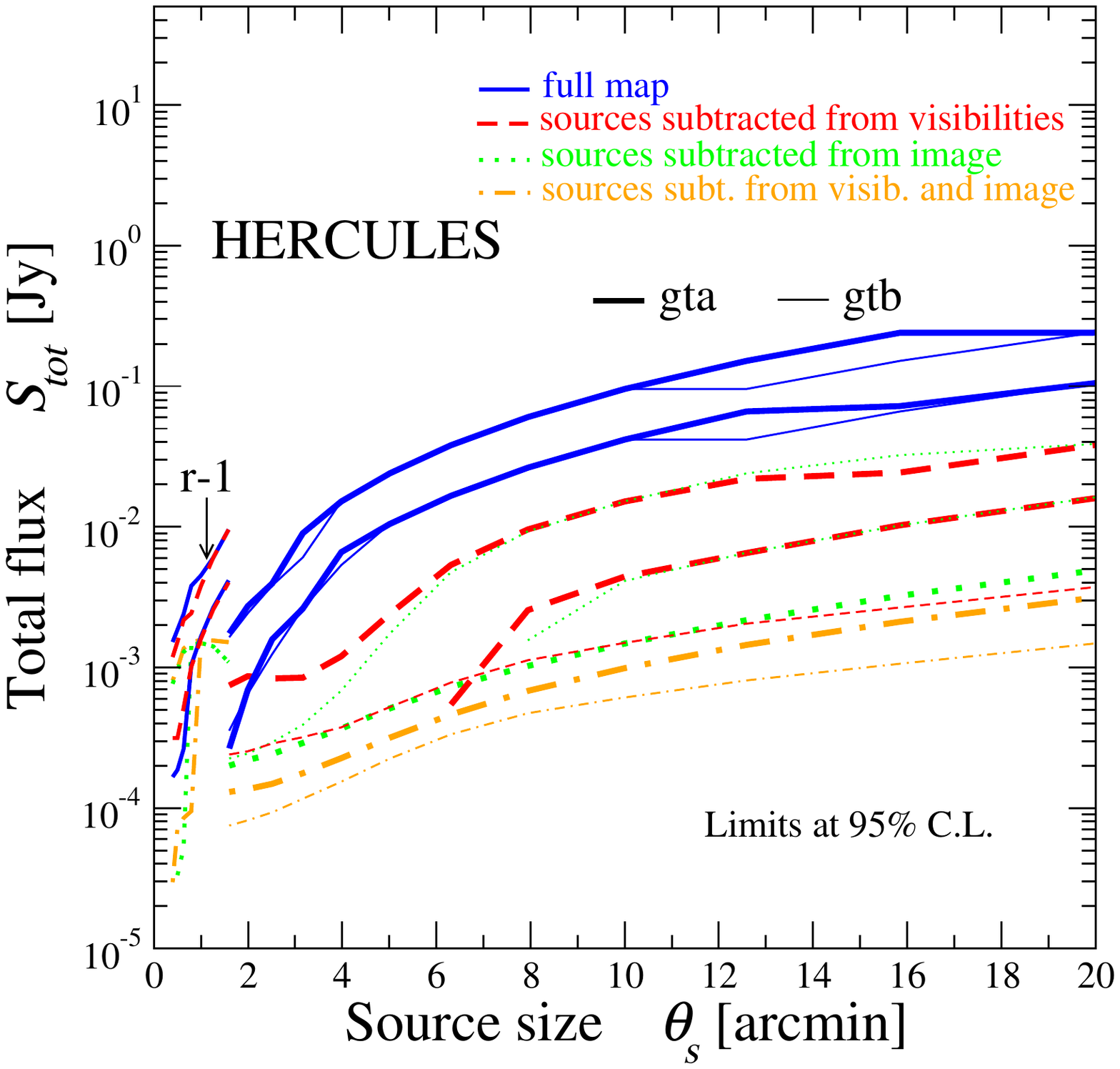}
 \end{minipage}
\hspace{-5mm}
 \begin{minipage}[htb]{6cm}
   \centering
   \includegraphics[width=\textwidth]{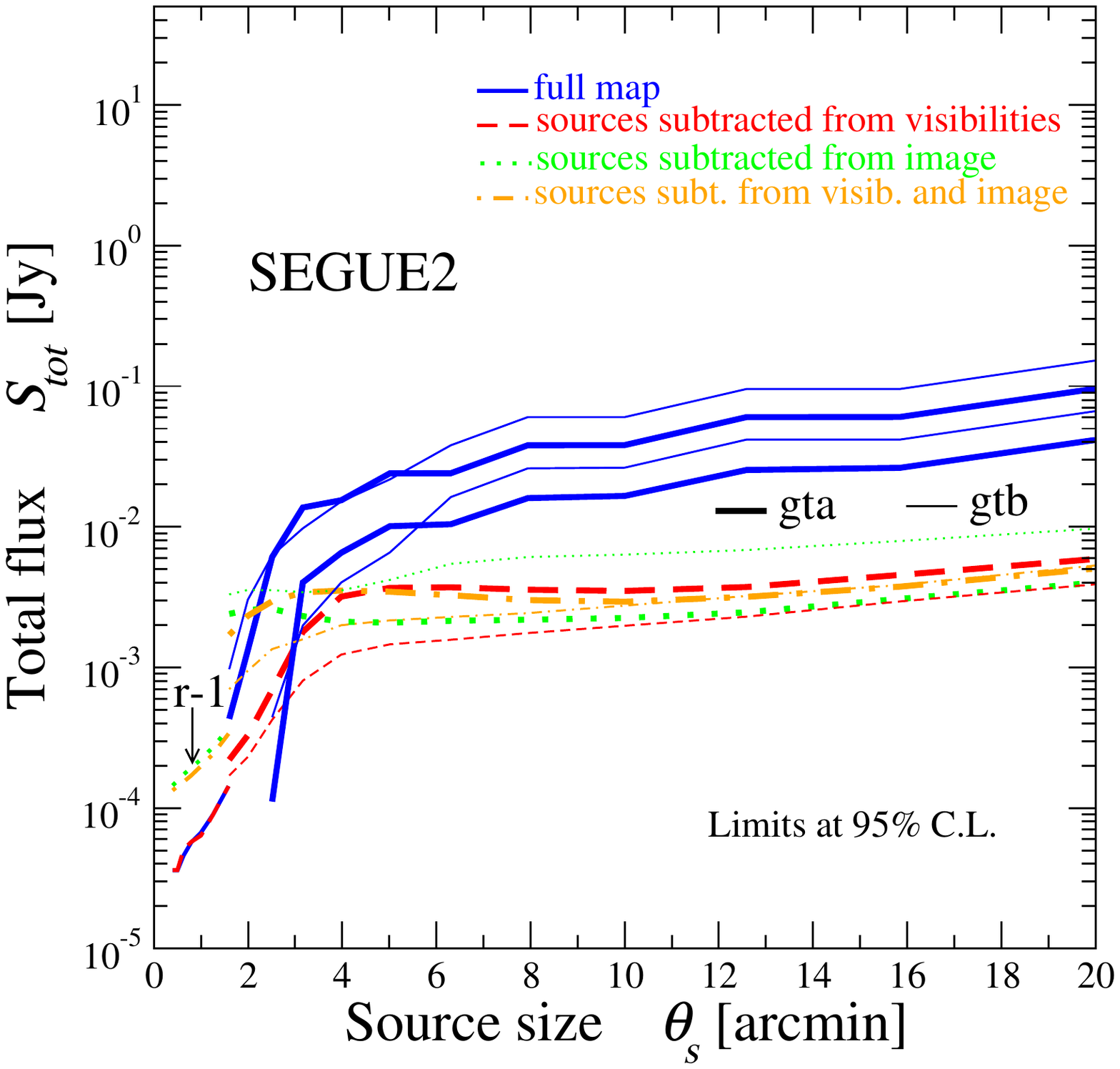}
 \end{minipage}
    \caption{{\bf Total flux}. 95\% C.L. observational limits on diffuse emission in the six targets, assuming a Gaussian flux density $S(\theta)=\frac{S_{tot}}{2\pi\,\theta_s^2}\,e^{-(\theta/\theta_s)^2/2}$. The presence of two different lines with the same style implies an evidence above 2-$\sigma$, and, in that case, both an upper and a lower limits can be derived. On the contrary, if only one curve is shown, it refers to the upper limit. The curves at small angles are derived from the $r_{-1}$ maps, while the bounds at $\theta_s$ above 1.5 arcmin come from the $gta$ (thick lines) and $gtb$ (thin lines) maps.}
\label{fig:Stot}
 \end{figure*} 

\section{Analysis}
\label{sec:ana}

\subsection{Statistical technique}
\label{sec:stat}
We will assume the likelihood for the diffuse emission of a given model to be described by a Gaussian likelihood:
\be 
\mathcal{L}=e^{-\chi^2/2} \;\;\; {\rm with} \;\;\; \chi^2=\frac{1}{N_{pix}^{beam}}\sum_{i=1}^{N_{pix}} \left(\frac{S_{th}^i-S_{obs}^i}{\sigma_{rms}^i}\right)^2\;,
\label{eq:like}
\ee
where $S_{th}^i$ is the theoretical estimate for the brightness (see Eq.~\ref{eq:Isynch}) in the pixel $i$, $S_{obs}^i$ is the observed brightness, and $\sigma_{rms}^i$ is the r.m.s. error derived as described in Section~\ref{sec:diffsubV}.
$N_{pix}$ is the total number of pixels in the area under investigations, while $N_{pix}^{beam}$ is the number of pixels in a synthesized beam.

In principle, we should have $\chi^2=\frac{1}{N_{pix}^{beam}}\sum_{i=1}^{N_{pix}}\sum_{j=1}^{N_{pix}}\left(S_{th}^i-S_{obs}^i\right)\left(\Sigma_{rms}\right)_{ij}^{-1}\left(S_{th}^j-S_{obs}^j\right)$, where $\Sigma_{rms}$ is the covariance matrix, which can be computed through a jackknife or bootstrap procedure.
In the estimate of the rms described in Section~\ref{sec:diffsubV}, the noise covariances between pixels are not considered.
However, in the image plane of interferometric images, a certain degree of correlation, even between non-adjacent pixels, is expected (because of the Fourier transformation).
On the other hand, after subtracting sources, we obtain pretty uniform rms map (see, e.g., Fig.~\ref{fig:annuli_rms}), and, varying the grid on which the computation is performed, this results remains stable. This means that the noise in uncorrelated pixels is analogous to the one in pixels having some correlations with other pixels. In other words, the covariance is subdominant with respect to the variance and we can neglect off-diagonal terms in $\Sigma_{rms}$.
This is not totally obvious for the full maps (i.e., containing sources). We will nevertheless show few bounds also in the latter case with the goal of reporting the order of magnitude of the constraints and for illustrative purposes.

\noindent{\bf Detection}:
To test the possible detection of a diffuse emission, we employ a maximum likelihood method with the estimator $\lambda_d=-2\ln({\mathcal{L}_{null}/\mathcal{L}_{b.f.}})$ treated as a $\chi^2$ variable with one d.o.f. (following Wilk's theorem in the limit of large statistics). $\mathcal{L}_{null}$ and $\mathcal{L}_{b.f.}$ are the likelihoods of no signal (i.e. with $S_{th}=0$ in Eq.~\ref{eq:like}) and of the best-fitting model, respectively.
The statistical significance is given by $\sqrt{\lambda_d}\,\sigma$ which is the C.L. at which the null hypothesis (no signal) can be rejected.

On top of using all the possible pixels, we considered also a restricted region, which is where the signal-to-noise ratio is larger.
To derive the optimal target region, we select the set of pixels which maximizes an estimate of the signal-to-noise ratio defined by $\sum_i S_{th}^i/\sqrt{\sum_i \sigma_{rms,i}^2}$. To identify such pixel-set we implement an iterative algorithm analogous to the one described in the Appendix of \citep{Bringmann:2012vr}. In this way, we expect to have a very good approximation of the area in our datasets which are most sensitive to the diffuse signal.
On the other hand, the shape of the expected signal and the fact that we consider only one component (i.e., we do not ``marginalize" over some extra components), imply that the evidence in the optimal target region is not much different (and typically lower) than in the full-map. 
Rather, this is a cross-check to ensure we avoid spurious effects from crowded regions (where the rms can be larger). For the sake of brevity, however, we only show results with statistical analysis performed using all the pixels. As an example, the comparison between the two methods is reported in Table~\ref{table:det}.

We do not get any firm evidence of a spherical diffuse emission.
In Table~\ref{table:det}, we show the best-fitting flux and statistical significance at which the no-signal case can be rejected.
A Gaussian profile and a width corresponding to the extent of the stellar component are assumed.
The reported values are computed for the $gta$ maps with source subtracted in the UV-plane.
The subtraction is not totally successful in the targets observed with the H168 configuration (Fornax, Sculptor, Hercules), for which large residuals are still present in the maps.
This leads to a ``detection'' at high C.L., but it is clearly fake. We indeed checked that it disappears when the source regions are masked. 
Moreover, if we further suppress the source contribution by performing the subtraction in the image plane of the UV-subtracted map, no evidence is obtained. 
Clearly, the addition of an emission from a model with two free parameters slightly improves the fit, but this is never at large statistical significance.
The only cases which might be showing a very weak hint are BootesII and Segue2 for which a $\sim$1-$\sigma$ deviation is found (with similar C.L. when considering all the pixels or the optimal region only).

\noindent{\bf Constraints}: Bounds on a certain parameter $\Pi$ of the model are computed for a given set of the other parameters $\vec\Pi_0$ and ``profiling out" nuisance parameters (i.e., they are taken to maximize $\mathcal{L}$ and can be different for different values of $\Pi$ and $\vec\Pi_0$). Therefore constraints are estimated though a profile likelihood technique where $\lambda_c(\Pi_x)=-2\ln[{\mathcal{L}(\vec\Pi_0,\Pi_x)/\mathcal{L}(\vec\Pi_0,\Pi_{b.f.})}]$ follows a $\chi^2$-distribution with one d.o.f. and with one-sided probability given by $P=\int^{\infty}_{\sqrt{\Pi_c}}d\chi\,e^{\chi^2/2}/\sqrt{2\,\pi}$. $\Pi_{b.f.}$ denotes the best-fit value for the parameter under investigation.
In other words, a one-sided 95\% C.L. upper limit on a parameter is obtained by increasing the signal from its best-fit value
until $\lambda_c=2.71$.

Results concerning the constraints on the theoretical models discussed above are presented in the next Section.

\begin{table*}
\centering
\begin{tabular}{|l|c|c|c|c|}
\hline
dSph & $S_{best-fit}$  & $\sqrt{\lambda_d}\,\sigma$ & $S_{max}$ & $S_{70}$  \\
name & [mJy] &  &  [mJy] &  $[10^{-14} {\rm W\,m}^{-2}]$ \\
\hline
Carina & 1.0  & 1.2 $\sigma$ (0.5 $\sigma$) & 2.7 & 1.9 \\ 
Fornax & 25  & 6.3 $\sigma$ (5.4 $\sigma$) & 43 & 30 \\ 
Sculptor & 50  & 11 $\sigma$ (9 $\sigma$)& 81 & 57\\ 
BootesII & 0.71 & 1.9 $\sigma$ (2.3 $\sigma$) & 1.4 & 0.96\\ 
Hercules & 6.3 & 3.7 $\sigma$ (3.3 $\sigma$) & 12 & 8.5\\ 
Segue2 &  0.32 & 0.35 $\sigma$ (0.1 $\sigma$) & 2.0 & 1.4\\ 
\hline
\end{tabular}
\caption{Total flux. Values are reported in the case of a Gaussian spherical diffuse emission, assuming a source size corresponding to the extent of the stellar component $\theta_s=\theta_*$, and considering the $gta$ map with source subtracted in the UV-plane. Second column shows the best-fitting total surface brightness. The statistical significance is reported in the third column considering the full map (optimal-region). Column 4 shows the 95\% C.L. bound derived from data. The corresponding bound at $70\,\mu$m emission (inferred using FIR-radio correlation, see text) is in column 5.}
\label{table:det}
\end{table*}

\begin{figure*}
\vspace{-2.cm}
\hspace{-0mm}
   \centering
 \begin{minipage}[htb]{6.cm}
   \centering
   \includegraphics[width=\textwidth]{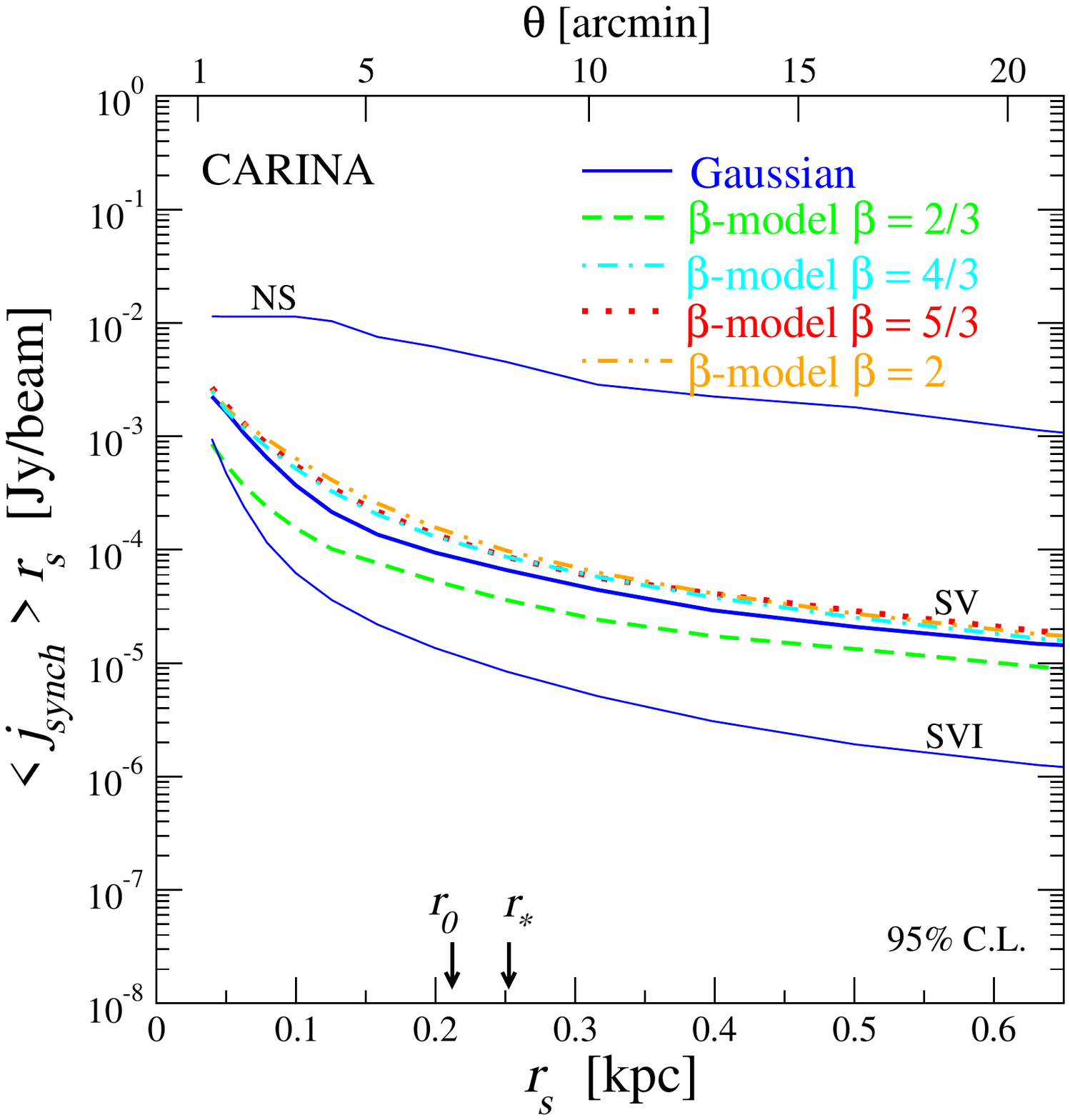}
 \end{minipage}
\hspace{-5mm}
 \begin{minipage}[htb]{6.cm}
   \centering
   \includegraphics[width=\textwidth]{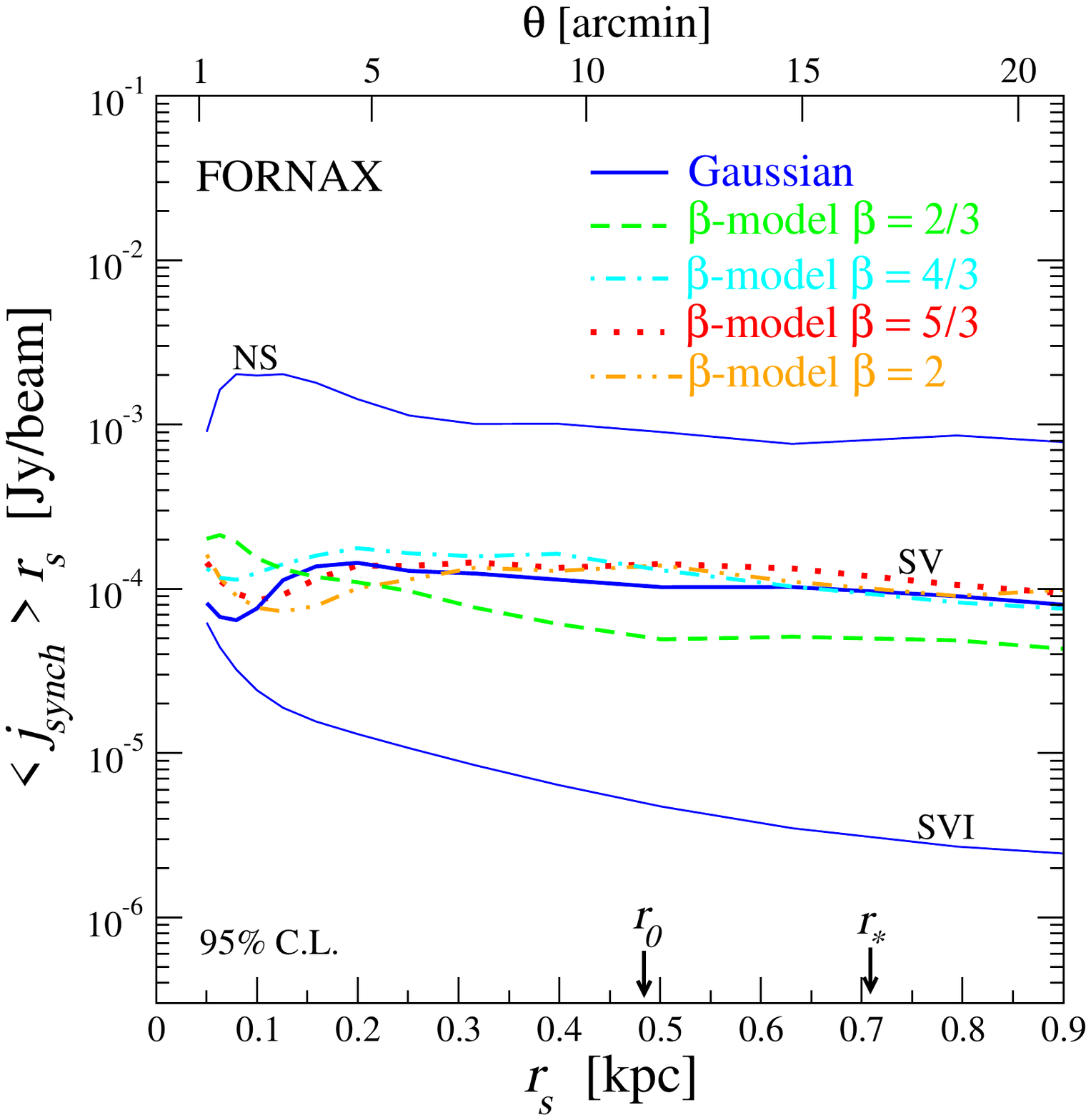}
 \end{minipage}
\hspace{-5mm}
 \begin{minipage}[htb]{6.cm}
   \centering
   \includegraphics[width=\textwidth]{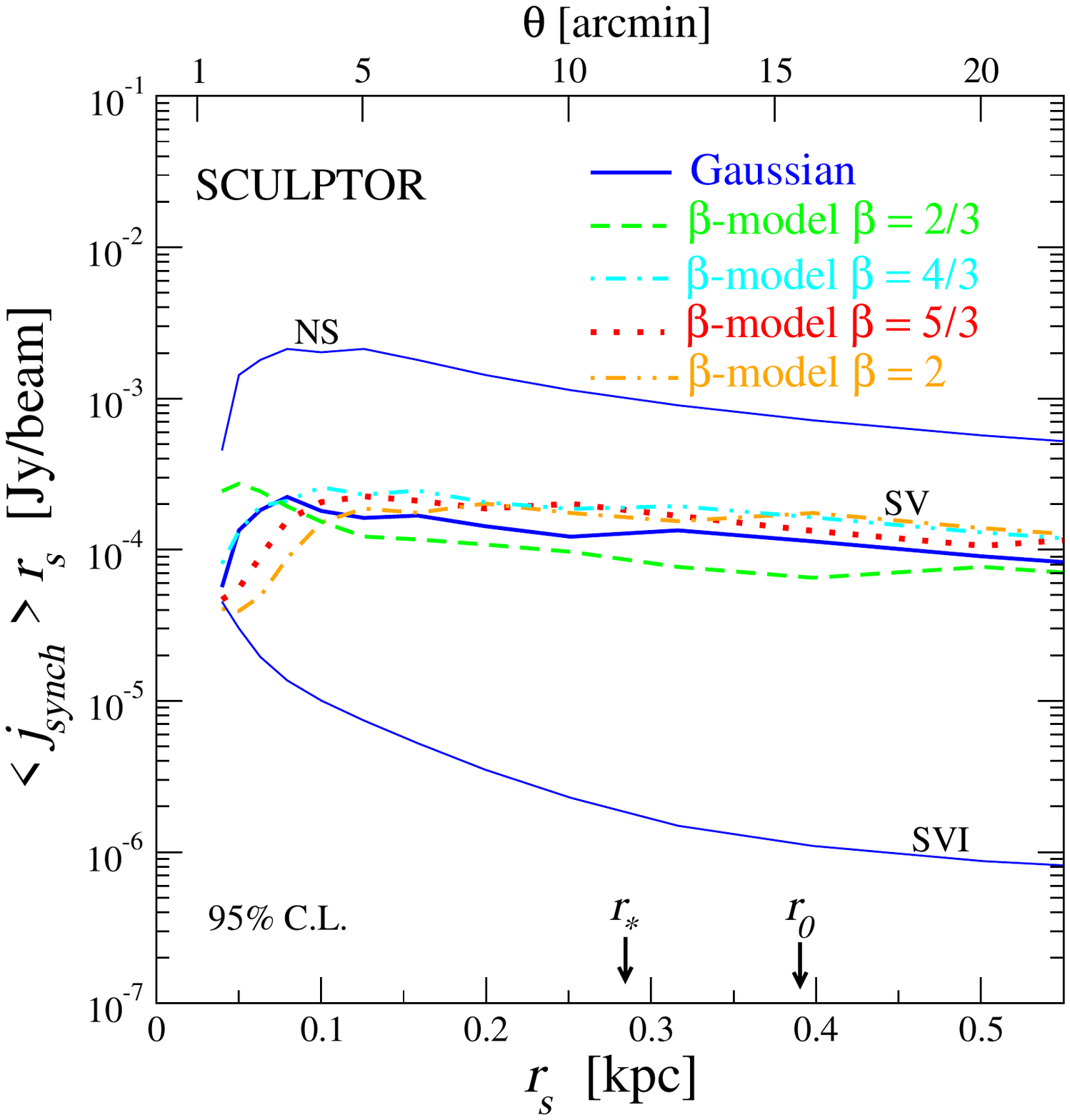}
 \end{minipage} \vspace{-2.5cm}\\ 
\hspace{-0mm}
 \begin{minipage}[htb]{6.cm}
   \centering
   \includegraphics[width=\textwidth]{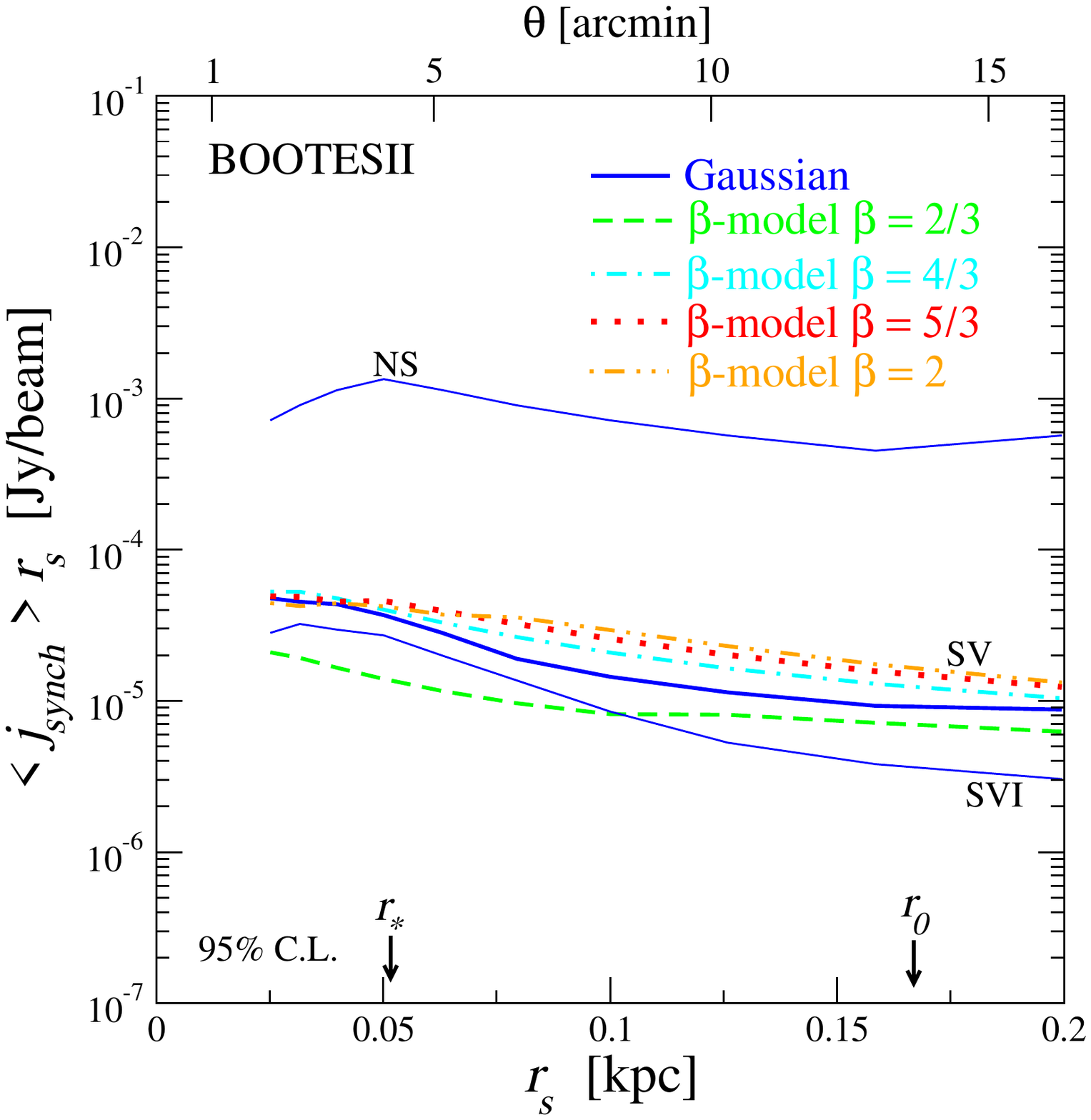}
 \end{minipage}
\hspace{-5mm}
 \begin{minipage}[htb]{6.cm}
   \centering
   \includegraphics[width=\textwidth]{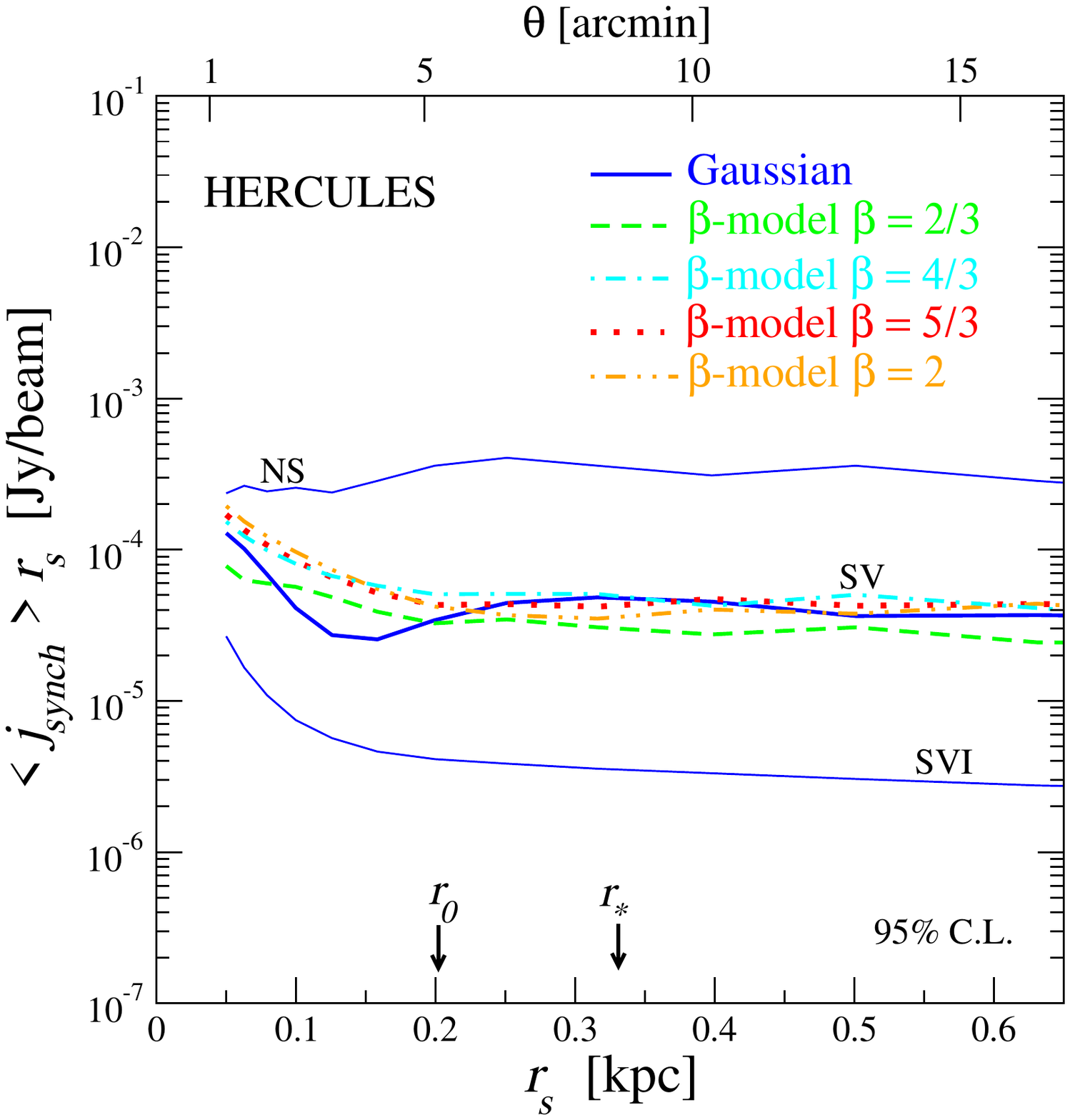}
 \end{minipage}
\hspace{-5mm}
 \begin{minipage}[htb]{6.cm}
   \centering
   \includegraphics[width=\textwidth]{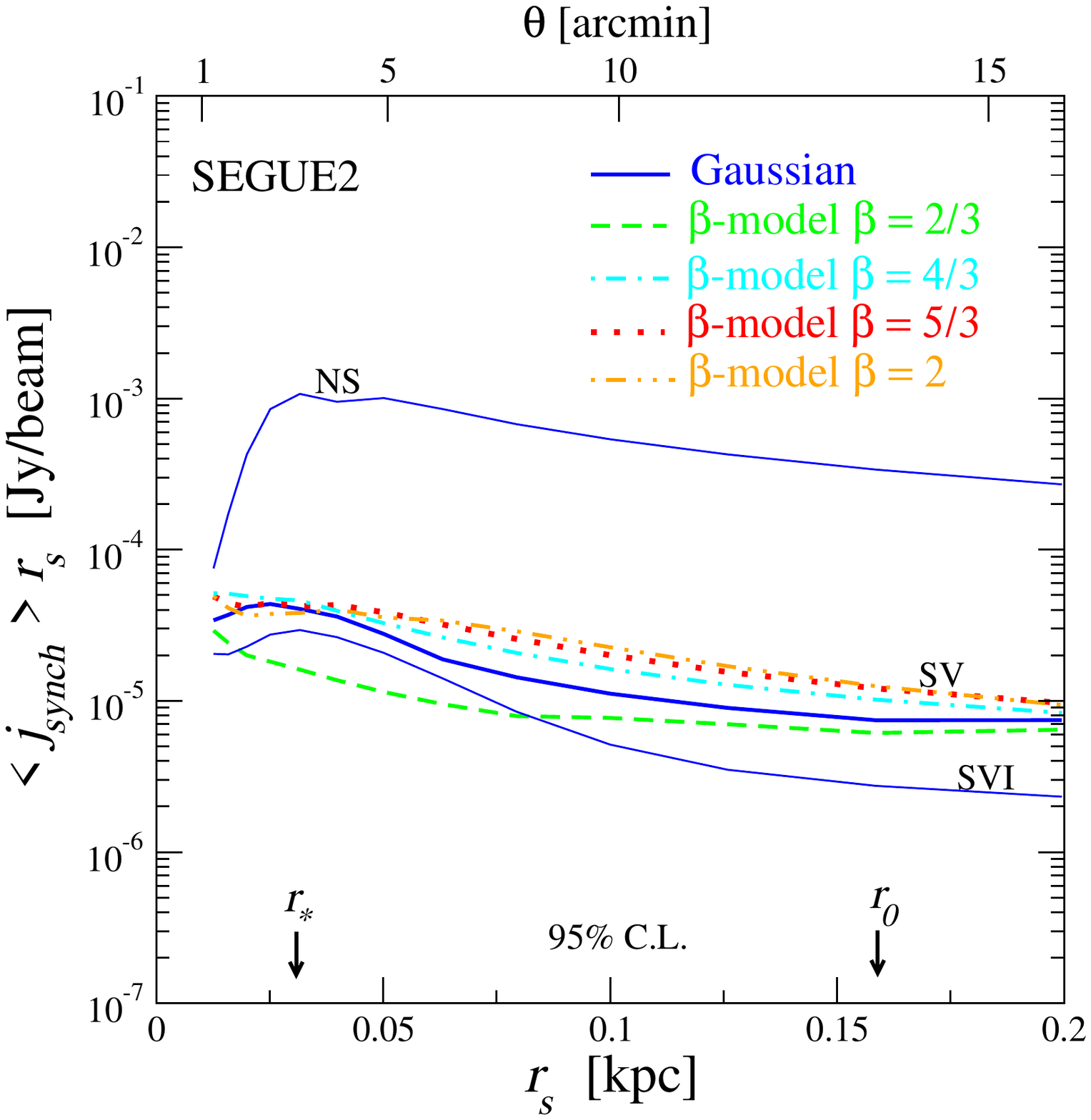}
 \end{minipage}\\
    \caption{{\bf Emissivity}. 95\% C.L. observational upper limits on the emissivity of diffuse emission. We show $\langle j_{synch}\rangle=3\,r_s^{-3}\int^{r_s}_0 dr\,r^2 j_{synch}(r/r_s)$ versus spatial extension $r_s$, for few different models $j_{synch}$ described in the text. Bounds are computed on maps with source subtraction performed in the visibility plane (SV, thick lines), except for the Gaussian model where, in addition, the cases with no source subtraction (NS, thin) and with subtraction on both visibility and image planes (SVI, thin) are shown. We report also the half-light radius of the stellar distribution $r_*$ (see Table~\ref{tableB}) and the DM halo scale $r_0$ for an NFW profile (from the central value of the fit in \citep{Martinez:2013els}).}
\label{fig:j0}
 \end{figure*} 

\subsection{Bounds}
\label{sec:bounds}

We derive the upper limits on the diffuse emission assuming spherical symmetry and taking the diffuse radio emission to be centered at the optical center of the dSph.
The uncertainty in the centroid position of the dSphs considered here is typically estimated to be below the arcmin level.
Since our sensitivity is rather homogeneous on a much larger scale, we do not expect a significant variation of the bounds due to possible misalignment between the assumed center of the spherical distribution of our models and the real dSph center.
For similar reasons, we also expect only mild modifications of our bounds in case of departure from spherical symmetry (e.g. ellipticity).
In case of a positive detection (which is unfortunately not our case) both effects should instead be accounted for, in order to have a robust determination of the model parameters.

Constraints on the total dSph flux $S_{tot}$ are shown in Fig.~\ref{fig:Stot} as a function of the size of the emission $\theta_s$, while the angular profile is constrained in Fig.~\ref{fig:j0} (see also Fig.~\ref{fig:annuli_f15}).
In Fig.~\ref{fig:Stot}, we compare the bounds that can be obtained by employing the different subtraction methods described in Section~\ref{sec:diff}, on the three types of maps ($r_{-1}$, $gta$, $gtb$) introduced in Section~\ref{sec:obs}.
The brightness is modelled with a Gaussian: $S(\theta)=S_{tot}/(2\pi\,\theta_s^2)\,\exp[-(\theta/\theta_s)^2/2]$.

For small source-sizes, the $r_{-1}$ map is the most constraining one, while only the tapered images $gta$ and $gtb$ can probe scales from 1 to 15 arcmin.
For illustrative purposes, the reported angular range in Fig.~\ref{fig:Stot} (as in other figures below) extends to slightly larger region, although, as described in Section~\ref{sec:maxsize} a full simulation (for each model) would be required to assess the actual sensitivity at scales $\gtrsim15'$.

The source subtraction clearly reduces the total flux. On the other hand, the trend of this variation among the different cases can significantly vary from dSph to dSph. Indeed, the impact of the source subtraction depends on the quality of the image, on the beam model, and on the number and brightness of sources near the dSph center.

The presence of two different lines with the same colour and style implies an evidence above 2-$\sigma$ (which is the C.L. chosen for the reported bounds). In this case, both an upper and a lower 95\% C.L. limits can be derived. On the contrary, if only one curve is shown, it refers to the upper limit.
As already mentioned, the source subtraction procedure often leaves some non-negligible level of residuals, which can lead to a spurious detection of a diffuse emission.
Therefore in Fig.~\ref{fig:Stot}, some targets show lower bounds not only in the maps including sources (solid lines), but also after source-subtraction in the Fourier (dashed) or image (dotted) planes. When both subtraction procedures are combined (dash-dotted), no significant detection is found, as already stated above. 

From the most conservative case (no subtraction) to the most aggressive one (subtraction in both UV and image planes), the upper bound can significantly vary, up to two or three orders of magnitude.

\begin{figure*}
\vspace{-3.cm}
   \centering
 \begin{minipage}[htb]{8cm}
   \centering
   \includegraphics[width=\textwidth]{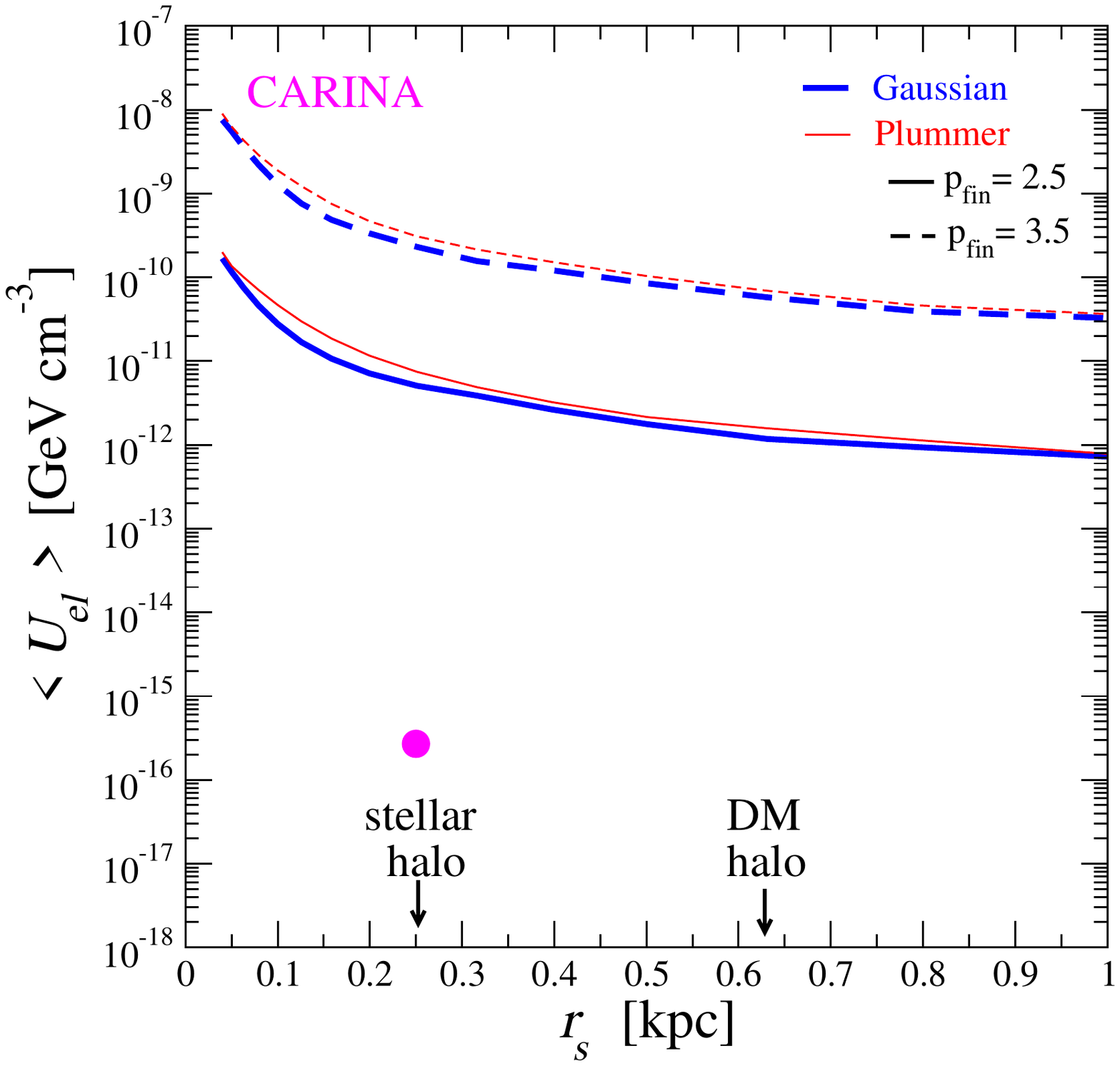}
 \end{minipage}
\hspace{0mm}
 \begin{minipage}[htb]{8cm}
   \centering
   \includegraphics[width=\textwidth]{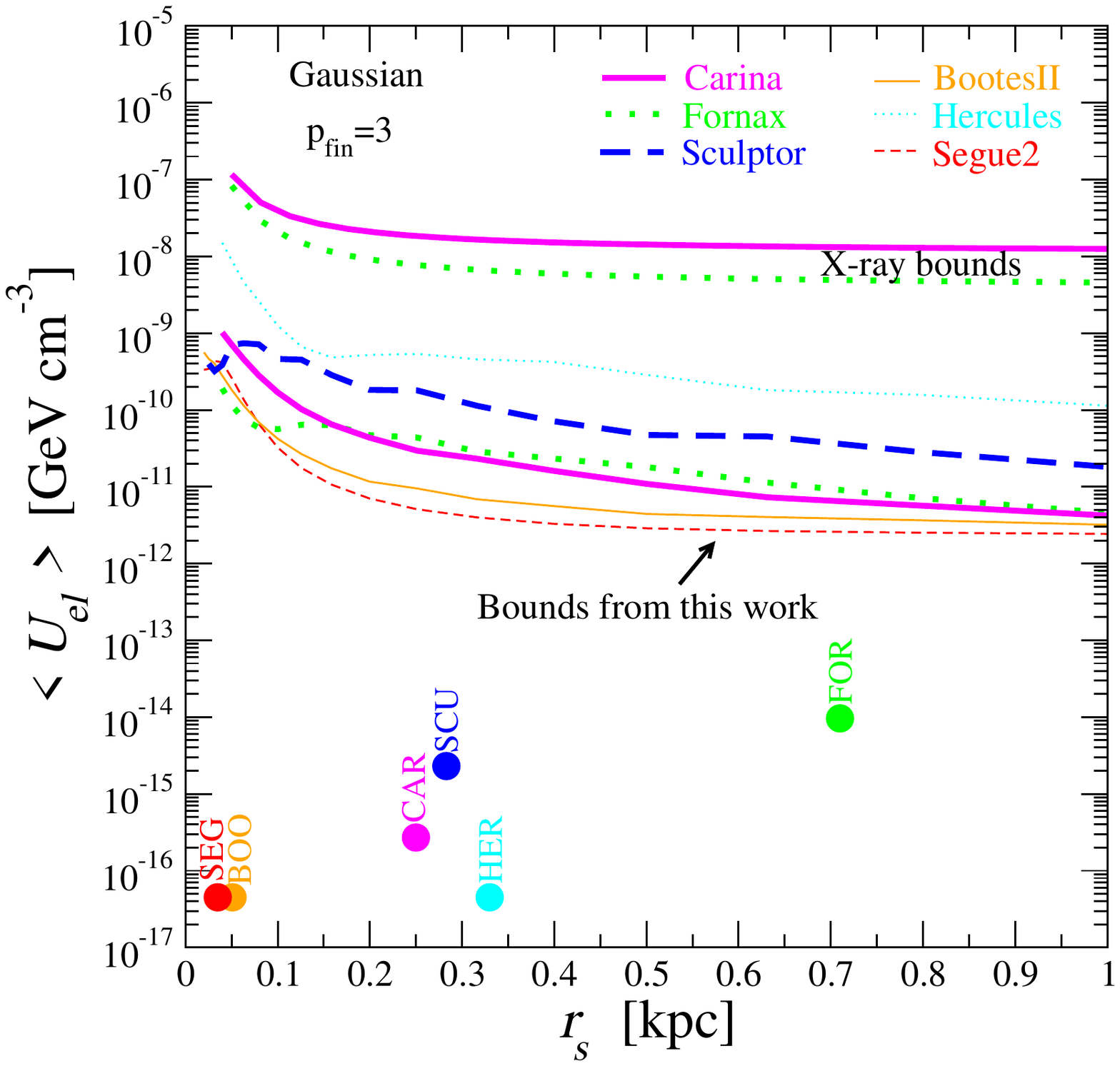}
 \end{minipage}
    \caption{{\bf Electron equilibrium distribution}. 95\% C.L. upper bounds on spatial average $\langle U_{el}\rangle=3\,r_s^{-3}\int^{r_s}_0 dr\,r^2 U_{el}(r/r_s)$ versus spatial extension $r_s$, obtained from the $gta$ maps with source subtracted from visibilities. $U_{el}(r)$ is the total equilibrium distribution of CR electrons and positrons in the dSph: $U_{el}(r)=\int_{0.1\,{\rm GeV}}^{1\,{\rm TeV}}dE\,E\,n_e(E,r)$. The spectrum of $n_e$ is assumed to be a power law. In the left-hand panel, we show the effect of varying the spectral index and the spatial profile in the Carina case (other targets show similar scalings). In the right-hand panel, $p_{fin}=3$ and Gaussian spatial profile are assumed. The upper lines show the corresponding X-ray bounds for the Carina and Fornax dSphs obtained following \citep{Jeltema:2008ax}. Circles show the expected CR density from the dSph SFR reported in column 8 of Table~\ref{tableB}.}
\label{fig:n0vsrs}
 \end{figure*} 

In Fig.~\ref{fig:j0}, we show the bounds on the emissivity of the diffuse signal averaged over the emission region $\langle j_{synch}\rangle=3\,r_s^{-3}\int^{r_s}_0 dr\,r^2 j_{synch}(r/r_s)$, as a function of the physical size $r_s$ of the emission. We compare different spatial profiles described in Section~\ref{sec:CR}. The impact of the profile on the bounds of the spatially-averaged emissivity is mild, while obviously, the different models can lead to bounds which can be locally quite different.
We highlight the physical sizes corresponding to the half-light radius of the stellar distribution $r_*$ (see Table~\ref{tableB}) and to the DM halo scale $r_0$ (taken from~\citep{Martinez:2013els} in the case of an NFW profile), which can be considered as the expected sizes of a possible extended emission in dSphs.

The constraints are derived focusing on the $gta$ maps with sources subtracted in the UV-plane (SV cases). However, for completeness, in the case of a Gaussian spatial profile, we report the curves obtained from the $gta$ maps without subtracting sources (labelled with NS) and with the source subtraction performed both in the UV and image planes (labelled with SVI).

In Fig.~\ref{fig:n0vsrs}, we derive constraints for the spatially averaged equilibrium distribution $\langle U_{el}\rangle$ of CR electrons and positrons in the dSph. 
The spectrum is assumed to be a power law. 
In these plots, as for the following ones, we will consider the $gta$ maps, with source subtraction performed in the UV-plane only, as our reference (conservative) images.
As benchmark models for the magnetic field, we assume a strength $B_0$ given by the maximum value among the ones quoted in columns 4-6 of Table~\ref{tableB} and a spatial profile given by  $B=B_0\,e^{-r/r_*}$. 

We show the impact of the spatial profile and spectral index in the case of the Carina dSph (left-hand panel).
As for Fig.~\ref{fig:j0}, the profile has a mild effect on the spatially averaged bounds.
The choice of the spectrum is instead relevant, although the re-scaling of the bounds can be easily computed for the case of a power law.

In the right-hand panel of Fig.~\ref{fig:n0vsrs}, we compare the bounds on $\langle U_{el}\rangle$ with the estimates from SF reported in column 8 of Table~\ref{tableB} (shown with filled circles). They indicate that if the estimates for the CR density associated with late-time SFR, $U_{el}^{SFR_0}$, are correct, we need a few order-of-magnitude improvement in the observational sensitivity (i.e., to reach a fraction of $\mu$Jy to nJy), to probe the CDS emission, which is thus possibly achievable by the SKA. The emission related to SF in UDS would require a very high sensitivity even for the SKA. 

\begin{figure*}
\vspace{-3.cm}
   \centering
 \begin{minipage}[htb]{8cm}
   \centering
   \includegraphics[width=\textwidth]{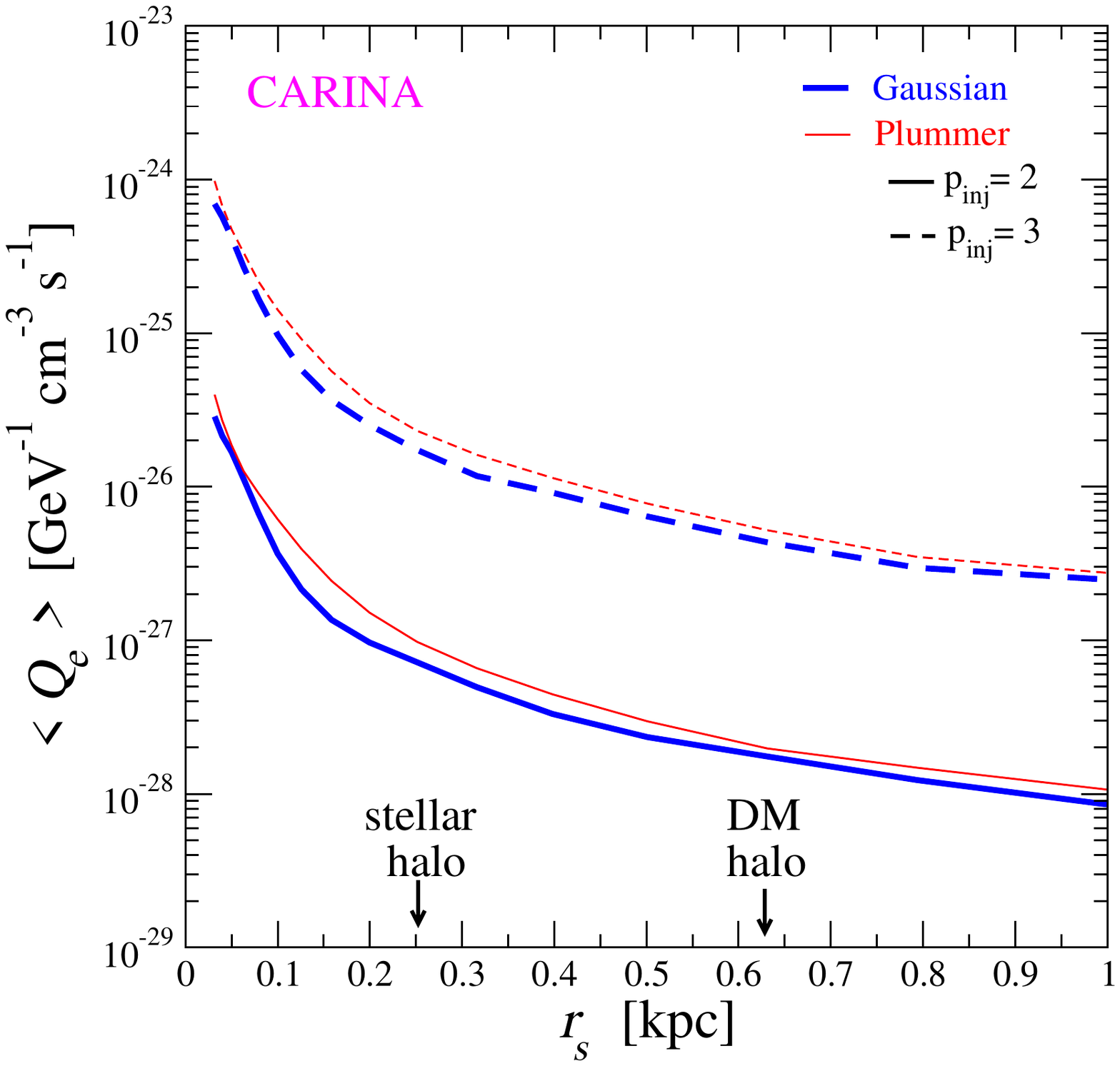}
 \end{minipage}
\hspace{0mm}
 \begin{minipage}[htb]{8cm}
   \centering
   \includegraphics[width=\textwidth]{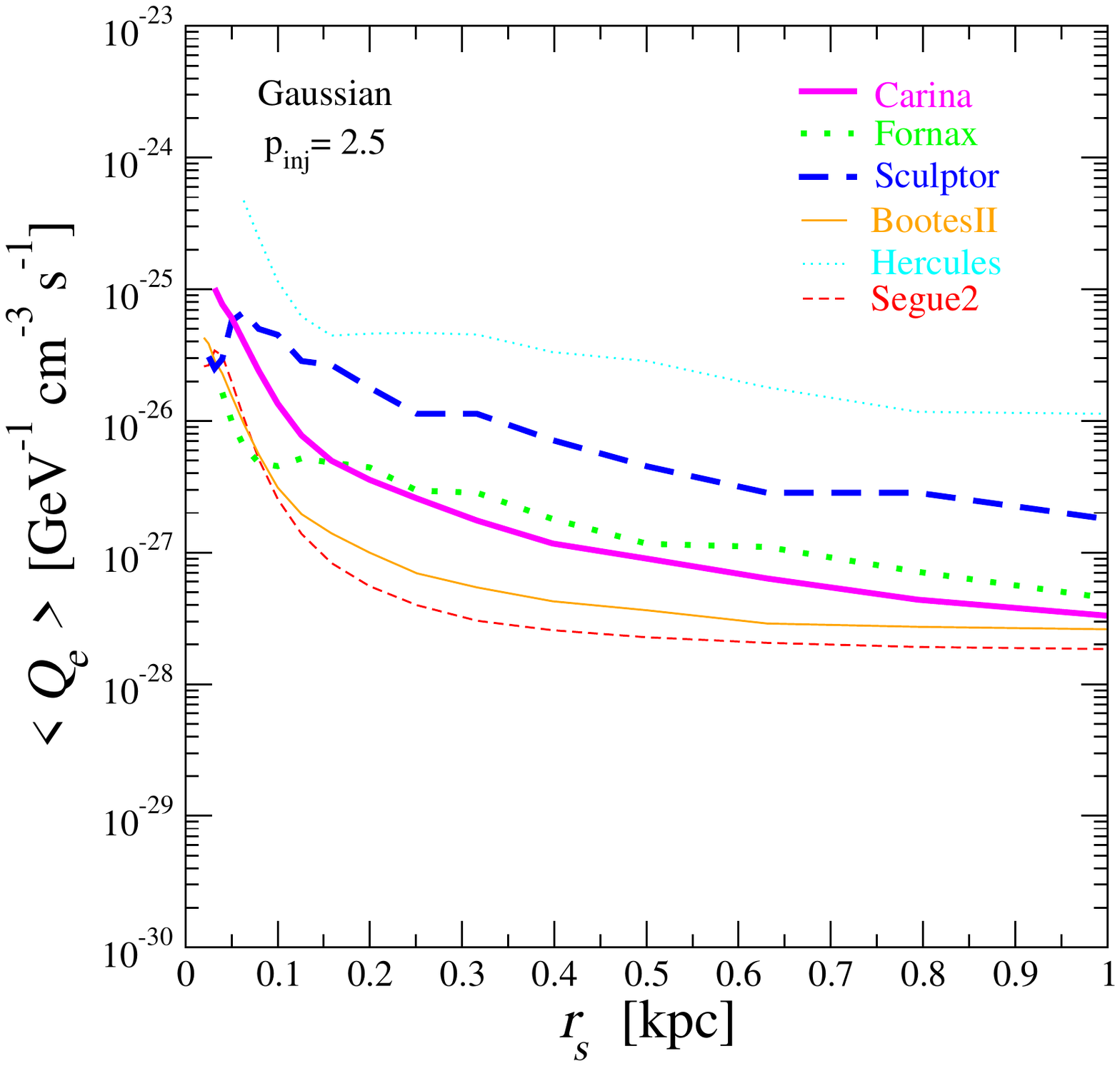}
 \end{minipage}
    \caption{{\bf Electron injection distribution}. 95\% C.L. upper bounds on the spatial average $\langle Q_{e}\rangle=3\,r_s^{-3}\int^{r_s}_0 dr\,r^2 Q_{e}(r/r_s)$ versus spatial extension $r_s$, obtained from the $gta$ maps with source subtracted from visibilities. $Q_{e}(r)$ is the injection distribution of CR electrons and positrons in the dSph, with the spectrum assumed to be a power law. The limits are computed in the loss at injection-place scenario described in Sec.~\ref{sec:turb}. In the left-hand panel, we show the effect of varying the spectral index and the spatial profile in the Carina case (other targets show similar scalings). In the right-hand panel, $p_{inj}=2.5$ and Gaussian spatial profile are assumed.}
\label{fig:A0vsrs}
 \end{figure*}

Fig.~\ref{fig:A0vsrs} is similar to Fig.~\ref{fig:n0vsrs}, but now we compute bounds on the injection distribution of CR electrons and positrons $Q_e$, rather than on the equilibrium distribution.
The limits are derived in the loss at injection-place scenario described in Section~\ref{sec:turb} and assuming a power-law for the injection spectrum.

Note that the bounds in Figs.~\ref{fig:n0vsrs} and \ref{fig:A0vsrs} are somewhat dependent on the chosen extrema for the energy integration of the spectrum, in particular on $E_{min}$. However, since the spectrum is a power-law, it is straightforward for the reader to derive the bound with a choice of $E_{min}$ different from 0.1 GeV.

\begin{figure*}
\vspace{-3cm}
   \centering
 \begin{minipage}[htb]{8cm}
   \centering
   \includegraphics[width=\textwidth]{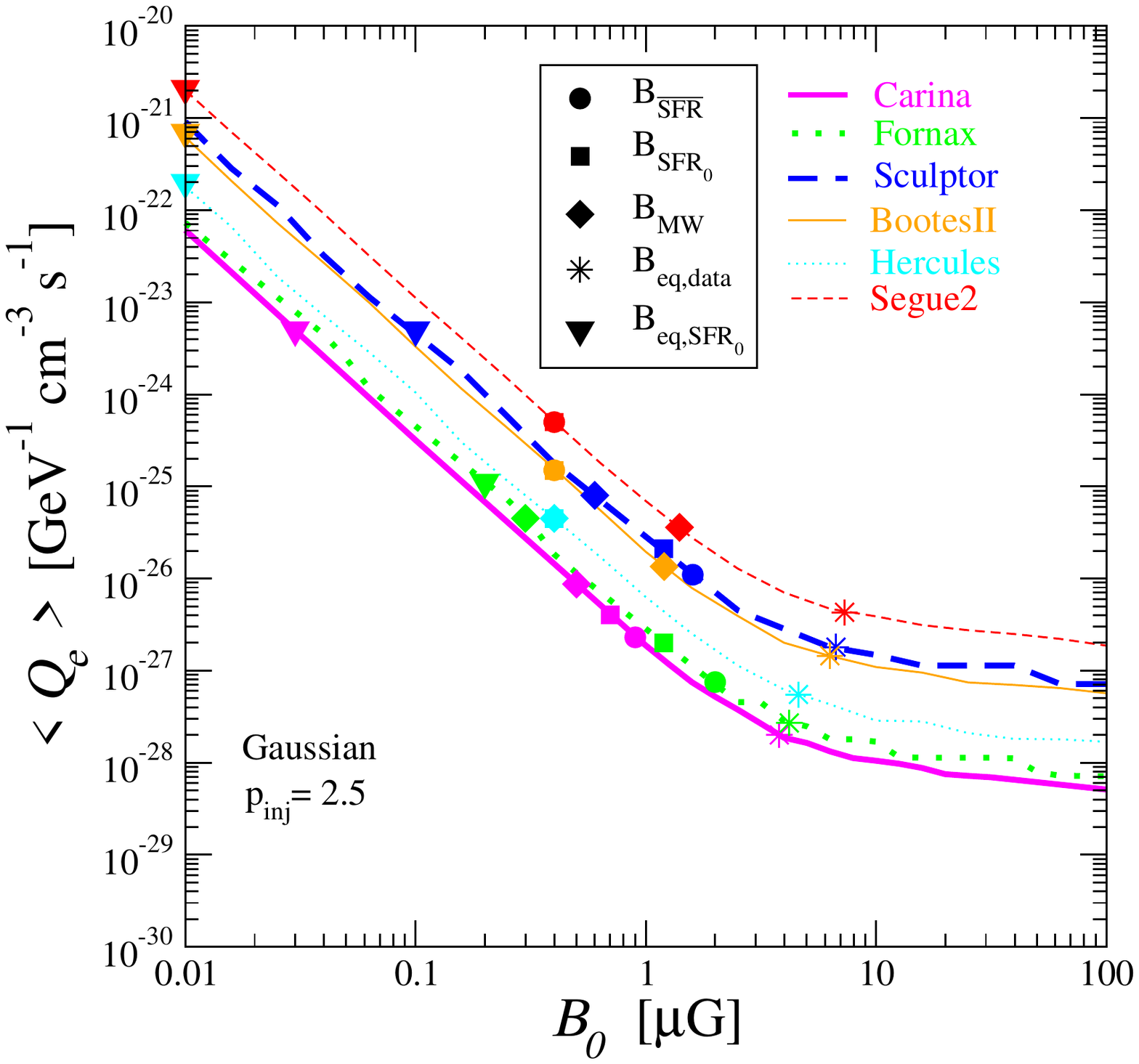}
 \end{minipage}
\hspace{0mm}
 \begin{minipage}[htb]{8cm}
   \centering
   \includegraphics[width=\textwidth]{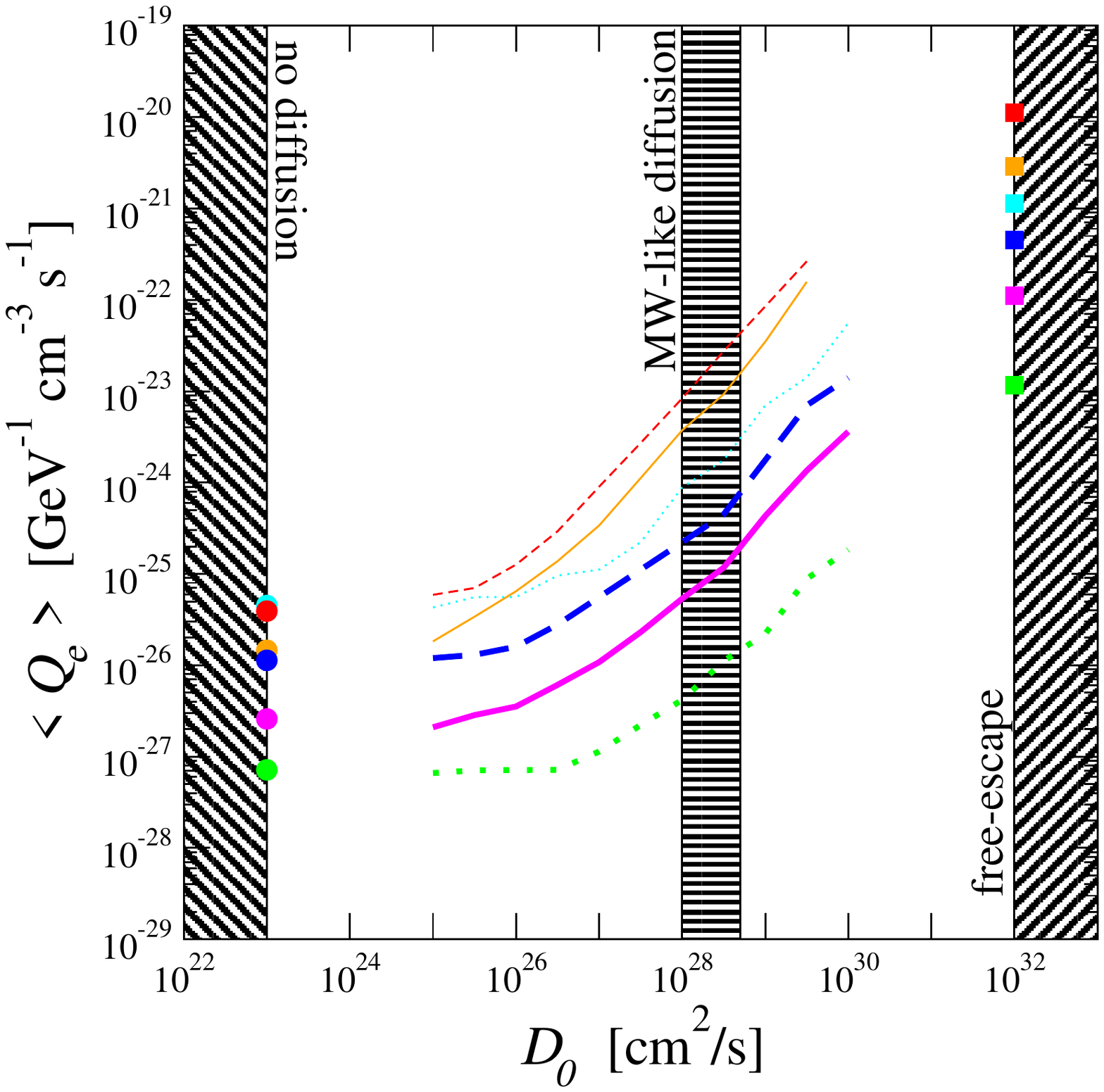}
 \end{minipage}
    \caption{{\bf Magnetic field and Diffusion}. 95\% C.L. upper bounds on  $\langle Q_e\rangle=3\,r_s^{-3}\int^{r_s}_0 dr\,r^2 Q_e(r)$ (with $Q_e(r)=\int_{E_{min}}^{E_{max}}dE\ E\ A_0\,\exp(-(r/r_s)^2/2) \,(E/{\rm GeV})^{-2.5}$ and $r_s$ taken to be equal to the stellar profile size) versus magnetic field strength $B_0$ (left) and diffusion coefficient $D_0$ (right). The bounds are obtained from the $gta$ maps with source subtracted from visibilities.
In the left-hand panel, we assume radiation at injection place and estimates of $B$ from Table~\ref{tableB} are overlaid.  
In the right-hand panel, we assume $D=D_0\,(E/{\rm GeV})^{1/3}\,e^{r/r_*}$. Square points show the limits for free-escape, while circles are the bounds in the loss at injection-place case. }

\label{fig:A0vsB0}
 \end{figure*}

Fig.~\ref{fig:A0vsB0} shows how different scenarios for the magnetic field strength and turbulence can change the constraints on the CR injection distribution of Fig.~\ref{fig:A0vsrs}.
The impact of the magnetic field strength is discussed in the left-hand panel of Fig.~\ref{fig:A0vsB0}.
The scaling of the curves follows from the scaling of the emissivity.
The magnetic field affects both the synchrotron power via Eq.~\ref{eqjsynch} and the CR energy losses via Eq.~\ref{eq:eloss}. 
When synchrotron losses are subdominant with respect to IC losses, the emissivity scales as $j_{synch}\propto B^{-(p_{fin}+1)/2}$, following the scaling of the power.
At large $B$, the curves flatten because synchrotron radiation becomes the dominant energy-loss mechanism and an increase in $B$ shows up in an approximately equal increase in both $P_{synch}$ and $dp/dt$, leaving $j_{synch}\sim$ const.

The observational bounds on the strength of $B$ under the equipartition assumption (shown in Table~\ref{tableB}) are $B\lesssim$ few $\mu G$, while expectations from theoretical arguments leads to about the $\mu G$-scale. Taking the latter estimate for the strength, the increase in sensitivity needed to probe a signal with CR density at equipartition with such magnetic field, roughly scales with the fourth power of the ratio between the current equipartition bound and the expected magnetic field strength. This scenario is thus within the reach of the SKA, and, in some cases, also of its precursors ASKAP and MeerKAT.

Note also that the bounds in Fig.~\ref{fig:n0vsrs} would be significantly stronger under the equipartition assumption (and close to $\langle U_{el}^{SFR_0}\rangle $ in the case of the Fornax and Sculptor dSphs), exactly because the magnetic field would be significantly larger.

The right-hand panel of Fig.~\ref{fig:A0vsB0} shows the impact of diffusion effects on the derived bounds.
A diffusion similar to the one observed in the MW (see discussion in Section~\ref{sec:turb}) makes the bound weaker by about one order of magnitude for the largest dSphs (Carina, Fornax, Sculptor, and Hercules) and by about two orders of magnitude for the smallest systems (BootesII and Segue2) with respect to the loss at injection-place case shown in the previous plots. We assumed a Kolmogorov spectrum, with the diffusion coefficient exponentially increasing outside the stellar region: $D=D_0\,(E/{\rm GeV})^{1/3}\,e^{r/r_*}$.

In the free-escape scenario, the bounds worsen by about four and five orders of magnitude, respectively. 
We remind the reader that, as discussed in Section~\ref{sec:turb}, a free-escape is probably extreme and too pessimistic, and the allowed range can shrink by a factor of few taking into account CR self-confinement.

The uncertainty associated with spatial diffusion is thus very relevant, as expected. This is indeed due to the smallness and low level of activity (at present time) of dSphs.

\subsubsection{FIR-radio correlation}
\label{sec:FIR}
A tight correlation between global radio and FIR flux of normal star-forming galaxies has been observed to exist over many orders of magnitude in luminosities and up to intermediate redshifts~\citep{Appleton:2004sr}. The correlation is probably connected to the fact that both radio and FIR emissions are related to the SFR of the object.
Indeed, the radio emission mainly comes from synchrotron radiation of CR electrons accelerated in SN remnants, and the FIR flux is mainly due to dust reprocessing of UV photons from young stars.
On the other hand, such a tightness of correlation for very different systems has still to be understood.
Empirically, the FIR-radio correlation can be written as:
\be
q_{IR}=\log_{10}\left[ \left(\frac{S_{IR}}{3.75\cdot 10^{12} {\rm W\,m}^{-2}}\right)/\left(  \frac{S_{1.4{\rm GHz}}}{{\rm W\,m}^{-2}{\rm Hz}^{-1}}  \right) \right]\;
\ee
where $q_{IR}$ has been found, for normal galaxies at $70\,\mu$m, to be $q_{70}=2.15\pm0.16$~\citep{Appleton:2004sr}.

Recently, \citep{Roychowdhury:2012bc} studied the FIR-radio correlation in samples of faint star-forming dwarf galaxies, finding good agreement with the $q_{70}$ value of \citep{Appleton:2004sr}.
However, it is not guaranteed that such relation holds also in the case of old dSph galaxies considered here.
On the other hand, by assuming it, we can use the bounds we derived for the total radio flux to infer bounds on the infrared emission.
They are reported in Table \ref{table:det}, where we use a spectral index of $-0.8$ to scale results at 2 GHz to 1.4 GHz, which leads to $S_{70}\simeq 7\cdot 10^{-15}\,(S_{2{\rm GHz}}/{\rm mJy}) \,{\rm W\,m}^{-2}$.

\subsubsection{Comparison with X-ray bounds}
\label{sec:xray}

The peak of the synchrotron emission at 2 GHz is produced by electrons with energy from few GeV to few tens of GeV, depending on the magnetic field.
The same population of non-thermal electrons inevitably gives rise to IC radiation via their interaction with the CMB photons.
Such emission falls in the X-ray frequency range, namely in the keV-MeV energy band. Indeed the peak of this IC emission occurs for a photon energy $E_\gamma\simeq (E_e/{\rm GeV})^2$ keV, where $E_e$ is the electron energy.

In Fig.~\ref{fig:n0vsrs}, we compare the bounds that can be obtained on the non-thermal electron density from current X-ray data with the constraints derived in this work.
For the former, we consider the analysis of \citep{Jeltema:2008ax}, which made use of XMM-Newton archival data for the Carina and Fornax dSph targets (on top of Ursa Minor which however is not part of our dSph sample). The derived flux limits in the 0.5-8 keV band are, respectively, $2.1\cdot10^{-5}$ and $10^{-5}$ photons cm$^{-2}$ s$^{-1}$ for an aperture of 6' radius~\citep{Jeltema:2008ax}.
Following this analysis, we can constrain $U_{el}$ for the models discussed above.
We found that X-ray bounds are few orders of magnitude weaker than the constraints from synchrotron emission derived in this work.
In Fig.~\ref{fig:n0vsrs}, we compare X-ray (upper curves) and radio (lower curves) bounds in the case of a power-law spectrum with $p_{fin}=3$ for the equilibrium electron distribution. 
The radio data of this project are more constraining than current X-ray data for a magnetic field strength larger than $\sim 0.05\,\mu$G.

\section{Conclusions}
\label{sec:concl}

Local Group dSph galaxies are our closest neighbours.
The knowledge about dSphs is however quite limited, as these objects are small, quiescent and dim.
DSphs have been recognized as important probes for cosmology.
Their structure, chemical composition and kinematics pose indeed important challenges to our current understanding of structure formation~\citep{Mateo:1998wg,Bullock:2009au,McConnachie:2012vd}. 
In addition, dSphs can also be key probes in the search for a particle DM signature (see Paper III).

At present, no evidence for the presence of a thermal or non-thermal plasma from dSphs has been reported in the literature.
Deep observations are thus required in order to possibly probe the non-thermal emission associated with particle DM or to the very-low level of dSph SF.

In this paper, we made use of deep mosaic radio observations of a sample of six local dSphs, three ``classical" ones (Carina, Fornax, and Sculptor), and three ``ultra-faint" dSphs (BootesII, Segue2, and Hercules) to investigate the presence of diffuse synchrotron emissions in the dSph interstellar medium. 
We collected data with the ATCA telescope in an array configuration specifically designed to seek an extended (few arcmin-scale) signal. The resulting maps have a sensitivity around 0.05 mJy at 2 GHz.
On top of the image from the compact array, we simultaneously obtained long-baseline data to map discrete sources (see Paper I for more details). 
The confusion limit is one of the greatest obstacles to be overcome in the search for a few-arcmin radio diffuse emission. 
Indeed, for arcmin synthesized beams, the nominal confusion level at GHz frequency is around few hundreds of $\mu$Jy, so well within current radio-telescope sensitivities. 
High-resolution maps and a proper source subtraction are thus required.

We performed an accurate procedure for the subtraction of small-scale sources.
It has been done by subtracting the Fourier transform of high-resolution sources from the visibility plane.
This procedure allowed to reduce the confusion noise and to gain a factor of few in sensitivity (depending on the target and the related quality of data). The sensitivity was brought closer to its nominal rms value, especially in the cases of Carina and BootesII.
We also described how to possibly further subtract sources in the image plane by means of the SExtractor package~\citep{Bertin:1996fj}.

With the study of radio diffuse components in dSphs, we aim at addressing open questions about the dSph environment (especially for what concerns the magnetic properties) and its activity and CR acceleration mechanisms.
No significant detection of a diffuse emission has been singled out from the ATCA data, and this allowed us to constrain a number of dSph properties.

First, we discussed the general bounds that can be obtained from the radial distributions of the observed surface brightness and the noise. They are approximately at the level of 1 mJy/arcmin$^2$ and can be straightforwardly exploited to constrain models involving a spherical diffuse emission in the observed dSphs (including WIMP-induced emissions).

Assuming some general and analytic functional forms, we derived bounds on the dSph total flux and emissivity, presented in Figs.~\ref{fig:Stot} and \ref{fig:j0}.
They are constrained at the level of about 1 mJy and 0.1 mJy/beam, respectively, for a source size of the order of the stellar profile extent (we remind here that the synthesized beam of the adopted maps is about 1 arcmin$^2$).

Assuming the dominant radio emission in dSphs to be due to synchrotron radiation associated with CR electrons accelerated in processes related to star formation (the DM interpretation is more extensively discussed in Paper III), the SFR of dSphs plays a crucial role in setting the brightness of the emission.
We discussed how to relate both the CR density and the magnetic field strength to the SFR inferred from the observed colour-magnitude diagram in CDS (while for the ultra-faint cases we have to mostly rely on extrapolations).
Although the sensitivity of current observations is above the level of the expected emission, we found that, in the case of CDS, the next-generation of radio telescopes could start probing the presence of a SF-induced diffuse synchrotron radiation.

The derived bounds depend on the magnetic field model, see Fig.~\ref{fig:A0vsB0}. It affects both the size of the radiated synchrotron power and the spatial diffusion of CR electrons.
We accurately modelled the CR transport in dSphs by developing a new numerical scheme based on the Crank-Nicolson algorithm (described in the Appendix).
We found that the impact of diffusion on the expected emission from dSphs can be dramatic. Indeed, due to the small size of dSphs and the probable low level of turbulence, CR electrons can in principle escape the dSph before radiating a significant amount of synchrotron power. 
Future polarization measurements of background sources with the SKA will be crucial to understand the magnetic properties of dSphs and to reduce the degree of uncertainty in the expected signal.

For each dSph, we derived limits on the magnetic field strength under the equipartition assumption. Physical arguments suggest a strength of the magnetic field at the level of $\mu$G (CDS) or a fraction of $\mu$G (UDS), while the equipartition between CR and magnetic density leads to an upper limit of few $\mu$G (see Table~\ref{tableB}).

We also discussed the connection of radio emission to FIR and X-ray observations.
Using the FIR-radio correlation observed for star-forming galaxies, we translated the radio upper limits into bounds for the dSph emission at $70\,\mu$m in Table~\ref{table:det}.
Observations in the X-ray band can probe the IC emission due to scattering with the CMB photons of the same GeV-population of electrons possibly producing a synchrotron radiation at GHz-frequency.
For a magnetic field strength larger than $\sim 0.05\,\mu$G, the current radio bounds are however significantly more constraining. 

To conclude, we presented the first study dedicated to the systematic search for a diffuse radio emission in dSphs making use of interferometric observations.
In this paper, we have shown this technique to be a relevant strategy to be pursued for addressing the puzzling history of dSphs and the fundamental nature of DM.
The discussed analysis pipeline can provide a benchmark case for near-future follow-ups with improved sensitivity, to be undertaken with the SKA and its precursors.

\section*{Acknowledgements}
We thank Piero Ullio for insightful discussions during the early stages of the project. 

S.C. acknowledges support by the South African Research Chairs Initiative of the Department of Science and Technology and National Research Foundation and by the Square Kilometre Array (SKA).
S.P. is partly supported by the US Department of Energy, Contract DE-FG02-04ER41268.
M.R. acknowledges support by the research grant {\sl TAsP (Theoretical Astroparticle Physics)} funded by the Istituto Nazionale di Fisica Nucleare (INFN).

The Australia Telescope Compact Array is part of the Australia Telescope National Facility which is funded by the Commonwealth of Australia for operation as a National Facility managed by CSIRO.

\appendix

\section{Solution for spherically-symmetric transport equation}
\label{sec:app}
Here we describe the numerical solution adopted for the transport equation.
We make a change of variable and consider a spatial logarithmic scale (using $\tilde r=\log (r/r_0)$) to better describe a possible overdensity at the dSph center (mostly motivated by the connection with the possible DM-induced emission described in Paper III). Eq.~\ref{eq:transp} can be thus rewritten (expressing also everything in terms of $E$ instead of $p$) as:
\bea 
\frac{\partial n_e}{\partial t}&=&
 \frac{e^{-2\,\tilde r}}{r_0^2}\left[D( \tilde r, E)\frac{\partial n_e}{\partial \tilde r}+\frac{\partial }{\partial \tilde r}(D( \tilde r, E)\frac{\partial n_e}{\partial \tilde r}) \right] \nonumber \\
 & -&\frac{\partial}{\partial E}(\dot E( \tilde r, E)\,n_e)+
  Q( \tilde r, E) \;.
\label{eq:transpapp}
\eea
Note that (for the moment) we do not consider the stationary limit.

Eq.~\ref{eq:transpapp} has been finite-differenced by means of the Crank-Nicolson scheme:
\bea
\frac{\partial n_i}{\partial t}&=&\frac{n_i^{t+\Delta t}-n_i^t}{\Delta t}=\frac{\alpha_1\,n_{i-1}^{t+\Delta t}-\alpha_2\,n_i^{t+\Delta t}+\alpha_3\,n_{i+1}^{t+\Delta t}}{2\Delta t}\nonumber \\
&+&\frac{\alpha_1\,n_{i-1}^t-\alpha_2\,n_i^t+\alpha_3\,n_{i+1}^t}{2\Delta t}+Q_i\;,
\label{eq:CNdiff}
\eea
which implies a tridiagonal system of equations (and we dropped the subscript $e$ for clarity, i.e., $n\equiv n_e$):
\bea
& &\frac{\alpha_1}{2}\,n_{i-1}^{t+\Delta t}+(1+\frac{\alpha_2}{2})\,n_i^{t+\Delta t}-\frac{\alpha_3}{2}\,n_{i+1}^{t+\Delta t} \nonumber \\
& &=\frac{\alpha_1}{2}\,n_{i-1}^t+(1-\frac{\alpha_2}{2})\,n_i^t+\frac{\alpha_3}{2}\,n_{i+1}^t+Q_i\,\Delta t\;.
\label{eq:CNtrid}
\eea

This kind of numerical method has been adopted for the solution of the transport equation in the Milky Way, e.g., by the publicly available codes GALPROP~\citep{Strong:1998pw} and DRAGON~\citep{Evoli:2008dv}. The main differences in our case are that we have a 2D propagation (in $r$ and $E$) due to the spherical symmetry (instead of 3D or 4D as for the Galaxy) and we consider a logarithmic scale for the spatial grid. For further details on the stability of the generalization of the above described Crank-Nicolson scheme to a multidimensional case, see, e.g., the GALPROP manual\footnote{http://galprop.stanford.edu/download/manuals/galprop\_v54.pdf}.
We apply the so-called ADI (alternating direction implicit) method, in which the implicit updating scheme is alternately applied to the $r$- and $E$-operators in turn, keeping the other coordinate fixed.

The $\alpha$-coefficients for the finite-differencing scheme in $r$ can be derived from:
\bea 
& &\frac{e^{-2\,\tilde r}}{r_0^2}\left[D( \tilde r, E)\frac{\partial n_e}{\partial \tilde r}+\frac{\partial }{\partial \tilde r}(D( \tilde r, E)\frac{\partial n_e}{\partial \tilde r}) \right] \rightarrow \frac{e^{-2\,\tilde r_i}}{r_0^2} \\
& & \times\left[(D +\frac{\partial D}{\partial \tilde r})_{|\tilde r_i}\frac{n_{i+1}-n_{i-1}}{2\,\Delta \tilde r}+D_{|\tilde r_i} \frac{n_{i+1}-2\,n_i+n_{i-1}}{\Delta \tilde r^2} \right]\;,\nonumber
\label{eq:rprop}
\eea
which leads to:
\bea 
\frac{\alpha_1}{\Delta t}&=&\frac{e^{-2\,\tilde r_i}}{r_0^2}\left[-\frac{D +\frac{\partial D}{\partial \tilde r}}{2\,\Delta \tilde r}+\frac{D}{\Delta \tilde r^2} \right]_{|\tilde r_i}\\
\frac{\alpha_2}{\Delta t}&=&\frac{e^{-2\,\tilde r_i}}{r_0^2}\frac{2\,D_{|\tilde r_i}}{\Delta \tilde r^2}\\
\frac{\alpha_3}{\Delta t}&=&\frac{e^{-2\,\tilde r_i}}{r_0^2}\left[\frac{D +\frac{\partial D}{\partial \tilde r}}{2\,\Delta \tilde r}+\frac{D}{\Delta \tilde r^2} \right]_{|\tilde r_i}\;,
\label{eq:alphar}
\eea

where we assumed a constant step $\Delta \tilde r$.
Neumann (Dirichlet) boundary conditions $\partial n_e/\partial \tilde r=0$ ($n_e=0$) has been set at the center $i=0$ (farthest boundary $i=N$). The $\alpha$-coefficients at $i=0$ (i.e., at $\tilde r_{min}$ very close to the center) turn out to be $\alpha_1=0$, $\alpha_2=4\,e^{-2\,\tilde r_{min}}D_{|\tilde r_{min}}/(r_0\,\Delta \tilde r)^2$, and $\alpha_3=\alpha_2$. 

The finite-differencing scheme for the $E$-propagation is analogous to the one adopted in \citep{Strong:1998pw} (see Table 1 of GALPROP manual).

In our runs we typically start with a large time-step $\Delta t=10^{11}$ years and perform a number of iterations to obtain a stable solution  on this large scale (more in detail, we stop when the fractional change of $n_e$ in a time $\Delta t$ is below 0.1\% for each point of the grid). Using such solution as $n^t$ of Eq.~\ref{eq:CNtrid}, we then reduce $\Delta t$ by a factor of 2 and iterate again. This is repeated until $\Delta t=10$ years is reached (which is a time-step much smaller than any time-scales of the process, in particular of energy losses), where we get our final solution. The convergence is ensured by requiring $n_e$ to become constant in time and the time-scale $\tau_c=n_e/(\partial n_e/\partial t)$ to be larger than diffusive and energy loss times-scales at each grid-point (typically $\tau_c>10^{10}$ years).

We also cross-checked our numerical solution against analytic solutions in the cases with only spatial-diffusion terms and with only the energy-loss term, and against the semi-analytic solution which makes use of Green's functions (see, e.g., \citep{Colafrancesco:2005ji,Colafrancesco:2006he}) for the full equation but with spatially constant $D$ and $\dot E$. The advantages of the Crank-Nicolson solution with respect to the latter is given by the much shorter computational time needed and by the possibility of having $D(r)$ and $\dot E(r)$.


\begin{thebibliography}{99}

\bibitem[Abdo et al. 2010]{Abdo:2010}
Abdo et al.\ , 2010, A\&A, 523, A46. 

\bibitem[Appleton et al. 2004]{Appleton:2004sr}
  Appleton P.~N. et al.,
  2004, Astrophys.\ J.\ Suppl.\, 154, 147
  [astro-ph/0406030].

\bibitem[Bertin \& Arnouts 1996]{Bertin:1996fj}
  Bertin E., Arnouts S.,
  1996, Astron.\ Astrophys.\ Suppl.\ Ser.\, 117, 393.

\bibitem[Blandford \& Eichler 1987]{Blandford:1987pw}
  Blandford R., Eichler D.,
  1987, Phys.\ Rept.\, 154, 1.

\bibitem[Blasi 2013]{Blasi:2013rva}
  Blasi P.,
  2013, The Astronomy and Astrophysics Review, 21, 70 [arXiv:1311.7346 [astro-ph.HE]].

\bibitem[Boylan-Kolchin$,$ Bullock \& Kaplinghat 2011]{BoylanKolchin:2011de}
  Boylan-Kolchin M., Bullock J.~S., Kaplinghat M.,
  2011, MNRAS, 415, L40
  [arXiv:1103.0007 [astro-ph.CO]].

\bibitem[Bringmann et al. 2012]{Bringmann:2012vr}
  Bringmann T., Huang X., Ibarra A., Vogl S., Weniger C.,
  2012, JCAP, 1207, 054
  [arXiv:1203.1312 [hep-ph]].

\bibitem[Briggs 1995]{Briggs:thesis}
Briggs D.~S., 1995, PhD Thesis, \
New Mexico Institute of Mining and Technology.

\bibitem[Brown et al. 2013]{Brown:2013xna}
  Brown T.~M. et al.,
  2013, arXiv:1310.0824 [astro-ph.CO].

\bibitem[Bullock$,$ Kravtsov \& Weinberg 2000]{Bullock:2000wn}
  Bullock J.~S., Kravtsov A.~V., Weinberg D.~H.,
  2000, ApJ, 539, 517
  [astro-ph/0002214].

\bibitem[Bullock et al. 2009]{Bullock:2009au}
  Bullock J.~S., et al.,
2009, The Astronomy and Astrophysics Decadal Survey, Science White Papers, no. 32, 
[arXiv:0902.3492 [astro-ph.CO]].

\bibitem[Calura$,$ Lanfranchi \& Matteucci 2008]{Calura:2008bb}
  Calura F., Lanfranchi G. L., Matteucci F.,
  2008, A\&A, 484, 107 
 [arXiv:0801.2547 [astro-ph]].

\bibitem[Carretti et al. 2013]{Carretti:2013sc}
  Carretti E. et al.,
  2013, Nature, 493, 66
  [arXiv:1301.0512 [astro-ph.GA]].

\bibitem[Cesarsky 1980]{Cesarsky:1980pm}
  Cesarsky C.~J.,
  1980, Ann.\ Rev.\ Astron.\ Astrophys.\, 18, 289.

\bibitem[Chyzy et al. 2011]{Chyzy:2011sw}
  Chyzy K.~T., Wezgowiec M., Beck R., Bomans D. J.,
 2011, A\&A, 529, A94
  [arXiv:1101.4647 [astro-ph.CO]].

\bibitem[Colafrancesco$,$ Profumo \& Ullio 2006]{Colafrancesco:2005ji}
  Colafrancesco S., Profumo S., Ullio P.,
  2006, A\&A, 455, 21
  [astro-ph/0507575].

\bibitem[Colafrancesco$,$ Profumo \& Ullio 2007]{Colafrancesco:2006he}
  Colafrancesco S., Profumo S., Ullio P.,
  2007, PRD, 75,  023513
  [arXiv:astro-ph/0607073].

\bibitem[Coleman \& de Jong 2008]{Coleman:2008kk}
  Coleman M.~G., de Jong J.~T.~A.,
  2008, arXiv:0805.1365 [astro-ph].

\bibitem[Cordes:2002wz]{Cordes:2002wz}
  Cordes J.~M., Lazio T.~J.~W.,
  2002, astro-ph/0207156.

\bibitem[deBoer et al. 2012]{deBoer:2012dv}
  de Boer T.~J.~L. et al.,
  2012, A\&A, 539, A103  [arXiv:1201.2408 [astro-ph.CO]].

\bibitem[Dolphin et al. 2005]{Dolphin:2005mv}
  Dolphin A.~E., Weisz D.~R., Skillman E.~D., Holtzman J.~A.,
  2005, [astro-ph/0506430].

\bibitem[Evoli et al. 2008]{Evoli:2008dv}
  Evoli C., Gaggero D., Grasso D., Maccione L.,
  2008, JCAP, 0810, 018
  [arXiv:0807.4730 [astro-ph]].

\bibitem[Gaensler et al. 2005]{Gaensler:2005qj}
  Gaensler B.~M. et al.,
  2005, Science, 307, 1610
  [astro-ph/0503226].

\bibitem[Grcevich \& Putman 2009]{Grcevich:2009gt}
  Grcevich J., Putman M.~E.,
  2009, ApJ, 696, 385
  [arXiv:0901.4975 [astro-ph.GA]].

\bibitem[Hernandez$,$ Gilmore \& Valls-Gabaud 2000]{Hernandez:2000qz}
  Hernandez X., Gilmore G., Valls-Gabaud D.,
  2000, MNRAS, 317, 831
  [astro-ph/0001337].

\bibitem[Jeltema \& Profumo 2008]{Jeltema:2008ax}
  Jeltema T.~E., Profumo S.,
  2008, ApJ, 686, 1045 
 [arXiv:0805.1054 [astro-ph]].

\bibitem[Klein 2012]{Klein:2012} 
Klein, U.\ 2012, Dwarf Galaxies: 
Keys to Galaxy Formation and Evolution, 23.

\bibitem[Longair 2011]{Longair}
  Longair M. S., 2011,
  'High Energy Astrophysics',
  Cambridge University Press.

\bibitem[McConnachie 2012]{McConnachie:2012vd}
  McConnachie A.~W.,
 2012, AJ, 144, 4 
  [arXiv:1204.1562 [astro-ph.CO]].

\bibitem[Mapelli et al. 2007]{Mapelli:2007}
Mapelli, M. et al.,\ 
2007, MNRAS, 380, 1127.

\bibitem[Mapelli et al. 2009]{Mapelli:2009}
Mapelli, M. et al.,\ 
2009, MNRAS, 396, 1771.

\bibitem[Martinez 2013]{Martinez:2013els}
  Martinez G.~D.,
  2013, arXiv:1309.2641 [astro-ph.GA].

\bibitem[Mateo 1998]{Mateo:1998wg}
  Mateo M.,
  1998, Ann.\ Rev.\ Astron.\ Astrophys.\, 36, 435
  [arXiv:astro-ph/9810070].

\bibitem[Mayer 2001]{Mayer:2001yf}
  Mayer L. et al.,
  2001, ApJ, 559, 754
  [astro-ph/0103430].

\bibitem[Momany et al. 2007]{Momany:2007}
Momany, Y. et al.,\ 2007, A\&A, 468, 973.

\bibitem[Monelli et al. 2012]{Monelli:2012}
Monelli, M. et al.,\ 2012, ApJ, 744, 157.

\bibitem[Natarajan et al. 2013]{Natarajan:2013dsa}
  Natarajan A. et al.,
  2013, PRD, 88, 083535
  [arXiv:1308.4979 [astro-ph.CO]].

\bibitem[Orban et al. 2008]{Orban:2008bb}
  Orban C., Gnedin O.~Y., Weisz D.~R., Skillman E.~D., Dolphin A.~E., Holtzman J.~A.,
  2008, ApJ, 686, 1030 [arXiv:0805.1058 [astro-ph]].

\bibitem[Parker 1979]{Parker:1979}
Parker E.~N., \ 1979, Oxford, 
Clarendon Press; New York, Oxford University Press, 1979, 858 p..  

\bibitem[Regis et al. 2014a]{Paper1}
Regis M., Richter L., Colafrancesco S., Massardi M., de Blok W.J.G., Profumo S., Orford N.,
2014, arXiv:1407.5479 [astro-ph.GA] (Paper I).

\bibitem[Regis et al. 2014b]{Paper3}
Regis M., Colafrancesco S., Profumo S., de Blok W.J.G., Massardi M., Richter L.,
2014, JCAP, 1410, 10, [arXiv:1407.4948 [astro-ph.CO]] (Paper III).

\bibitem[Roychowdhury \& Chengalur 2012]{Roychowdhury:2012bc}
  Roychowdhury S., Chengalur J.~N.,
2012, MNRAS, 423, L127 
  [arXiv:1204.3305 [astro-ph.CO]].

\bibitem[Rybicki \& Lightman 1979]{Rybicki}
  G. Rybicki and A.P. Lightman, 1979,
  'Radiative Processes in Astrophysics',
  John Wiley \& Sons Inc.

\bibitem[Sand et al. 2009]{Sand:2009ut}
  Sand D.~J. et al.,
  2009, ApJ, 704, 898
  [arXiv:0906.4017 [astro-ph.CO]].

\bibitem[Sault$,$ Teuben \& Wright 1995]{Sault:95}
Sault R.~J., Teuben P.~J., Wright M.~C.~H., 
1995, Astronomical Data Analysis Software and Systems IV, 77, 433.

\bibitem[Sault \& Wieringa 1994]{Sault:94}
Sault R.~J., Wieringa M.~H., \ 
1994, Astron.\ Astrophys.\ Suppl.\, 108, 585.

\bibitem[Spekkens et al. 2013]{Spekkens:2013ik} 
  Spekkens K., Mason B.~S., Aguirre J.~E., Nhan B.,
  2013, ApJ, 773, 61
  [arXiv:1301.5306 [astro-ph.CO]].

\bibitem[Strong \& Moskalenko 1998]{Strong:1998pw}
  Strong A. W., Moskalenko I. V.,
  1998, ApJ, 509, 212
  [astro-ph/9807150].

\bibitem[Tollerud et al. 2008]{Tollerud:2008ze}
  Tollerud E.~J., Bullock J.~S., Strigari L.~E., Willman B.,
  2008, ApJ, 688, 277
  [arXiv:0806.4381 [astro-ph]].

\bibitem[Tolstoy$,$ Hill \& Tosi 2009]{Tolstoy:2009jb}
  Tolstoy E., Hill V., Tosi M.,
  2009, Ann.\ Rev.\ Astron.\ Astrophys.\, 47, 371
  [arXiv:0904.4505 [astro-ph.CO]].

\bibitem[Walker 2013]{Walker:2012td}
  Walker M.~G.,
  2013, Planets, Stars and Stellar Systems.~Volume 5: Galactic Structure and Stellar Populations, 1039
 [arXiv:1205.0311 [astro-ph.CO]].

\bibitem[Weisz et al. 2014]{Weisz:2014} 
Weisz, D.~R. et al.,\ 
2014, ApJ, 789, 147.

\bibitem[Weinberg et al. 2013]{Weinberg:2013aya}
  Weinberg D.~H. et al.,
  2013, arXiv:1306.0913 [astro-ph.CO].

\bibitem[Windhorst$,$ Mathis \& Neuschaefer 1990]{Windhorst:1990}
  Windhorst R. A., Mathis D., Neuschaefer L., 
  1990, ASP Conf. Ser., Evolution of the universe of galaxies, p. 389.

\bibitem[Zang \& Meurs 2001]{Zang:2001}
 Zang Z., Meurs E.~J.~A.,
2001, ApJ, 556, 24.


\end{thebibliography}
\end{document}